\documentclass[sts]{imsart}

\RequirePackage{amsthm,amsmath,amsfonts,amssymb}
\RequirePackage[numbers]{natbib}
\RequirePackage[colorlinks,citecolor=blue,urlcolor=blue]{hyperref}
\RequirePackage{graphicx}
\usepackage{pifont}
\usepackage{color}
\usepackage{subfig}
\usepackage{epstopdf}
\usepackage{natbib}
\usepackage{bm}
\usepackage{enumerate}
\usepackage{caption}

\startlocaldefs

\newtheorem{definition'}[theorem]{Definition}

\newcommand{\iidsim}{\overset{iid}{\sim}}

\newcommand{\Norm}{\text{N}} 
\newcommand{\Unif}{\text{U}} 
\newcommand{\DP}{\text{DP}} 
\newcommand{\GP}{\text{GP}} 

\newcommand{\1}{\mathbf{1}} 

\newcommand{\one}{\mathbf 1\,}

\endlocaldefs

\begin{document}

\begin{frontmatter}
\title{Bayesian dependent mixture models: A predictive comparison and survey}
\runtitle{Bayesian dependent  mixture models: A predictive comparison and survey}

\begin{aug}
\author[A]{\fnms{Sara}~\snm{Wade}\ead[label=e1]{sara.wade@ed.ac.uk}},
\author[B]{\fnms{Vanda}~\snm{In\'acio}\ead[label=e2]{vanda.inacio@ed.ac.uk}}
\and
\author[C]{\fnms{Sonia}~\snm{Petrone}\ead[label=e3]{sonia.petrone@unibocconi.it}}
\address[A]{Sara Wade is Lecturer (Assistant Prof.), School of Mathematics,
	University of Edinburgh, Edinburgh, UK \printead[presep={\ }]{e1}.}
\address[B]{Vanda In\'acio is Lecturer, School of Mathematics, University of Edinburgh, Edinburgh, UK\printead[presep={\ }]{e2}.}
\address[C]{Sonia Petrone is Full Professor, Department of Decision Sciences, Bocconi University, Milano, Italy \printead[presep={\ }]{e3}.}
\end{aug}

\begin{abstract}
For exchangeable data, 
mixture models are an extremely useful tool for density estimation due to their attractive balance between smoothness and flexibility. When additional covariate information is present, mixture models can be extended for flexible regression by modeling the mixture parameters, namely the weights and atoms, as functions of the covariates. 
These types of models are interpretable and highly flexible, allowing non only the mean but the whole density of the response to change with the covariates, which is also known as density regression. 
This article reviews Bayesian covariate-dependent mixture models 
and highlights which data types can be accommodated by the different models along with the methodological and applied areas where they have been used. 
In addition to being highly flexible, these models are also 
numerous; we focus on nonparametric constructions 
and broadly organize them into  three categories: 1) joint models of the responses and covariates, 2) conditional models with single-weights and covariate-dependent atoms, and 3) conditional models with covariate-dependent weights. The diversity and variety of the available models in the literature raises the question of how to choose among them for the application at hand. We attempt to shed light on this question through a careful analysis of the predictive equations for the conditional mean and density function as well as predictive comparisons in three simulated data examples. 
\end{abstract}


\end{frontmatter}

\section{Introduction}\label{sec:intro}
Advances in data acquisition have led to numerous 
challenges for modern data and statistical analysis. In a supervised context with the aim of studying the relationship between the response variables and covariates, such challenges include 
high-dimensionality, mixed non-Gaussian data types, structured dependence, nonlinearity, and more. 
While the linear regression model is the standard tool in supervised 
settings due to its simplicity, ease of interpretation, straightforward computations, and desirable asymptotic properties, it cannot cope with such challenges, leading to inadequate fitting of the data and poor predictive inference.

To relax the linearity assumption, a flexible approach consists in representing the regression function as a linear combination of (adaptive) basis functions. Indeed, most standard nonparametric methods, such as splines \citep[see, e.g.,][for book-length reviews]{Denison,Wood2017}, 
wavelets \cite{V09}, neural networks \cite{Neal96,goodfellow2016deep}, regression trees \cite{BFOS84,CGM10}, kernel regression \cite[][Chapter 8]{Scott2015}, and Gaussian processes \cite{Ras}, can be represented in this fashion. Such methods can potentially approximate a wide range of regression functions,  
yet are also limited in the sense that they only allow for flexibility in the regression function. Extensions to location-scale regression models where both the mean and variance are covariate-dependent and flexibly modeled have also been considered \citep[e.g.,][among many others]{Rodriguez2014,Pratola2020}. 
Alternatively, quantile regression includes covariate dependence for specified quantiles  \citep[e.g.,][both from a Bayesian viewpoint]{Reich2010,Waldmann2013}.
When trying to go beyond the the notion that the effect of the covariates is restricted to change some particular functional(s) of the response variable distribution, \emph{density regression} arises as a natural option. Under such an approach, the entire density of the response variable is allowed to change as a function of the covariates. Further, and importantly in practice, by using a density regression model, all inferences are coherent (in opposition to using different approaches to analyse different functionals, e.g. multiple quantiles). 
The need for this flexibility afforded by density regression models is evident in many modern datasets, which present nonstandard features, such as non-Gaussianity, multi-modality, or skewness and tail behavior, that may change across the covariate space.

To achieve flexible density regression, mixture models are attractive tools. 
They are commonly used for density estimation 
due to their ability to approximate a large class of densities and their attractive balance between smoothness and flexibility in modeling local features. When additional covariate information is present, mixture models can be extended for density regression in one of two ways. The first approach, termed the \textit{joint approach}, is closely related to classical kernel regression methods and involves  modeling the joint density of the response and covariates with a mixture model. 
The second approach, called the \textit{conditional approach}, directly models the conditional density by 
allowing the mixing distribution, namely the mixture weights and atoms, to depend on the covariates. 
 Conditional models are often referred to as dependent mixture models in statistics and are also known as mixtures of experts in machine learning (\cite{Jacobs91},\cite{JJ94} and Chapter~12 of \cite{Fruhwirth2019} for a recent review) or smooth mixtures of regressions in econometrics \citep{GK07}.

\begin{figure}[!t]
 \begin{center}
  \subfloat[Kernel density estimates]{\includegraphics[width=0.5\textwidth]{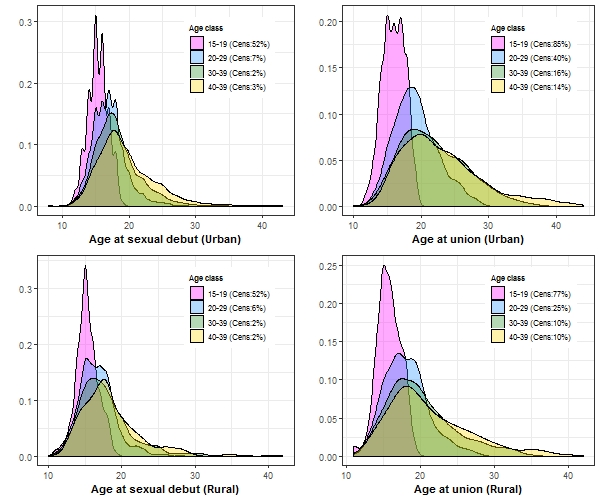}\label{fig:kde}} \\
  \subfloat[Spline regression]{\includegraphics[width=0.5\textwidth]{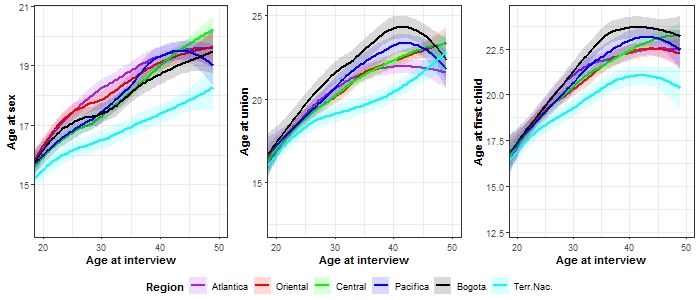}\label{fig:spline} }\\
  \subfloat[Joint relationship]{\includegraphics[width=0.5\textwidth]{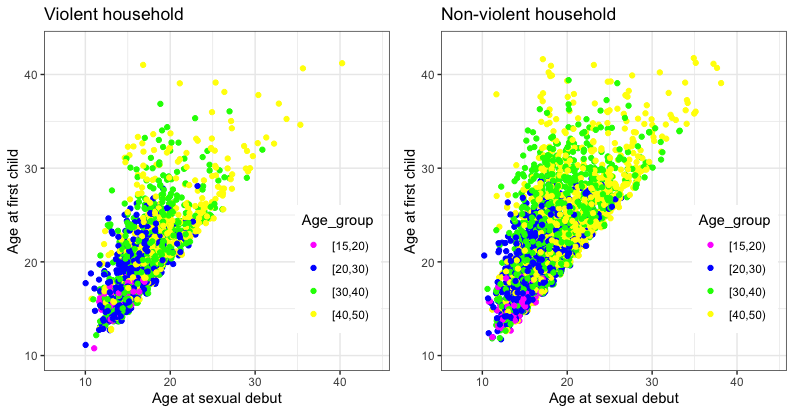}\label{fig:scatter} }
  \caption[]{Exploratory analysis of Colombian women data: (a) Kernel density estimate of (non-censored) ages at events conditioned on cohort (i.e. age at interview, in groups) and area of residence (urban or rural); (b) Relation (smoothed regression) between (non-censored) ages at events and age at interview; (c) Joint relationship between age at sexual debut and age at first child conditioned on cohort and violent upbringing.}
  \label{fig:descr_densities}
   \end{center}
\end{figure} 

A compelling application is presented in \cite{wade2022colombian} that aims to study Colombian women's life choices, in particular, women's fertility and partnership history and its interplay with employment given background information related to their family of origin (e.g. region of residence, type of area, disciplining methods, presence of domestic violence). Such a study is important to identify and quantify critical situations and help in planning targeted interventions to improve the welfare of women, especially in a state such as Colombia which has experienced ongoing conflict since 1948. However, the data (from the Demographic and Health Survey (DHS)  2010, \url{http://www.dhsprogram.com/}) present challenges to modeling and analysis. Specifically, the mixed multivariate response includes both binary variables (employment status) and ages at event that are subject to censoring and constraints, and the mixed covariates contain numerical and categorical variables. Moreover, an exploratory analysis (Figure \ref{fig:descr_densities}) highlights the need for an approach that can capture varying right-skewness in the ages-at-event depending on the covariates (Figure \ref{fig:kde}), nonlinearity (Figure \ref{fig:spline}), and the non-Gaussian joint relationship between the ages-at-event that also varies with covariates (Figure \ref{fig:scatter}). Dependent mixture models provide the flexibility required to capture such behavior as well as many other challenges of modern, complex datasets.

 In this article we provide an overview of the various proposals of dependent mixture models, focusing on Bayesian approaches and extensions based on the Dirichlet process, and how such models can be adapted to the variety of data types encountered in modern applications. 
 The literature on this subject is rich but somewhat fragmented; thus our aim is to provide a contribution to the subject by unifying existing literature. Due to the numerous constructions of dependent mixture models, choosing among them for the application at hand can be a daunting task. Ideally, the chosen model should have good approximation properties to a large class of data-generating covariate-dependent densities and posterior consistency properties. These types of properties are explored for specific models based on the joint approach \cite{HBP} and the conditional approach \cite{Bar,Norets,pati2013posterior}. 
Posterior consistency is an interesting frequentist property that should be minimally satisfied;
however, it studies the behavior of the random conditional densities as the sample size goes to infinity. In practice, the sample size is finite, and a study of posterior consistency properties may hide what happens in the finite case.
This is a general theoretical issue, and it raises an important question: how do we choose among the different proposals of nonparametric models and priors from a \textit{Bayesian} perspective? 
We aim to shed light on this issue by adopting a natural approach from a Bayesian perspective that consists of a detailed study of properties based on \textit{finite} samples.  
In particular, we carefully examine features of the model and prior and their effects on the predictive mean and density estimate and the corresponding uncertainty for some new covariate values. 
In addition, we provide a comparative study of the predictive performance of existing models, including advantages and disadvantages depending on specific aspects of the observed data. This is important to aid researchers and practitioners in selecting and constructing models with efficient estimators and improved prediction. These two aspects greatly distinguish our work from the the recent review article of \cite{Quintana2022} on the dependent Dirichlet process and related models. Our work also has a greater focus on which data types can be accommodated by the different model constructions and the methodological and applied areas where they have been used. 

The outline of this article is as follows. We begin with a review of Bayesian mixture models, followed by a review of extensions for density regression, providing a unifying framework for the models of interest. Throughout, we highlight how modern data types and challenges can be accommodated. As this section clearly shows, the number of proposals and model choices is large and varied. Thus, to decide among the various choices in practice, a detailed understanding of properties of these models is needed. The next section is devoted to a predictive comparison of the methods through simulated data examples. Finally, we provide a final discussion and directions for future research.

\section{From finite to infinite mixture models}\label{sec:mixtures}
The form of a mixture model is given by
\begin{equation}\label{mm1}
f(y\mid P) = \int k(y;\theta)\text{d}P(\theta),
\end{equation}
where the mixing measure $P$ is a probability measure on the parameter space $\Theta$, $k(y;\theta)$ is a fixed parametric probability mass or density function, often referred to as the kernel, defined on  $\mathcal{Y}\times\Theta$, with $\mathcal{Y}$ denoting the sample space. Note that the kernel may be univariate or multivariate and may also contain a global parameter common to all mixture components (e.g., the scale parameter in a location mixture of univariate normal distributions), but here for notation simplicity, it is omitted. Throughout this article, we consider the case when the mixing measure  $P$ is discrete, i.e.,
\begin{equation*}
P = \sum_{j=1}^{J}\omega_j\delta_{\tilde{\theta}_j},
\end{equation*}
where the atoms $\{\tilde{\theta}_j\}_{j=1}^{J}$ take values in $\Theta$ and the weights $\{w_j\}_{j=1}^{J}$ are non-negative, sum to one, and represent the probability of belonging to each mixture component. The mixture model can then be expressed as a convex combination of kernels
\begin{equation}\label{mm2}
f(y\mid P) = \sum_{j=1}^{J}\omega_j k(y;\tilde{\theta}_j).
\end{equation}
Obviously, the nature of $f$ will depend on the nature of the kernel, and the choice of an appropriate kernel depends on the underlying sample space. If the underlying density function is defined on the whole real line, a normal kernel is the most popular choice, whereas a (skew) $t$  or skew-normal distribution may provide robustness to outliers and asymmetry 
\citep{fruhwirth2010bayesian, lee2014finite}. On the positive half line, mixtures of gamma, Weibull or lognormal distributions are a possibility, while on the unit interval, a beta kernel may be used. For discrete sample spaces, 
mixtures of Bernoulli or multinomial distributions, known as latent class models, are appropriate for categorical data  \citep{goodman1974exploratory,blei2003latent}, and for ordinal data, latent variable approaches based on a logistic or probit transformation may be used \citep{kottas2005nonparametric,deyoreo2018bayesian}. For count data, kernel choices include Poisson  \citep{karlis2005mixed, krnjajic2008parametric}, negative-binomial  \citep{wu2019nonparametric,liu2022}, and rounded continuous kernels \citep{canale2011bayesian,canale2017robustifying}. 
Mixed data of different types can be modeled, for instance, by assuming conditional independence and combining appropriate kernels through a product operation  or through a latent variable approach  \citep{cai2011mixture, norets2018adaptive}.

 The unknown parameter in the mixture model formulation is the mixing measure $P$ and placing a prior distribution on $P$ is equivalent to placing a prior distribution on its constituents. Broadly speaking, there are three classes of models, depending on whether (i) $J$ is finite and known, (ii) $J$ is finite and unknown, and (iii) $J$ is infinite. In any case, the prior for the mixing measure $P$ induces a prior on the density $f(y
\mid P)$.

\indent Let us start by case (i) where the number of mixture components, $J$, is fixed.  In this case, the prior consists of a prior distribution on the collections of weights $(\omega_1,\ldots,\omega_{J})$ and atoms $(\tilde{\theta}_1,\ldots,\tilde{\theta}_J)$. 

Typically, for conjugacy reasons, the prior on $\omega$ is chosen to be a Dirichlet distribution with parameter vector $(\gamma_1,\ldots,\gamma_J)$, with a usual choice being $\gamma_1 =\ldots =\gamma_J = \gamma$, where a small $\gamma$ value encourages sparsity in the weights, and in the extreme case when $\gamma \rightarrow 0$, all prior mass is placed on the vertices of the simplex, with all weight on a single component. Model selection tools or information criteria can be used to compare the resulting mixture models under different choices of $J$ and thus to select the most appropriate number of mixture components. An alternative approach is to use the so-called overfitted mixtures \citep{Rousseau2011} where the idea is to saturate the model with a large number of components $J$, which can be regarded as an upper bound on the number of occupied mixture components or clusters. The problem with a large $J$ value is that different components that are very similar and hence redundant may be introduced, leading to a degrading of the model performance. Some form of sparsity is therefore essential in order to effectively regularise and prune the extra, redundant, components. With this in mind, \cite{Rousseau2011} propose a prior distribution for the weights that is still a Dirichlet distribution but the 
values of $\gamma_1,\ldots,\gamma_J$ are specified in such a way that the resulting distribution favours either emptying or merging the extra redundant components.

In turn, the atoms $\{\tilde{\theta}_j\}_{j=1}^{J}$ are typically assumed to be independently and identically distributed (iid) from a base measure, say $P_0$.
A popular choice for $P_0$ is the conjugate prior to the kernel, which has the main advantage of computational convenience. The hyperparameters of the base measure $P_0$ can either be specified subjectively based on prior knowledge of the component-specific parameters; set hierarchically, inferred with additional hyperpriors; or set empirically, being data-dependent. An exception to the case of iid atoms 
is considered in \cite{Petralia2012}, where 
a joint prior for the atoms is proposed that introduces dependence among them; 
the resulting class of repulsive mixtures only place components close together if it results in a substantial improvement in model fit. Regardless of whether the atoms are iid from the base measure or not, the finite mixture  can be equivalently written in a hierarchical way. Let $(y_1,\ldots,y_n)$ be the data and let $(\theta_1,\ldots,\theta_n)$ be continuous latent subject-specific parameters. The model in \eqref{mm2} can be hierarchically written as 
\begin{align*}
&y_i\mid\theta_i\overset{\text{ind.}} \sim k(y_i;\theta_i),\\
&\theta_i\mid P \overset{\text{iid}} \sim P,\quad i=1,\ldots,n,\\
& P = \sum_{j=1}^{J}\omega_j\delta_{\tilde{\theta}_j},\\
&(\omega_1,\ldots,\omega_J)\sim \text{Dirichlet}(\gamma_1,\ldots,\gamma_J),\\
&\tilde{\theta}_j\sim P_0,\quad j = 1,\ldots, J.
\end{align*}
Instead of introducing the latent parameters $\theta_i$, one may equivalently rewrite the finite mixture model in terms of latent discrete allocation variables, say $s_i\in\{1,\ldots,J\}$, for $i=1,\ldots, n$, with $s_i=j\Leftrightarrow \theta_i= \tilde{\theta}_j$, that is, if $s_i=j$ then observation $y_i$ belongs to the $j$th mixture component which is parametrized by $\tilde{\theta}_j$. Hierarchically, we can express this as
\begin{align*}
&y_i \mid\tilde{\theta}_1,\ldots, \tilde{\theta}_J, s_i \overset{\text{ind.}}\sim k(y_i;\tilde{\theta}_{s_i}),\quad i=1,\ldots,n,\\
&\Pr(s_i=j\mid \omega_1,\ldots, \omega_J)=\omega_j,\quad j = 1,\ldots, J,\\
& (\omega_1,\ldots,\omega_J)\sim \text{Dirichlet}(\gamma_1,\ldots,\gamma_J),\quad \tilde{\theta}_j\sim P_0.
\end{align*}
If we marginalise over $\theta_i$ (in the first case) or $s_i$ (in the second case), for $i=1,\ldots, n$, we recover the mixture formulation in \eqref{mm2}.

On the other hand, in case (ii), the number of mixture components $J$ is unknown and therefore a prior distribution is placed on it (see, among many others, \cite{Richardson1997}, \cite{Nobile2007}, and \cite{Miller2018}). 
In this case, the collection of unknown parameters also include $J$ and the prior distribution is constructed hierarchically as follows
\begin{align*}
&J\sim p(J),\\
&\omega_1,\ldots, \omega_J\mid J \sim \text{Dirichlet}(\gamma_1,\ldots,\gamma_J),\\
&\tilde{\theta}_1,\ldots,\tilde{\theta}_J\mid J \sim P_0.
\end{align*}
Such a model is often referred to as a mixture of finite mixtures \citep{Miller2018}. Possible  prior distributions for the number of components are, for instance, a Poisson, a (discrete) uniform, or a geometric distribution. Obviously, conditional on the value of $J$, the model can also be written hierarchically as in case (i) where the number of components is fixed. Posterior inference is typically carried out using reversible-jump Markov chain Monte Carlo (MCMC) algorithms but these can be difficult to implement efficiently in practice. Recently, \cite{Miller2018} showed that many of the essential properties of Dirichlet process mixtures are also exhibited by mixture of finite mixtures and therefore the powerful methods developed for posterior inference in Dirichlet process mixtures (see more at the end of this section) can also be directly applied to these class of models, simplifying their computational implementation.
It is worth mentioning that extensions of repulsive priors to the case where $J$ is unknown have also been proposed (see, for instance, \cite{Xie2020}).

Finally, in case (iii), we have infinite mixture models which correspond to $J=\infty$. Unarguably, the Dirichlet process (DP) \citep{Ferg,Ferguson1974} is the most commonly used prior for $P$ in the Bayesian nonparametric literature and has many desirable properties including easy elicitation of its parameters, conjugacy, large support, and posterior consistency \citep[][Chapter 4]{Ghosal2017}. Consequently, here we focus on the DP, providing an overview of its properties and constructions which form the basis of extensions for the dependent mixtures in Section \ref{sec:depmix}.  Of course, other nonparametric priors \citep{LP10} may also be considered, and in fact, many of these extensions in Section \ref{sec:depmix}  include priors beyond the DP.

The DP is characterised by two parameters: a positive scalar parameter, $\alpha$, and a distribution on $\Theta$, $P_0$. We write $P\sim\text{DP}(\alpha,P_0)$ to denote that $P$ follows a DP prior. For any measurable set $A$ of $\Theta$, the following holds
\begin{align*}
E\{P(A)\} & = P_0(A),\\
\text{Var}\{P(A)\} &= \frac{P_0(A)\{1-P_0(A)\}}{\alpha + 1},
\end{align*}
and hence the interpretation of $P_0$ as the centring or base distribution and $\alpha$ as the precision parameter. 
Realisations from a DP are discrete distributions with probability one, even if $P_0$ is continuous. This becomes immediately evident from the constructive definition of the DP as a stick-breaking process \cite{Sethuraman94}. 
Any $P\sim\text{DP}(\alpha, P_0)$ can be represented as
\begin{equation*}
P=\sum_{j=1}^{\infty}\omega_j\delta_{\tilde{\theta}_j},
\end{equation*}
where the atoms $\tilde{\theta}_j$ are generated from the base distribution $P_0$, that is,
\begin{equation*}
\tilde{\theta}_j\overset{\text{iid}}\sim P_0,
\end{equation*}
independently from the weights $\omega_j$, where
\begin{equation*}
\omega_1 = v_1,\quad
\omega_j = v_j\prod_{j^{\prime} < j}(1-v_{j^{\prime}}), \quad v_j\overset{\text{iid}}\sim \text{Beta}(1,\alpha).
\end{equation*}
More general stick-breaking constructions are reviewed and given in \cite{IJ}.

Since $P$ is discrete with probability one, this implies ties among the $\theta_i\overset{\text{iid}}\sim P$. 
Let $k_n$ denote the number of unique values among the 
($\theta_1,\ldots,\theta_n$) and let $(\theta_1^*, \ldots, \theta^{*}_{k_n})$ denote the unique values. In the stick-breaking representation, $(\theta_1^*, \ldots, \theta^{*}_{k_n})$ correspond to $k_n$ different values of $\tilde{\theta}_j$, drawn from $P_0$, where a $\tilde{\theta}_j$ with large $\omega_j$ has better chances to be among the $(\theta_1^*, \ldots, \theta^{*}_{k_n})$. The predictive distribution of the latent subject-specific parameters is given by the P\'olya urn scheme \citep{BM},
\begin{align}
\mathbf{\theta}_1 &\sim P_0, \nonumber \\
\mathbf{\theta}_{n+1} \mid \theta_1,\ldots,\theta_n &\sim \frac{\alpha}{\alpha+n}P_0+ \sum_{j=1}^{k_n} \frac{n_{n,j}}{\alpha+n} \delta_{\theta_j^*}, \label{eq:urn}
\end{align}
where $n_{n,j}=\sum_{i=1}^n \1(\theta_i=\theta_j^*)$ is the number of `observations' that are equal to the $j$th unique value. For ease of notation, we drop the subscript $n$ from $(k_n, n_{n,j})$ when the sample size is understood. An existing value $\theta_{j}^{*}$ will be drawn for $\theta_{n+1}$ with probability proportional to $n_j$, while a new value will be drawn from $P_0$ with probability proportional to $\alpha$. A popular metaphor, the Chinese restaurant process, essentially describes the same model as the Polya urn.

Random partition models define the distribution of the partition of $n$ subjects into $k$ clusters (see \cite{Q1}). The DP implicitly defines a random partition model, through the joint distribution of the latent allocation variables $ (s_1,\ldots,s_n)= \rho_n$, where, with a slight abuse of notation, we use the same notation $s_i=j$ in this case to denote that $\theta_i$ is equal to $j$th unique value observed $\theta_j^{*}$, for $i=1,\ldots,n$ and $j=1,\ldots, k$. The Polya urn characterization of the DP implies that
\begin{align*}
p(\rho_n)= \frac{\Gamma(\alpha)}{\Gamma(\alpha+n)}\alpha^k\prod_{j=1}^{k}  \Gamma(n_{j}),
\end{align*}
where we highlight that due to assumptions of exchangeability and invariance with respect to cluster labels, the prior on the partition only depends on the latent allocation variables $s_1,\ldots,s_n$ through the cluster sizes $n_1,\ldots,n_k$.

In model \eqref{mm1} when $P\sim\text{DP}(\alpha, P_0)$, the resulting model is known as a Dirichlet process mixture and this type of model was first introduced and studied by \cite{Lo}. As in cases (i) and (ii), the model can also be written hierarchically in a similar way, i.e., 
\begin{align*}
&y_i\mid\theta_i\overset{\text{ind.}} \sim k(y_i;\theta_i),\\
&\theta_i\mid P \overset{\text{iid}} \sim P,\quad i=1,\ldots,n,\\
&P\sim\text{DP}(\alpha, P_0).
\end{align*}
Integrating out the $(\theta_1, \ldots, \theta_n)$, we have that given $P$, the $y_i$ are independent with density
\begin{equation}\label{eq:dp_dens}
f(y\mid P)= \int_{\Theta} k(y; \theta) dP(\theta)= \sum_{j=1}^\infty w_j k(y; \tilde{\theta}_j). 
\end{equation}

As noted for instance in \cite{Neal2000}, DP mixtures can equivalently be obtained by taking the limit as $J$ goes to infinity of a finite mixture model with $J$ components where the weights are assigned a prior of the form
\begin{equation*}
(\omega_1,\ldots,\omega_J)\sim\text{Dirichlet}(\gamma/J,\ldots,\gamma/J).
\end{equation*}
The DP mixture model in \eqref{eq:dp_dens} for density estimation is very flexible and it combines the nice features of mixture modeling with strong theoretical properties of nonparametric priors. In particular, posterior consistency of DP mixture models for univariate density estimation is studied in \cite{GGR99,GV01,GV07,T06,WLP07,Petrone2010}. Results for multivariate density estimation appear later in \cite{WG08,WG10,T11}. A variety of samplers for efficient posterior simulation have been proposed over the years. MCMC approaches include: (a) algorithms relying on the Polya urn representation 
(e.g., \cite{Neal2000}), (b) algorithms based on the stick-breaking representation of the DP that truncate the infinite sum to a finite value \citep{IJ}, (c) retrospective sampling techniques \citep{Papaspiliopoulos2008}, and (d) slice sampling methods (e.g., \cite{Kalli2011}). Strategies (c) and (d) avoid infinite computations without deterministically truncating the stick-breaking representation as in (b). Methods beyond MCMC techniques have also been proposed. For example, \cite{Carvalho2010} developed particle learning methods for estimation of general mixtures, including DP mixtures whereas, in an attempt to scale DP mixture models to large volumes of data, variational approximations were proposed originally in \cite{Blei2006}.

\section{Dependent mixture models}\label{sec:depmix}

\subsection{Joint modeling approach} \label{sec:joint}
A simple extension of  mixture models for density estimation to covariate-dependent density estimation augments the observations to include the covariates. Let $y$ denote a univariate response variable and let $x \in \mathcal{X}$ be a p-dimensional vector of covariates (note that the methodology can also be applied to multivariate responses). The joint  density of $y$ and $x$ is modeled  through
\begin{equation}\label{jointdensity}
f(y,x\mid P) =  \int k(y,x;\theta)\text{d}P(\theta), 
\end{equation}
where $k(y,x;\theta)$ is an appropriate kernel density. For example, assuming a DP for the random mixing measure, $P\sim\text{DP}(\alpha,P_0)$, we can write the joint density as

\begin{equation*}
f(y,x\mid P) = \sum_{j=1}^{\infty} \omega_j k(y,x;\tilde{\theta}_j),
\end{equation*}
where $\tilde{\theta}_j\overset{\text{iid}}\sim P_0$, independent of the weights that arise from the stick-breaking construction. 
Inference is carried out  for the joint density, through any of the available samplers for posterior simulation for DP mixture models, and conditional density estimates are obtained as a by-product. 
In particular, the model for the conditional response density can be written as
\begin{align*}
f(y\mid x, P) & = \frac{f(y,x\mid P)}{f(x\mid P)} = \frac{\sum_{j=1}^{\infty} \omega_j k(y,x;\tilde{\theta}_j)}{\sum_{j^{\prime}=1}^{\infty} \omega_{j^{\prime}} k(x;\tilde{\theta}_{j^{\prime}})}\\
& = \sum_{j=1}^{\infty} \omega_j^{*}(x)k(y\mid x;\tilde{\theta}_j),
\end{align*}
where 
\begin{equation}
\omega_j^{*}(x) = \frac{\omega_jk(x; \tilde{\theta}_j)}{\sum_{j^{\prime} =1}^{\infty} \omega_{j^{\prime}}k(x; \tilde{\theta}_{j^{\prime}})}. \label{eq:joint_dw}
\end{equation}
Thus the model for the joint density in \eqref{jointdensity} implicitly defines a model for the conditional response density which admits a representation as a mixture of the conditional response kernel densities with covariate-dependent mixture weights. We note that this approach is more meaningful if the covariates can be considered as random variables, and can problematic for fixed covariates, for instance, binary treatment allocation variables in clinical trial studies. A practical appealing feature of this approach is that covariates with values missing (completely) at random can be easily handled through an extra simple step of imputing these missing values from the marginal distribution of the covariates, during the MCMC algorithm. Of course, the same is also true if the response contains missing values but this is not distinctive of this approach as it can also be easily handled by approaches that target the conditional distribution of the response directly (as in Section \ref{sec:cond_appr}).

The mean regression function implied by the joint model is given by
\begin{equation*}
E(Y\mid x, P) = \sum_{j=1}^{\infty}\omega_j^{*}(x)  E(Y\mid x, \tilde{\theta}_j),
\end{equation*}
where $E(Y\mid x, \tilde{\theta}_j)$ is the conditional mean of the $j$th component.
Analogous expressions can be derived for the conditional variance and quantile functions. This approach was first introduced by \cite{MEW}, who assumed a multivariate normal kernel within component for a continuous response and continuous covariates and use a DP prior for $P$. Note that in this case, the conditional mean of each component, $E(Y\mid x, \tilde{\theta}_j)$, is a linear regression function and the fact that the weights are covariate dependent, leading to a locally weighted mixture of linear regressions, is key to allow estimation of nonlinear regression relationships and general density shapes for the conditional response distribution.

\paragraph*{Predictive structure.} In the supervised setting, our aim is prediction of the response given a new covariate value $x_{n+1}$ and the data $\mathcal{D} = \lbrace(x_1,y_1), \ldots, (x_n,y_n)\rbrace$. 
To shed further insight on this predictive distribution, 
it is helpful to integrate out the unknown mixing measure $P$ and parameterize in terms of the random partition $\rho_n$. For notational simplicity, we focus on the multivariate normal kernel, which we rewrite as a marginal normal kernel for $x$ and a normal linear regression kernel for $y$ given $x$; we also assume the base measure $P_0$ is the conjugate prior, in order to analytically marginalize the  cluster-specific parameters $(\theta_1^*,\ldots, \theta_{k}^*)$. The predictive distribution is based on a covariate-dependent urn scheme, such that
conditioned on the partition $\rho_n$ and $(x_1,\ldots, x_{n+1})$, the cluster allocation $s_{n+1}$ of a new subject with covariate $x_{n+1}$  is determined as
\begin{align*}
    s_{n+1}\mid \rho_n,x_{1:n+1} \sim \frac{\omega_{k+1}(x_{n+1})}{c_0} \delta_{k+1} +\sum_{j=1}^{k} \frac{\omega_j(x_{n+1})}{c_0} \delta_j,
\end{align*}
where 
\begin{align*}
 \omega_{k+1}(x_{n+1}) &= \frac{\alpha}{\alpha+n}\int k(x_{n+1}; \theta) d P_0(\theta), \\
\omega_j(x_{n+1}) &= \frac{n_j}{\alpha + n} \int k(x_{n+1}; \theta)  d P_0(\theta | x_j^*) , 
\end{align*}
with $x_j^* = \lbrace  x_i : s_i = j \rbrace$ containing the covariates in cluster $j$,
$P_0(\theta | x_j^*)$ representing the posterior of  $\theta_{j}^*$ in cluster $j$, and $c_0 = p(x_{n+1}|\rho_n, x_{1:n})$ being the normalizing constant.  This generalizes the P\'olya urn scheme in \eqref{eq:urn} by allowing the cluster allocation probability to depend on the covariates. From this covariate-dependent urn scheme, the predictive distribution of the response is obtained. The predictive mean for the response given a new covariate value $x_{n+1}$ is:
\begin{align}
 &E(Y\mid x_{n+1},  \mathcal{D}) = \sum_{\rho_n} \sum_{s_{n+1}=1}^{k +1}  E(Y\mid x_{n+1},  \mathcal{D}, \rho_n, s_{n+1}) \nonumber \\
 &\quad \quad \quad \quad \quad \quad \quad \quad p(s_{n+1}| x_{1:n+1}, \rho_n) p(\rho_{n}|\mathcal{D}, x_{n+1}) \nonumber \\
 &\quad = \sum_{\rho_n} \left(\sum_{s_{n+1}=1}^{k +1}   \frac{\omega_{s_{n+1}}(x_{n+1})}{c}  \tilde{x}_{n+1} \widehat{\beta}_{s_{n+1}}\right) p(\rho_{n}|\mathcal{D}), \label{eq:jdp_mean}
\end{align}
where $c = p(x_{n+1}| x_{1:n})$,  $\widehat{\beta}_j$ is the posterior mean of the linear regression coefficients given the data in cluster $j$, and $\tilde{x}= (1,x^\prime)$. 
Similarly, the predictive density evaluated at $y$ is:
\begin{align*}
 &f(y\mid x_{n+1},  \mathcal{D}) \\
 &\quad = \sum_{\rho_n} \left(\sum_{s_{n+1}=1}^{k +1}   \frac{\omega_{s_{n+1}}(x_{n+1})}{c} h(y \mid \mathcal{D}_j^*) \right) p(\rho_{n}|\mathcal{D}),
\end{align*}
where 
$$ h(y \mid \mathcal{D}_j^*) = \int \phi(y ; \tilde{x}_{n+1}\beta, \sigma^2)  d P_0(\beta, \sigma^2 \mid \mathcal{D}_j^*), $$
with $\mathcal{D}_j^*= \lbrace (x_i, y_i) : s_i=j \rbrace$ containing the data in cluster $j$ and  $\phi(y;\mu,\sigma^2)$ denoting the density of the normal distribution, evaluated at $y$, with mean $\mu$ and variance $\sigma^2$.
These equations highlight how the model achieves flexible predictive inference by partitioning the data into clusters and fitting local linear regression models within each cluster. These local linear predictions are then averaged with dependent weights reflecting the similarity of $x_{n+1}$ to the covariates within each cluster, as measured by the marginal normal kernel, and further averaged to account for uncertainty in the partitioning structure.   It is important to emphasize that the posterior of the random partition, $p(\rho_{n}|\mathcal{D})$, is based on the joint likelihood; therefore, if the joint distribution is complex, many clusters may be required to fit it. This may result in local linear predictions based on small sample sizes and less efficient and more uncertain predictive inference \cite{Wade2014}, which is further examined in the comparative examples of Section \ref{sec:comparison}.  

\paragraph*{Further developments.}
Due to the difficulties associated with estimation of full covariance matrices, even for moderate $p$, 
\cite{SN}, who focus  on classification of a categorical response variable, modified the original approach of \cite{MEW} in two ways. 
First, the joint multivariate kernel is decomposed as the product of a marginal kernel on $\mathcal{X}$ and a conditional kernel on $\mathcal{Y}$ given $x$ (in this case, a multinomial logit kernel) and the parameter space consequently is expressed in terms of the parameters of the marginal and of the conditional kernels. Second, the authors considered the covariates to be independent within each component so that the covariance matrix of the marginal kernel is diagonal, improving scalability with $p$. These two modifications further allow for easy inclusion of discrete or other types of responses or covariates. Indeed, \cite{HBP} extended this approach to allow any standard generalized linear model to replace the multinomial logit kernel so to accommodate a greater variety of response types. 
A related method also capable of dealing with both continuous and discrete responses was proposed by \cite{DB2010}. 
The particular case of a binary response variable was considered by \cite{deyoreo2015}, but using a different strategy that relies on assuming that the binary response arises from  an underlying continuous random variable through discretization and this latent variable is jointly modeled with the (continuous) covariates through a multivariate Gaussian kernel. A similar approach for (multivariate) ordinal responses was developed by \cite{deyoreo2018a} (see also \cite{deyoreo2018b} for a dynamic extension). The model developed by \cite{Papageorgiou2019} also assumes that discrete variables, either responses or covariates, as discretised versions of continuous latent random variables, and can handle mixed scale covariates and discrete responses (with an emphasis on count responses).
Variable selection for the case of both a continuous response and covariates, and  conditional and marginal Gaussian kernels, using shrinkage prior distributions for the linear regression coefficients, was considered by \cite{Ding2021}. Finally, the covariate-dependent urn scheme implicitly defined by the joint model was examined by \cite{PD} and \cite{MQ}.

In addition, the decomposition of the multivariate kernel into the product of the marginal and conditional kernels allows for easy incorporation of local nonlinear regression models.  For instance, in machine learning, the conditional kernel is referred to as the expert, and the joint modeling approach for dependent mixtures is termed a generative or alternative mixture of experts  \citep{xu1994alternative}. Flexible experts, such as neural networks \cite{bishop1994,ambrogioni2017} or  Gaussian process regression models \citep{meeds2005alternative,yuan2008variational}, provide an effective tool for modeling highly nonlinear data, such as, in robotics.

Variations and extensions of the joint mixture model for density regression have been applied, among others, to causal inference \citep{Xu2018}, functional data analysis \citep{Rodriguez2009}, inverse dynamics \citep{abdulsamad2021variational}, Markov switching regression \citep{Taddy2009}, missing data \citep{Daniels2023}, point processes \citep{Taddy2012}, quantile regression \citep{Taddy2010}, survival analysis \citep{Poynor2017}, and time series \citep{deyoreo2017,Kalli2018,Heiner2022}.

As already alluded, modifications and alternatives to the DP prior for the random mixing measure $P$ have also been considered. For instance, the skewed Dirichlet process \citep{IOQ09}, which includes the DP as a particular case, is discussed in \cite{Q11}. Motivated by the fact that even for a moderate number of covariates, the clusters induced by the joint DP mixture model will be overwhelmingly determined by the covariates rather than the response, leading to a degrading of the predictive performance of the model, \cite{Wade2014} proposed to replace the DP prior for $P$ with an enriched DP \citep{Wade}, which by better modeling the random partition and allowing a nested clustering structure, overcomes this key disadvantage. This was further extended in \cite{pmlr-v108-gadd20a} with local generalized Gaussian process kernels for increased flexibility. In turn, \cite{Quinlan2018} used a finite Gaussian mixture to jointly model the continuous response and covariates with the components locations modeled with a repulsive distribution, whereas \cite{NP12} considered also a finite Gaussian mixture model but with both continuous and discrete responses and covariates (and similarly to some previously mentioned approaches discrete variables are handled through the use of latent variables).

\subsection{Conditional approach} \label{sec:cond_appr}
If our main interest is on the conditional density, then, in such a case, modeling also the marginal density of the covariates is an unnecessary complication. The conditional approach overcomes this by directly modeling the collection of conditional densities $ \lbrace f(y\mid x) \rbrace_{x \in \mathcal{X}}$. Mixture models for density estimation can be extended to define a flexible model for such a collection of conditional densities by allowing the mixing measure to depend on the covariates, i.e., 
\begin{align}
	f(y\mid x, P_x)= \int k(y;x,\theta) \text{d}P_x(\theta),  \label{eq:cond_mod2}
\end{align}
The question is then which prior to assign to the collection of  mixing measures $\{P_x: x\in \mathcal{X}\}$. Two possible choices are: (i) all $P_x$ are assumed to be identical, e.g., $P_x\equiv P\sim\text{DP}(\alpha, P_0)$ for all $x\in \mathcal{X}$, and (ii) all $P_x$ are assumed to be distinct and independent, e.g., $P_x\sim\text{DP}(\alpha, P_0)$, independently for each $x$. We seek a compromise between these two extreme choices as (i) is too restrictive and corresponds to maximum borrowing of strength across covariate values, and (ii) is wasteful and corresponds to no borrowing of strength. Indeed, \cite{CD11} lists some desirable properties of a prior for the collection of dependent mixture measures, which include: (1) increasing dependence between $P_x$ and $P_{x^{*}}$ as the distance between $x$ and $x^{*}$ decreases, (2) simple and interpretable expressions for the expectation and variance of each $P_x$ as well as the correlation between $P_x$ and $P_{x^{*}}$, and (3) efficient posterior simulation in a broad variety of applications.

\subsubsection{Early proposals} \label{sec:ep}
A first proposal to define a prior for a collection of random probability measures indexed by covariates was given by \cite{CR1}, where the focus was on discrete covariates, and dependence between the vector of random probability measures was introduced through the base measure of the DP. 
In particular, assuming $\mathcal{X}=\lbrace 1,\ldots,M \rbrace$ for some finite $M$, the law of the $M$-vector of random probability measures is
\begin{align} 
	P_1, \ldots, P_M\mid u_1, \ldots, u_M \sim \prod_{x=1}^M \DP(\alpha_x, P_0(\cdot;u_x)), \label{eq:px_cr}
\end{align}
where
\begin{equation*}
u_1, \ldots, u_M \sim H,
\end{equation*}
for some distribution $H$.
Note that in this construction the weights are allowed to vary with $x$, but are constructed independently across $x$, in accordance with the DP.
Thus, dependence is induced through the covariate-dependent atoms, where
\begin{equation*}
\tilde{\theta}_j(x)\mid u_x \overset{\text{ind.}}\sim P_0(\cdot;u_x).
\end{equation*}
This approach extends Antoniak's \cite{Ant} mixture of Dirichlet processes, 
and it was applied in regression and ANOVA settings \citep{CMS81}, for studying the search of an optimal drug dose \citep{Muliere}, and to address change point problems \citep{Mira}. In this type of approach, since the weights are independent across $x$, multiple observations at each covariate value are needed for inference. For example, in \cite{Muliere}, only a finite number of doses $x$ were possible, and the authors assume $u_x=\beta$ for all  $x \in \mathcal{X}$ and 
\begin{equation*}
\tilde{\theta}_j(x)\mid\beta \overset{\text{ind.}}\sim P_0\equiv \Norm(\tilde{x}\beta, \sigma^2),
\end{equation*}
where 
$\beta\sim H$ and $\Norm (\mu,\sigma^2)$ stands for a normal distribution with mean $\mu$ and variance $\sigma^2$. Association between $P_x$ and $P_{x^{*}}$ is thus attained via sharing common regression coefficients. Related approaches applied to regression for count data and for variable selection in survival analysis were explored by \cite{Carota2002} and \cite{Giudici2003}, respectively. In all these studies, however, the idea was to use (\ref{eq:px_cr}) to directly define a model for the collection of conditional distribution functions, not through a mixture as in \eqref{eq:cond_mod2}. A limitation of this approach is that the nature of the dependence is restricted to the form specified in the base measure. For a deeper discussion of the drawbacks of this approach, we refer to \cite{PR97}.

In turn, an early proposal for a mixture model of type (\ref{eq:cond_mod2}) defines the weights as constant functions of $x$ and assumes a standard linear regression kernel, i.e.,
\begin{align}
&f(y\mid x, P_x)  = \int \phi(y; \mu,\sigma^2)\text{d}P_x(\mu,\sigma^2) \label{eq:ddplin}\\
&\quad =\sum_{j=1}^\infty \omega_j \phi(y; \tilde{\mu}_j(x), \tilde{\sigma}_j^2),\quad \tilde{\mu}_j(x) = \tilde{x}\tilde{\beta}_j, \nonumber\\
& P_x = \sum_{j=1}^{\infty}\omega_j\delta_{(\tilde{x}\tilde{\beta}_j,\tilde{\sigma}_j^2)},\quad (\tilde{\beta}_j,\tilde{\sigma}_j^2)\overset{\text{iid}}\sim P_0. \nonumber
\end{align}
One can imagine a non-homogeneous population, where a subject's response behaviour may be described by one of the models in the infinite collection of linear regression models, and allocation to a specific component is independent of $x$. Note that this model simply corresponds  a DP mixture of normal linear regression models, that is,
\begin{equation*}
f(y\mid x, P) =  \int \phi(y; \tilde{x}\beta,\sigma^2)\text{d}P(\beta,\sigma^2),\quad P\sim\text{DP}(\alpha, P_0).
\end{equation*}
For an early overview of DP mixtures of linear models, with applications, we refer the reader to \cite{West}.

\subsubsection{General model} 
In \cite{Mac1} and in a more detailed technical report \cite{MacTR}, the dependent DP (DDP) was originally proposed as a prior for the collection of random probability mixing measures indexed by covariates. MacEachern was specifically interested in models that assumed that the marginal of $P_x$ is a DP, which was chosen because of the desired theoretical properties as well as the availability of computational procedures for inference, as discussed in Section \ref{sec:mixtures}. MacEachern modified the stick-breaking representation of the DP  to accommodate covariates and, in full generality, the DDP is specified as 
\begin{align} \label{ddpfull}
 P_x & = \sum_{j=1}^{\infty}\omega_j(x)\delta_{\tilde{\theta}(x)}, \nonumber \\   
 w_1(x) & = v_1(x),\\
 w_j(x) & = v_j(x)\prod_{j^{\prime} < j}\{1 - v_{j^{\prime}}(x)\}, \quad j>1,
\end{align}
where each $v_j(x)$ is a stochastic process on $\mathcal{X}$ with marginal distributions
$v_j(x) \sim \text{Beta}(1, \alpha(x))$, independent across $j$. The atoms $\tilde{\theta}_j(x)$ are also independent across $j$ and for each $j$, $\tilde{\theta}_j(x)$ is a stochastic process on $\mathcal{X}$ with marginal distribution $P_{0x}$. Additionally, the atoms $\{\tilde{\theta}_j(x)\}_{j\geq 1}$ are independent of the stick-breaking proportions $\{v_j(x)\}_{j\geq 1}$. The corresponding model for the conditional density is given by
\begin{equation*}
f(y\mid x, P_x) = \sum_{j=1}^{\infty}\omega_j(x)k(y;x,\tilde{\theta}_j(x)).
\end{equation*}
 This model is very general and includes as 
 particular cases many regression models, including, among others, fixed and random effects linear and generalized linear models and infinite mixtures of Gaussian process regression models.

Applications of models with fully flexible formulations for the weights and atoms are not as common in practice. Exceptions include, for example,  the model for spatial data, namely for point-referenced data, proposed by \cite{Duan} where both the weights and atoms rely on Gaussian process specifications.  This lack of proposals for fully flexibly models is due not only to to interpretability issues and computational complexities, but also due to the fact that desirable theoretical properties are still available with simpler constructions. In fact,  full weak support \cite{Bar} and desirable consistency properties \cite{pati2013posterior} 
are available for the general DDP model and also for two simplified versions which assume constant weights or constant atoms.

\subsubsection{Covariate-dependent atoms} \label{sec:thetax}
An important class of DDPs is the `single-weights' DDP, which defines the weights in accordance with the DP, i.e., the weights do not depend on covariates. This was the DDP model considered in the illustration of one of the two original articles proposing this prior \citep{MacTR} and, as the author mentions, in this class of models one merely replaces the atoms $\tilde{\theta}_j$ 
with stochastic processes $\tilde{\theta}_j(x)$, for $x\in\mathcal{X}$. For example, $\tilde{\theta}_j(x)$ might be a Gaussian process. The corresponding model for the conditional density takes the form
\begin{equation*}
f(y\mid x, P_x)= \sum_{j=1}^\infty \omega_j k(y;x,\tilde{\theta}_j(x)), \label{eq:sp_mod}
\end{equation*}
with
\begin{equation*}
P_x = \sum_{j=1}^{\infty}\omega_j\delta_{\tilde{\theta}_j(x)}.
\end{equation*}
In most cases, the kernel $k(y; x,\theta(x))$ is defined so that the regression function $E(y\mid x, P_x)$ is described by one of infinite collection of possible mean functions $\tilde{\theta}_j(x)$, with probability $w_j$. It is important to note that this probability of allocation to a specific component is independent of the covariates. Single-weights DDP mixture models are attractive and popular because posterior inference can be carried out using any of the established algorithms for DP mixture models, resulting in much simpler computations. In fact, the collection of dependent mixing measures  can also be marginalized in this setting, and the model can be parameterized in terms of the random partition $\rho_n$ and the unique cluster-specific stochastic processes. 

\paragraph*{Predictive structure.}
This formulation also helps to shed light on how predictions are constructed, where again the aim is prediction of the response given a new covariate value $x_{n+1}$. For notational simplicity, we focus on a continuous response with Gaussian kernel $ \phi(y; \mu_j^*(x), \sigma^{2\,*}_j)$, where  $\mu_j^*(x)$ and $\sigma^{2\,*}_j$ represent the mean function and variance of the $j$th cluster. As the weights do not depend on the covariates, the cluster allocation $s_{n+1}$ of a new subject follows the standard P\'{o}lya urn scheme:
\begin{align*}
  s_{n+1} \mid \rho_n, x_{n+1} \sim \frac{\alpha}{\alpha+n} \delta_{k+1} + \sum_{j=1}^k  \frac{n_j}{\alpha+n} \delta_j.
\end{align*}
Then, the predictive mean for the response given a new value $x_{n+1}$ is:
\begin{align}
 &E(Y\mid x_{n+1},  \mathcal{D}) = \sum_{\rho_n}  \left(\frac{\alpha}{\alpha+n} E[\mu^*_{k+1}(x_{n+1})] \right. \label{eq:ddpmean}\\
 & \quad \left. +\sum_{j=1}^{k} \frac{n_j}{\alpha+n} E[\mu^*_{j}(x_{n+1})\mid \mathcal{D}_{j}^*] \right) p(\rho_{n}|\mathcal{D}). \nonumber
\end{align} 
Thus, the local predictions are averaged with weights proportional to the cluster sizes, and further averaged to account for uncertainty in the partition structure.  Again, since the weights do not depend on $x$, flexibility in the cluster-specific mean functions is key to achieve  flexible, nonlinear predictions. Indeed, if the mean functions $\mu^*_j(x)$ are simply linear (as in \eqref{eq:ddplin}), it is easy to see that the predictive mean function in \eqref{eq:ddpmean} is also linear.  
Similarly, the predictive density evaluated at $y$ is:
\begin{align*}
 &f(y\mid x_{n+1},  \mathcal{D}) = \sum_{\rho_n}  \left(\frac{\alpha}{\alpha+n} h(y \mid x_{n+1}) \right. \\
 &\quad \quad \left. +\sum_{j=1}^{k} \frac{n_j}{\alpha+n} h(y \mid x_{n+1}, \mathcal{D}_{j}^*) \right) p(\rho_{n}|\mathcal{D}),
\end{align*}
where the cluster-specific predictive densities for a new and existing cluster, respectively, are:
\begin{align*}
h(y \mid x_{n+1}) &= \int \phi(y ; \mu(x_{n+1}), \sigma^2)  d P_0(\mu, \sigma^2 ), \\
h(y \mid x_{n+1}, \mathcal{D}_j^*) &= \int \phi(y ; \mu(x_{n+1}), \sigma^2)  d P_0(\mu, \sigma^2 \mid \mathcal{D}_j^*).    
\end{align*}
By mixing over Gaussian kernels, the predictive conditional densities can have flexible shapes. However, we highlight that the model partitions the data into clusters, where within each cluster the regression relationship can be modeled by a common local mean function with normal errors. In some cases when the regression kernel is not flexible enough, the inferred partition structure may depend on the covariates and poor prediction may result, as the cluster-specific predictions are averaged regardless of the covariate values. This is further explained and explored in the examples of Section \ref{sec:comparison}.  
\paragraph*{Further developments.}
These models have been successfully applied to address a wide range of problems from classical regression problems \citep{MacTR,Mac3} to ANOVA \citep{DeIorio} and including, among others, discriminant analysis \citep{DLC,Gutierrez2011}, dose-response studies \citep{Fronczyk2014,Fronczyk2014b,Fronczyk2017},  dynamic density estimation \cite{Rod_TS}, extreme value analysis \cite{Kottas2012}, functional \cite{DunHer} and longitudinal data analysis \cite{MullerLong,Quintana2016}, mediation analysis \cite{Devick2022}, multiple testing \cite{Gutierrez2019,Gutierrez2023}, multiple imputation for missing data \cite{Burgette2012}, multivariate count data \cite{Li2021}, marked point process intensities \cite{Xiao2015}, ordinal regression \cite{Bao2015}, quantile regression \cite{Kottas2009}, receiver operating characteristic curve analysis \cite{Inacio2013,Inacio2022}, spatial modeling \cite{Mac2, Kottas2008}, stochastic ordering \cite{Dunson2008SO}, survival analysis \cite{DeIorio09,JaraANOVA,Zhou2015, Xu2016,Xu2019,Shi2021,Xu2022}, and time series \cite{Caron2008,DiLucca2013}.
For a continuous response, a popular single-weights DDP model employs Gaussian process priors for the components' means: 
\begin{align}
	f(y\mid x, P_x) &= \sum_{j=1}^\infty \omega_j \phi(y; \tilde{\mu}_j(x), \tilde{\sigma}^2_j), \label{eq:ddp_gp} \\
 \tilde{\mu}_j(x) &\sim\GP(m_j,C_j),  \nonumber
\end{align}
where $\GP(m,C)$ denotes a Gaussian process with mean function $m$ and covariance function $C$.  
Standard covariance functions (e.g., the squared exponential) lead to smooth changes in the conditional density with $x$, favouring similarity in $f(y\mid x, P_x)$ and $f(y\mid x^{*}, P_{x^{*}})$ when $x$ and $x^{*}$ are close. Note that  \eqref{eq:ddp_gp} characterizes the conditional density using an infinite mixture of normal distributions where the components' mean functions 
vary nonlinearly with the covariates but the weights on the different mixture components remain constant as $x$ varies. 
This corresponds to a generalization of the popular Gaussian process regression model where the mean function is assigned a Gaussian process prior and the errors are Gaussian with zero mean and constant variance. In (\ref{eq:ddp_gp}), various choices are available for $m_j$ and $C_j$. For example, \cite{MacTR} studied the log area of Romanesque churches given the log perimeter, and \cite{Mac3} studied biology exam scores given previous exam scores, and in both applications, 
$m_j(x)=\tilde{x}\beta_j$ is assumed to be linear with an exponential variogram for the covariance function 
\begin{align*}
C_j(x, x^{*})= &(c_{0j}-c_{1j})\{1-\exp(-\tau_j \lVert x-x^{*}\rVert)\}\\
&+c_{1j} \1(\lVert x-x^{*}\rVert>0),
\end{align*}
where $c_{0j}$, $c_{1j}$, and $\tau_j$ are hyperparameters and, depending on the application and context, some may be assumed common across components. 
This model was also applied in \cite{Mac2}, where $x$ represents the spatial location of an observation. In this example, the Gaussian process priors were specified to have mean zero with a squared exponential covariance function, 
\begin{equation*}
C_j(x,x^{*})=c_j\exp(-\tau_j \lVert x-x^{*}\rVert^2).
\end{equation*}
Other response types can also be accommodated through a generalised Gaussian process framework. 
More recently, Gaussian process priors for the components' means were also employed by \cite{Xu2016,Xu2019,Xu2022} in applications involving survival analysis and clinical trials.

In \cite{DeIorio} the focus is on discrete covariates and the authors show that in this setting, the single-weights DDP is equivalent to a DP mixture of linear regression models under a transformation, say $\lambda$, of $x$ into a higher-dimensional space. This model is often referred in the literature as the ANOVA-DDP model. The general model for discrete covariates and a continuous response is
\begin{equation}
f(y\mid x, P_x)= \sum_{j=1}^\infty \omega_j \phi(y; \tilde{\beta}_j'\lambda(x), \tilde{\sigma}^2_j). \label{eq:ddp_transform}
\end{equation}
The most flexible choice of $\lambda$ transforms the $p$-dimensional discrete vector $x$ into a $M_1 \times \ldots \times M_p$-dimensional vector of zeros apart from a single element of one indicating the categories of the $p$ covariates, where $M_h$ is the number categories of the $h$th covariate. 
An extension to hierarchical models was also considered and an illustration involving a longitudinal continuous response, white blood cell count over time, with two discrete covariates, representing the levels of two cancer treatment drugs, was presented. Specifically, $y$ is indexed by an additional variable $t$, representing time, and the model is extended by replacing the local mean $\tilde{\beta}_j'\lambda(x)$ in (\ref{eq:ddp_transform}) with some specified function of $t$ and $\tilde{\beta}_j'\lambda(x)$. A similar extension was discussed in \cite{DLC}, who used the ANOVA-DDP model for classification based on longitudinal markers, where the response represents the level of a specific hormone over time and $x$ is a binary indicator for normal pregnancy. Both approaches incorporated  dependence in the random effects distribution across groups.

In general, the procedure used in (\ref{eq:ddp_transform}) of mapping $x$ to a high-dimensional vector may also be used for continuous covariates by defining an appropriate transformation function. In fact, models that define the component's mean function $\tilde{\mu}_j(x)$ through a Gaussian process, as in Equation (\ref{eq:ddp_gp}), can be represented in terms of models with mean functions of the form $\tilde{\beta}_j'\lambda(x)$ as in  \eqref{eq:ddp_transform}, since $\tilde{\mu}_j(x)$ can be equivalently written as $\tilde{\beta}_j'\lambda(x)$, where $\lambda(x)$ transforms $x$ into a possibly infinite dimensional space whose transformation is defined by the covariance function of the Gaussian process. More specifically, and omitting the components index, if $C$ is the covariance function, then $C(x_1,x_2)=\lambda(x_1)'\lambda(x_2)$. We refer the reader to Section 4.3 of \cite{Ras} for examples.

To accommodate continuous and discrete covariates, an appropriate transformation needs to be defined. For example, in \cite{DeIorio09}, flexible mean functions for discrete covariates, say $x_d$, and linear mean functions for the continuous covariates, say $x_c$, are used, so that $\tilde{\mu}_j(x)= \tilde{\beta}_{d,j}'\lambda(x_d)+\tilde{\beta}_{c,j}'x_c$. Instead, in \cite{JaraANOVA}, linear mean functions for both the discrete and continuous covariates are used, i.e. $\tilde{\mu}_j(x)= \tilde{x}\tilde{\beta}_j$; the resulting model is sometimes referred to in literature as the linear dependent Dirichlet process (LDDP). Both articles consider applications to survival analysis where the former studies the survival time for cancer patients given the dose level of a drug (discrete), estrogen receptor status (discrete), and tumor size (continuous), and the latter studies time to dental carry given information of dental hygiene (mostly binary apart from the age at the start of brushing). Differently from popular survival regression models, such as the Cox proportional hazards model or the accelerated failure time model (see, e.g., Chapters 3 and 5 of \cite{Collett2023} for a review of these models), which impose that survival curves from different covariate levels are not allowed to cross, a feature that is unrealistic many practical applications, the aforementioned two works allow survival curves to cross, or not, as the data dictate. As noted in \cite{JaraANOVA}, the LDDP mixture model can be interpreted as a mixture of parametric accelerated failure time regression models. Indeed, this model corresponds also to a generalization of earlier semiparametric approaches for the accelerated failure time model that assume a parametric component for the regression coefficients and a DP mixture model for the error distribution (e.g. \cite{Kuo1997}) by additionally mixing over the regression coefficients. Further, an ANOVA-DDP mixture model, considering both discrete and continuous covariates, and using linear mean functions, was also used by \cite{Richardson2018} in the context of modeling and predicting healthcare claims. For flexible interactions terms, an appropriate transformation is needed. Note that when the transformation is simply the identity function, i.e., $\lambda(x)=x$, so that the components' mean functions are linear, the model is equivalent to the mixture of linear regression models discussed in Section \ref{sec:ep}.  Although such a model may seem very flexible at a first glance, as highlighted by the predictive equations \eqref{eq:ddpmean}, 
the predictive mean and conditional density are greatly restricted. For instance, the mean regression structure is linear; we have a weighted combination of parametric regression functions, but without the local adjustment afforded by covariate-dependent weights. That is, the single-weights DDP mixture (of normals) model is flexible in terms of non-Gaussian response, but not in terms of regression relationships.
For increased model flexibility, in terms of the implied mean regression structure, higher-dimensional transformations of the continuous covariates are needed. In fact, \cite{DeIorio09} mentions including higher-order terms for the continuous covariates and \cite{JaraANOVA} comment that $\lambda(x_c)$ may be defined through B-splines basis. Indeed, \cite{Inacio2013} and \cite{Inacio2022} in the context of incorporating covariates in the receiver operating characteristic curve, used a single-weights DDP mixture model with a normal kernel and where the mean function is modeled through cubic B-splines basis, with the number of basis functions selected through model selection criteria. The resulting model can be regarded as a DP mixture of additive normal models. This strategy works best when there is only one continuous covariate. With two continuous covariates, we would need, in principle, to fit the model for all possible conceivable combinations of number of basis functions, which would imply fitting the model and computing associated model selection criteria, potentially quite a large number of times, which is impractical.

\subsubsection{Covariate-dependent weights} \label{sec:wx}
Motivated by the limited modeling flexibility of the single-weights DDP mixture model, a wealth of approaches have been proposed to allow the weights of the random mixing measure to depend on covariates. In general, and by opposition to single-weights dependent DP mixture models, computations tend to be more burdensome.  
The general model (\ref{eq:cond_mod2}) is usually simplified by assuming that the atoms do not depend on the covariates, i.e., 
\begin{align}
	f(y\mid x, P_x)= \sum_{j=1}^\infty \omega_j(x) k(y;x,\tilde{\theta}_j), \label{eq:sa_mod}
\end{align}
with 
\begin{equation*}
P_x= \sum_{j=1}^\infty \omega_j(x) \delta_{\tilde{\theta}_j},
\end{equation*}
where, for example, when $y$ is continuous and univariate, the kernel may correspond to a linear regression model $\phi(y;\tilde{x}\beta, \sigma^2)$ with $\theta = (\beta, \sigma^2)$ or other simple formulations.  Other response types require replacing the linear regression model with an appropriate kernel. For example, if the response is binary, ordinal, categorical, or counts, a generalized linear model is appropriate. We highlight that in machine learning, such models are termed 
discriminative mixtures of experts, with the regression kernel being the expert and the covariate-dependent weights referred to as the gating network.   

The main constraint with covariate-dependent weights is the need to specify a prior such that the weights are positive and $\sum_j \omega_j(x)=1$ for all $x\in \mathcal{X}$. 
Most proposals are based on the stick-breaking representation, while others utilize normalization or indicator functions to ensure this restriction is met.  
Stick-breaking constructions are motivated by the general DDP \cite{Mac1}, even if not all maintain marginal DP priors, and assume:
\begin{align*}
\omega_1(x) & =v_1(x),\\
\omega_j(x) &= v_j(x) \prod_{j'<j} \{1-v_{j'}(x)\}, \quad \text{for } j>1,
\end{align*}
where $0\le v_j(x)\le 1$ for all $j$ and $x$. Instead, approaches using normalization assume:
\begin{align*}
   \omega_j(x) &= \frac{\nu_j(x)}{\sum_{k=1}^\infty \nu_{k}(x)},
\end{align*}
where $\nu_j(x)\geq 0$ for all $j$ and $x$ and $\sum_{j=1}^\infty \nu_{j}(x)$ is finite almost surely. Alternatively, the dependent weights can be defined using indicator functions:
\begin{align*}
   \omega_j(x) &= \one(x \in  R_j),
\end{align*}
where $\mathcal{X}$ is partitioned into regions $R_1, R_2, \ldots$. 
Interestingly, the joint DP model in Section \ref{sec:joint} does not induce marginal DP priors for $P_x$ and implies normalized covariate-dependent weights as defined in \eqref{eq:joint_dw}. 
The various models available in the literature differ in the definition of the $v_j(x)$, $\nu_j(x)$, or regions $R_j$, and for each proposal, various model choices regarding hyperparameters and functional shapes are needed. Without loss of generality, we denote the additional parameters 
by the same symbol $\tilde{\psi}_j$ in all constructions. 
\paragraph*{Predictive structure.}
Note that in contrast to the single-weights DDP model, the implied prior on the random partition model (although not available in closed form) changes with the covariates, which is relevant when there is scientific interest in the underlying implied partition. Moreover, unlike the joint approach, the random partition structure is driven solely  by good approximation of the conditional density.  
However, as the random mixing measures cannot be marginalized,  expressions for predictions analogous to the joint model in \eqref{eq:jdp_mean} and single-weights model in \eqref{eq:ddpmean} are not available. Instead (focusing on the linear regression kernel), we write the predictive mean   as:
\begin{align}
 &E(Y\mid x_{n+1},  \mathcal{D}) = \int  \sum_{j=1}^\infty \omega_j(x)  \, \tilde{x}_{n+1}\tilde{\beta}_j  \, p(d\tilde{\psi}, d\tilde{\beta} |\mathcal{D}), \label{eq:dw_mean}
\end{align} 
where the integral is taken with respect to the posterior over the  parameters $\tilde{\psi} = (\tilde{\psi}_1, \tilde{\psi}_2, \ldots) $
of the dependent weights and the kernel coefficients $\tilde{\beta} = (\tilde{\beta}_1, \tilde{\beta}_2, \ldots) $. To address the infinite sum in \eqref{eq:dw_mean}, truncated approximations or slice sampling are typically employed. Similarly, the predictive density is:
\begin{align*}
 &f(y\mid x_{n+1},  \mathcal{D}) \\
 &= \int  \sum_{j=1}^\infty \omega_j(x)  \, \phi(y ; \tilde{x}_{n+1}\tilde{\beta}_j, \tilde{\sigma}^2_j)  \, p(d\tilde{\psi}, d\tilde{\beta}, d\tilde{\sigma}^2 |\mathcal{D}).
\end{align*}
Thus, such models build on  simple, interpretable local linear models, which are combined with local relevance that changes across the covariate space as determined by the dependent weights, to construct flexible shapes for the predictive regression function and conditional density.   

\paragraph*{Further developments.}
The form of the dependent weights plays an important role. 
One of the first approaches was developed by \cite{GS} who, for continuous covariates, proposed the order-based DDP that allows 
the ordering in the stick-breaking proportions to depend on the covariates, i.e., the $v_j$'s are reordered based on $x$. One way to accomplish this is to associate each pair $(v_j, \tilde{\theta}_j)$ with a random variable $\tilde{\psi}_j$, taking values in $\mathcal{X}$. For every $x$, the $\tilde{\psi}_j$'s are reordered based on their distance to $x$, and this ordering is then used to define a permutation of $(v_j, \tilde{\theta}_j)$. This construction ensures that $P_x$ is a DP at each covariate value. The authors successfully applied this idea to stochastic volatility and spatial modeling but did not discuss how to handle discrete covariates. Note that in the context of spatial modeling, and in contrast to the approaches of \cite{Mac2} and \cite{Duan}, this approach does not require replications to conduct inference.

In \cite{Dun1}, the kernel-stick breaking process was proposed, which defines
\begin{equation*}
v_j(x) = v_j C(x,\tilde{\psi}_j), \quad v_j\overset{\text{iid}}\sim\text{Beta}(1,\alpha),
\end{equation*}
for some bounded kernel $C:\mathcal{X}\times \mathcal{X} \rightarrow [0,1]$ and the kernel locations $\tilde{\psi}_j$ are sampled from a distribution, say $H$, defined on $\mathcal{X}$. A possibility is the Gaussian kernel
\begin{equation*}
C(x,\tilde{\psi}_j)=\exp\{-\lambda_j \|x-\tilde{\psi}_j \|^2\},\quad \lambda_j>0.
\end{equation*}
In this construction, the stick-breaking proportions are dampened by the distance between $x$ and the (random) locations $\tilde{\psi}_j$, so that if $\tilde{\psi}_j$ is very close to $x$, there is little down-weighting by the kernel and the weight can be relatively high.  
An epidemiological application was provided 
involving continuous covariates, and for incorporation of discrete covariates, adequate kernels must be specified. 
In the standard version of the kernel-stick breaking process, the parameters of the conditional kernel,  $\tilde{\theta}_j$, are sampled from a common baseline distribution, say $P_0$. For hierarchical data, a slightly more general version is obtained by placing a DP prior on $P_0$. 
This has been used in multi-task image processing \cite{An2008} and also  to model the distribution of random effects in a toxicological risk assessment application in \cite{Hwang2014}. A similar weight construction, but tailored to the spatial context, specifically for modeling hurricane surface wind fields, was developed by \cite{Reich}. Uniform and squared exponential kernels were used and the covariates considered were geographical coordinates. Still in the spatial context, \cite{Duan} extends the single-weights spatial model of \cite{Mac2} to allow different surface selection at different sites. Motivated by the fact that none of the spatial models described so far apply to areal data, that is, data that are observed within given boundaries (e.g. counties), \cite{Li2015} proposed an areally-referenced stick-breaking prior for a spatial random effects distribution. This corresponds to an adaptation to the areal data setting of the approach of \cite{Reich}, that is suitable for point-referenced data, by using a latent conditionally autoregressive model (on the logit scale) to define the kernel function. A  discrete areal data kernel function for use in the kernel stick-breaking process framework was  recently proposed in \cite{Warren2022}, where a hydraulic application is provided.

A closely related approach, termed the local DP, is given in \cite{CD11}, in which the kernel is defined as the indicator that $x$ belongs to the $r$-neighborhood centred at $\tilde{\psi}_j$. Specifically, let $\mathcal{L}_{x} = \{j: d(x,\tilde{\psi}_j) < r\}$ be a covariate-dependent set indexing the components whose location  $\tilde{\psi}_j$ falls within an $r$-neighborhood of $x$, where $d$ is some distance measure (e.g., the Euclidean distance) and $r$, which controls the neighborhood size, can be treated either as fixed or inferred with a hyperprior. Then,
\begin{equation*}
P_x=\sum_{j \in \mathcal{L}_x}v_j\prod_{j^{\prime}< j}(1-v_j)\delta_{\tilde{\theta}_j}.
\end{equation*}
The resulting weights for two distinct covariate values $x$ and $x^{*}$ will be similar if $x$ and $x^{*}$ are close. The authors proved that $P_x$ follows a DP marginally for each $x$, a property that the kernel stick-breaking process lacks, with dependence between $P_x$ and $P_{x^{*}}$ induced through the inclusion of shared stick-breaking weights and atoms within the region of overlap in the neighborhoods around $x$ and $x^{*}$. The idea of the local DP was extended by \cite{GS10}, who proposed the DP regression smoother, and which considers the kernel as the indicator that $x$ lies in a random subset $\tilde{\psi}$ of $\mathcal{X}$.

Another common method defines the covariate-dependent stick-breaking proportions by extending ideas in generalized linear models. In this case, 
\begin{equation*}
v_j(x)= l\{\tilde{\psi}_j(x)\},
\end{equation*}
where $l: \mathbb{R} \rightarrow [0,1]$ is a monotone, differentiable link function and $\tilde{\psi}_j(x)$ is a random, real-valued function on $\mathcal{X}$. The function $l(\cdot)$ is commonly chosen to be the probit or logit link function, and $\tilde{\psi}_j(x)$ may be defined as a simple linear function, as a linear combination of basis functions, or through a Gaussian process prior.  
For example, \cite{Rod} use a probit link function, with the resulting model being referred as the probit-stick breaking process. The authors consider four possibilities for $\tilde{\psi}_j(x)$ depending on the application at hand: 1) for classic regression problems with continuous covariates, $\tilde{\psi}_j(\cdot)$ has a Gaussian process prior with a constant mean and the squared exponential covariance function, 2) for spatial and temporal applications, $\tilde{\psi}_j(\cdot)$ is a Gaussian Markov random field, 3) for discrete covariates, $\tilde{\psi}_j(\cdot)$ has a multivariate Gaussian distribution with a constant mean and identity covariance matrix, 4) in applications with both continuous and discrete covariates, they assume $\tilde{\psi}_j(x)$ is a linear function of the continuous covariates with slopes that depend on the value of the discrete covariates. Posterior inference is tractable and can be performed through data augmentation by introducing latent normal variables and borrowing tricks from probit regression \citep{Albert1993}, but the number of latent variables that need to be updated can be huge, as this is a function of both the sample size and number of components. By comparison, the kernel stick-breaking process has the advantage that $v_j(x)$ is defined through a finite dimensional parameter $\tilde{\psi}_j$ and a known kernel function, so that the number of computations may be much more reasonable. \cite{Chung} also use a probit link function but assume that $\tilde{\psi}_j(x)$ is a linear function of the absolute value of $x$ and an important focus of this work is variable selection to discard unimportant covariates, while allowing estimation of posterior inclusion probabilities. In turn, \cite{Ren} employ a logistic link function and basis function expansion of $\tilde{\psi}_j(x)$ in terms of squared exponential basis functions, leading to the so-called logistic stick-breaking process. Recently, \cite{Rigon2021} also used a logit stick-breaking prior for density regression, which relies on a representation of the stick-breaking prior via sequential logistic regressions and leverages the P\'olya-Gamma data augmentation for logistic regression \citep{Polson2013}, which might improve the mixing of the MCMC chains compared to the probit stick-breaking prior. This representation also facilitates the implementation of several computational methods (MCMC via Gibbs sampling, expectation-maximization algorithms, and mean field variational Bayes). Both the probit and logit stick-breaking priors can deal with continuous and discrete covariates, under appropriate specification of $\tilde{\psi}_j(x)$.   
Applications of the probit and logistic stick-breaking priors include stochastic volatility models and spatial models for count data \cite{Rod}, spatial models for clustered ordered (periodontal)
data \cite{Bandyopadhyay2016}, epidemiological studies \cite{Chung,Rigon2021}, image segmentation \cite{Ren}, and insurance loss prediction \cite{huang2020bayesian}. 

However, in this stick-breaking construction of the dependent weights, understanding how the chosen kernel, basis functions, link function, and other hyperparameters of the stick proportions then influence the dependent weights can be challenging, making such choices difficult.  
Motivated by this,  
\cite{Antoniano2014} proposed defining the dependent weights directly through normalization: 
\begin{align*}
f(y\mid x, P_x) &= \sum_{j=1}^{\infty}\omega_j(x)k(y\mid x, \tilde{\theta}_j),\\
\omega_j(x) &= \frac{\omega_j k(x\mid \tilde{\psi}_j)}{\sum_{j^{\prime}=1}^{\infty}\omega_{j^{\prime}} k(x\mid \tilde{\psi}_{j^{\prime}})}.
\end{align*}
The covariate-dependent weight $\omega_j(x)$ represents the probability that an observation with a covariate value $x$ is allocated to the $j$th regression component . Such probability can be decomposed into the unconditional probability $\omega_j$ that an observation, regardless of the value of the covariate, comes from parametric regression model $j$, and $k(x\mid \tilde{\psi}_j)$ describes how likely it is than an observation generated from regression model $j$ has a covariate value of $x$. The parametric kernel $k(x\mid \tilde{\psi}_j)$ can be defined to accommodate different types of covariates. Moreover, the form of the dependent weights coincides with that of the joint model in \eqref{eq:joint_dw}, yet this model has the advantage that the random partition of the data is based on the conditional density of interest. Due to the challenging features of the dataset described in the Introduction, \cite{wade2022colombian} extended this approach to accommodate mixed responses with censored, constrained, and binary traits. A similar normalized construction of the dependent weights is provided in \cite{NIPS2012_8f1d4362} using the normalized gamma process representation of the DP, and \cite{rao2009spatial} employ this idea with box kernels for spatial applications. This idea is further extended in \cite{10.1214/17-BA1072} based on normalized compound random measures, where the weights are proportional to the jumps of underlying L\'evy process multiplied by a random score function, with in application  to forensic analysis in \cite{10.1214/20-AOAS1334}. We note that due to the dominance of neural networks in classification tasks, in machine learning the covariate-dependent weights, i.e. gating networks, are commonly defined through neural networks with soft-max outputs \cite[see e.g.][]{etienam2020ultra}. That is, they are defined through normalization as 
\begin{align*}
\omega_j(x) &= \frac{\exp( h_j(x\mid \psi))}{\sum_{j^{\prime}=1}^{J} \exp(h_{j^\prime}(x\mid \psi))},
\end{align*}
where $h_j$ is the $j$th component of the feedforward neural network mapping the covariate space to $\mathbb{R}^J$, with $J$ being the fixed, finite number of components, and all weights and biases of the network are contained in $\psi$.

An alternative idea is to define the dependent weights as indicator function, $\omega_j(x) = \one(x \in  R_j)$, which corresponds to (randomly) partitioning the covariate space into regions and fitting local regression models with each region. For example, \cite{gramacy2008bayesian} use trees to partition  the covariate space into axis-aligned rectangular regions with local Gaussian process regression models. More flexible partitioning approaches have also been considered, such as Voronoi tessellations in \cite{pope2021gaussian}. Such approaches can effectively  capture discontinuities and nonstationarities but are not suited to multimodaility, skewness, and general shapes for the conditional density.  

Lastly, when the space $\mathcal{X}$ indexes discrete time, $\mathcal{X}=\{1,\ldots,T\}$, \cite{Gutierrez2016} proposed a time dependent mixture model or, in other words, a model for dynamic density estimation, where the sequence of stick proportions $\{v_j(x), x=1,\ldots, T\}$ has a Markov chain structure, which guarantees that $P_x$ marginally is a DP. An application to air quality monitoring was provided. In a related proposal, \cite{Mena2016}, instead of a Markov chain, consider a diffusion process (namely, a Wrights--Fisher diffusion) for $v_j(x)$, where again $x$ denotes time. Still in the context of time course data, and motivated by a functional proteomics application; specifically, by the need to analyze protein activation over time after an intervention, \cite{Nieto2012} proposed the time series DDP. Sequential dependence is achieved by introducing a sequence of latent random variables that link $v_j(x)$ and $v_{j}(x+1)$ and thus $\{\omega_j(x), \omega_j(x+1)\}$. Another time-dependent nonparametric prior, the stick-breaking autoregressive process, with marginal DP distributions, was proposed in \cite{Griffin2011}.

\subsection{Other approaches}
Another important class of models extends the random partition model and the urn scheme of the DP to depend on covariates. For these models, obtaining a representation in terms of (\ref{eq:cond_mod2}) can be far from straightforward.
Reversely, deriving an expression for the random partition model and urn scheme induced by (\ref{eq:cond_mod2}) can also be difficult. An exception is when the random partition model and urn scheme correspond to the joint model of $y$ and $x$ (see \cite{PD}),  in which deriving a representation in terms of (\ref{eq:cond_mod2}) is straightforward, and vice versa.

\cite{MQ} and \cite{muller2011product} developed a general class of covariate-dependent random partition models that modify product partition models by multiplying by a similarity function: 
\begin{align*}
	p(\rho_n|x_{1:n}) \propto \prod_{j=1}^{k} c(S_j)  g(x_j^*),
\end{align*} 
where $S_j= \lbrace i \in \lbrace 1, \ldots, n \rbrace : s_i=j \rbrace$. 
In product partition models, the term $ c(S_j)$ is called the cohesion function, and for example, $c(S_j)=\Gamma(n_j)$ for the DP. The similarity function, $g( \cdot) \geq 0$, captures the closeness of covariates, where large values indicate high similarity. 
The covariate-dependent random partition model of the joint approach is a special case, satisfies marginalization and scalability properties, and is easier from a computational perspective; thus, in examples, it is the focus of the authors. A nice application of product partition models to functional clustering is given in \cite{Page2015}.  In \cite{MQP12}, the covariate-dependent random partition model is extended to allow variable selection, whereas in \cite{page2016} it is extended to the spatial setting. Other approaches that constrain the random partition model by removing inadmissible partitions  can be found in \cite{WWP12} for curve fitting and \cite{Balocchi2022,Wehrhahn2020} for spatial applications.

Proposals that modify the urn scheme to depend on the covariates include \cite{RG02,Dahl08,orbanz2008nonparametric,BF11,Dahl2017}, to mention a few.
For example, the probability that a new subject is allocated to $j$th cluster may be altered to depend on the covariates in that cluster, so that
\begin{align*}
	p(s_{n+1}|s_{1:n},x_{1:n+1}) &\propto \left\lbrace   \begin{array}{cc}  g(x_{n+1}|x_j^*) & \text{if } s_{n+1}=j \\ \alpha & \text{if } s_{n+1}=k+1 \end{array} \right. .
\end{align*}
The function $g(x_{n+1}|x_j^*)$ is a measure of the similarity of $x_{n+1}$ to the covariates in the $j$th cluster and may be defined through a distance or  kernel function. 
In most proposals, analytical expressions for the induced prior over the number of clusters and cluster sizes are lost, posing challenges for model interpretation and hyperparameter selection; instead, the Ewens-Pitman attraction model of \cite{Dahl2017} maintains key properties of the random partition model of the DP, and one of its most widely used generalizations, the Pitman-Yor process \citep{PitmanYor}, albeit at increased computational cost. 
The distance-dependent Chinese restaurant process \cite{BF11} was used for image segmentation in computer
vision \cite{Ghosh2011} and to model geometric variability in spinal images \cite{seiler2013random}. A recent application of the Ewens-Pitman attraction model \cite{Dahl2017} to cluster heterogeneous populations
while considering individuals' treatment histories, motivated by the need of inferring drug combination effects on mental health in people with HIV is given in \cite{Jin2022}. 
For graph structured data, such as imaging data, modifications of the urn scheme based on the Potts model \cite{stoehr2017review} include   \cite{orbanz2008nonparametric,da2016bayesian,  lu2020bayesian}, and an application to extract regions of interest for disease diagnosis is presented in \cite{xian2022bayesian}.

In \cite{DunPP}, the random covariate-dependent probability measure $P_x$ is defined through a weighted mixture of $n$ independent random probability measures with weights constructed through kernel functions centered at the observed covariate values
\begin{equation*}
P_x=\sum_{i=1}^n \frac{w_i C(x,x_i)}{\sum_{i'=1}^n w_{i'} C(x,x_{i'})} P_i,
\end{equation*}
where $P_i \overset{\text{iid}}\sim \DP(\alpha, P_0)$ and $i=1,\ldots,n$ indexes subjects in the sample. The authors applied this approach for density regression. 
However, because the prior of $P_x$ depends on the sample size and observed covariates, it is unappealing from a Bayesian perspective and lacks desirable marginalization and updating properties (see \cite{DunBNP} for more details). For instance, the kernel stick-breaking process reviewed in the previous section shares some of the appealing characteristics of the kernel weighted specification but without the sample dependence.

Other proposals along the lines of (\ref{eq:cond_mod2}) focus exclusively on discrete categorical covariates where, for example, $x$ might indicate the hospital among, say $M$, hospitals, where a patient was treated. An interesting proposal for the law of $P_x$, in this setting, that allows to borrow strength across $P_1,\ldots,P_M$, is the hierarchical DP of \cite{Teh}, which assumes $P_x \mid P_0 \overset{\text{iid}}\sim \text{DP}(\alpha, P_0)$, for each $x=1,\ldots,M$, and models the random base measure $P_0$ nonparametrically, where $P_0 \sim \text{DP}(\gamma, H)$. In words, the multiple group specific distributions are assumed to be drawn from a common DP whose base measure is in turn a draw from another a DP. This allows the different distributions to share the same set of atoms but have distinct sets of weights. Recently, \cite{Diana2020} proposed the hierarchical dependent DP which combines the hierarchical and the (single-weights) dependent DP, and can be used as a prior for the mixing measure in the generic context where density estimation is to be performed jointly across different groups and in the presence of covariates. An application to model bird migration patterns in the UK motivated the development of the methods.
A further development is the nested DP (\cite{Rod}) where the model is given as  $P_x \mid Q \overset{\text{iid}}\sim Q$ and $ Q \sim \DP( \alpha, \DP(\gamma, H))$. By opposition to the hierarchical DP, for the nested DP, the different distributions have either the same atoms with the same weights or completely different atoms and weights.
Alternative proposals are given by \cite{MQR}, \cite{WM03}, and \cite{KGS11}, just to mention a few. In this setting, $x$ is just a label and the distance between two covariate values has no meaning. 

\section{Predictive comparison: strengths and drawbacks}\label{sec:comparison}
In this section, we provide three illustrative examples, constructed to highlight the drawbacks and strengths in predictive performance of the three general constructions, namely, the joint approach of Section \ref{sec:joint}; the conditional approach with covariate-dependent atoms and single-weights of Section \ref{sec:thetax}; and the conditional approach with covariate-dependent  weights of Section \ref{sec:wx}. For illustrative purposes, we consider a continuous response and covariates in all examples. For the joint approach, the specific models considered include the joint DP mixture model  (joint DP) \cite{MEW} and the joint EDP mixture model  (joint EDP) \cite{Wade2014}, in both cases with a linear regression kernel for $y|x$ and a marginal multivariate normal kernel for $x$. For the conditional approach with dependent atoms, we consider the single-weights DDP with a linear regression kernel  (LDDP) and the single-weights DDP that combines a cubic B-splines basis expansion with a linear regression kernel (LDDP-BS) \cite{Inacio2022}. 
For the conditional approach with dependent weights, the models included are two logit stick-breaking constructions \cite{Rigon2021} with linear dependence in the stick proportions (LSBP) and with nonlinear dependence through natural cubic splines (LSBP-NS), as well as the normalized weights (NW) of \cite{Antoniano2014}; in all three, a linear regression kernel is considered. We use B-splines in the single-weights approach and natural splines in the logit stick-breaking prior because this is the configuration used by the authors in the original cited articles. In both, we assume an additive splines expansion, with  interior knots placed at the quantiles of the covariates for the latter and no interior knots for the former. For the normalized weights, we employ a Gaussian kernel in all examples. 

To assess the predictive performance, we consider the error in the predictive regression function and conditional density, as well as the soundness of the uncertainty in the predictive quantities. Specifically, the regression error is measured by the root mean square error and the density error is computed based on the approximate $\ell_1$ distance between the predictive and true conditional density  averaged across all test points. In uncertainty quantification, we desire tight credible intervals (CIs) that cover the truth at the nominal level; thus, we also report the empirical coverage of the 95\% CIs of the predictive regression function, as well as the average length, along with visual comparisons.

\subsection{Example 1: Drawbacks of the Joint Approach}

\begin{figure*}[!h]
    \centering
    \subfloat[Joint DP: Covariate Space]{\includegraphics[width=0.3\textwidth]{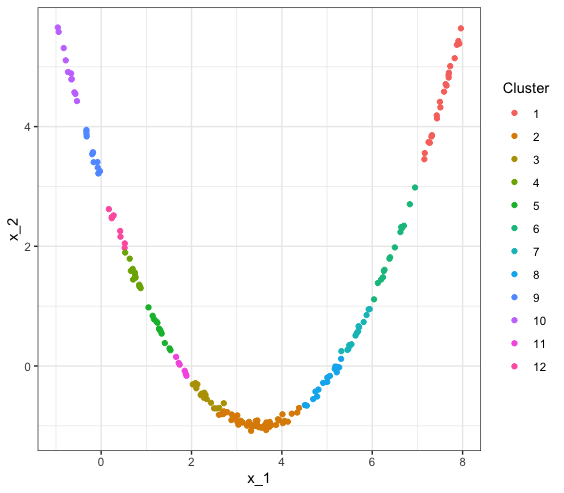}\label{fig:ex1_jointdp_clusters1}}
    \subfloat[Joint DP: Regression space]{\includegraphics[width=0.3\textwidth]{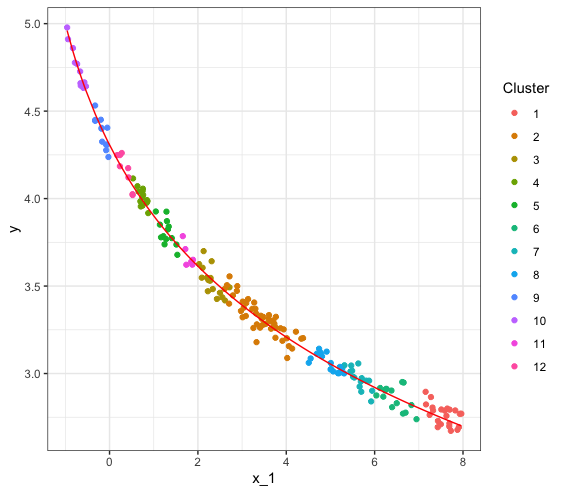}\label{fig:ex1_jointdp_clusters2}}
    \subfloat[Joint EDP: Regression space]{\includegraphics[width=0.3\textwidth]{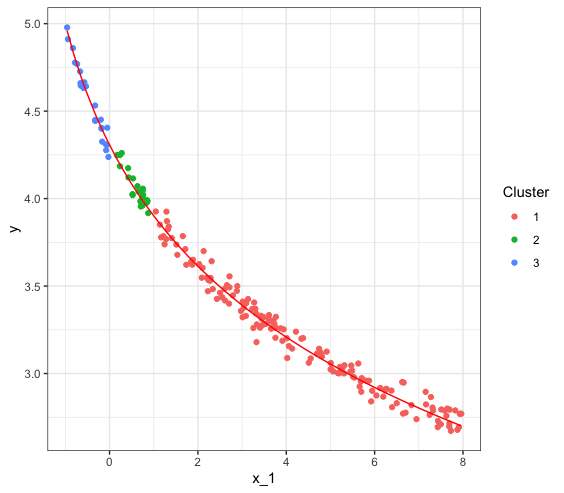}\label{fig:jointedp_clusters}}\\
    \subfloat[NW: Regression space]{\includegraphics[width=0.3\textwidth]{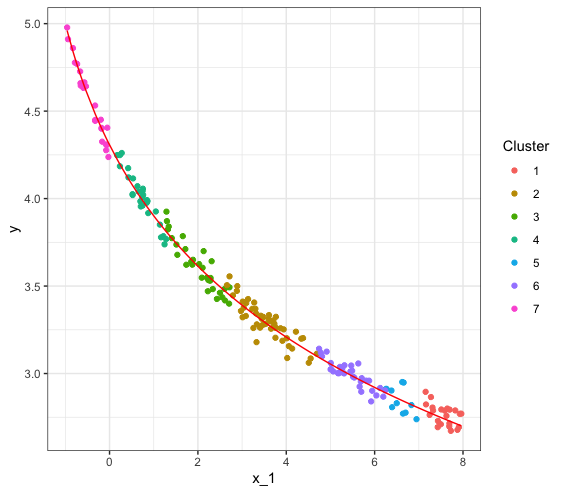}\label{fig:nwreg_clusters}}
    \subfloat[LDDP-BS: Regression space]
    {\includegraphics[width=0.26\textwidth]{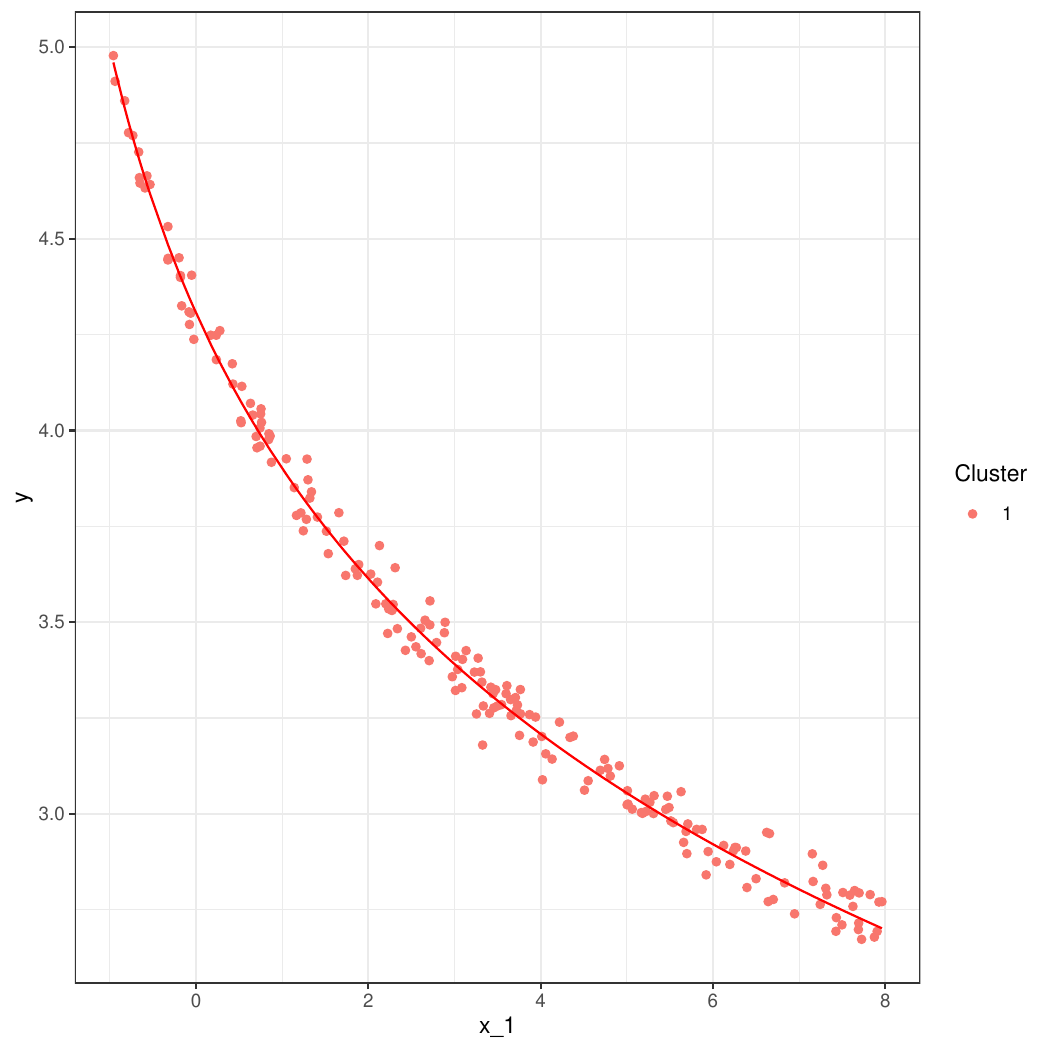}\label{fig:lddpbs_clusters}}
    \subfloat[LDDP: Regression space]
    {\includegraphics[width=0.26\textwidth]{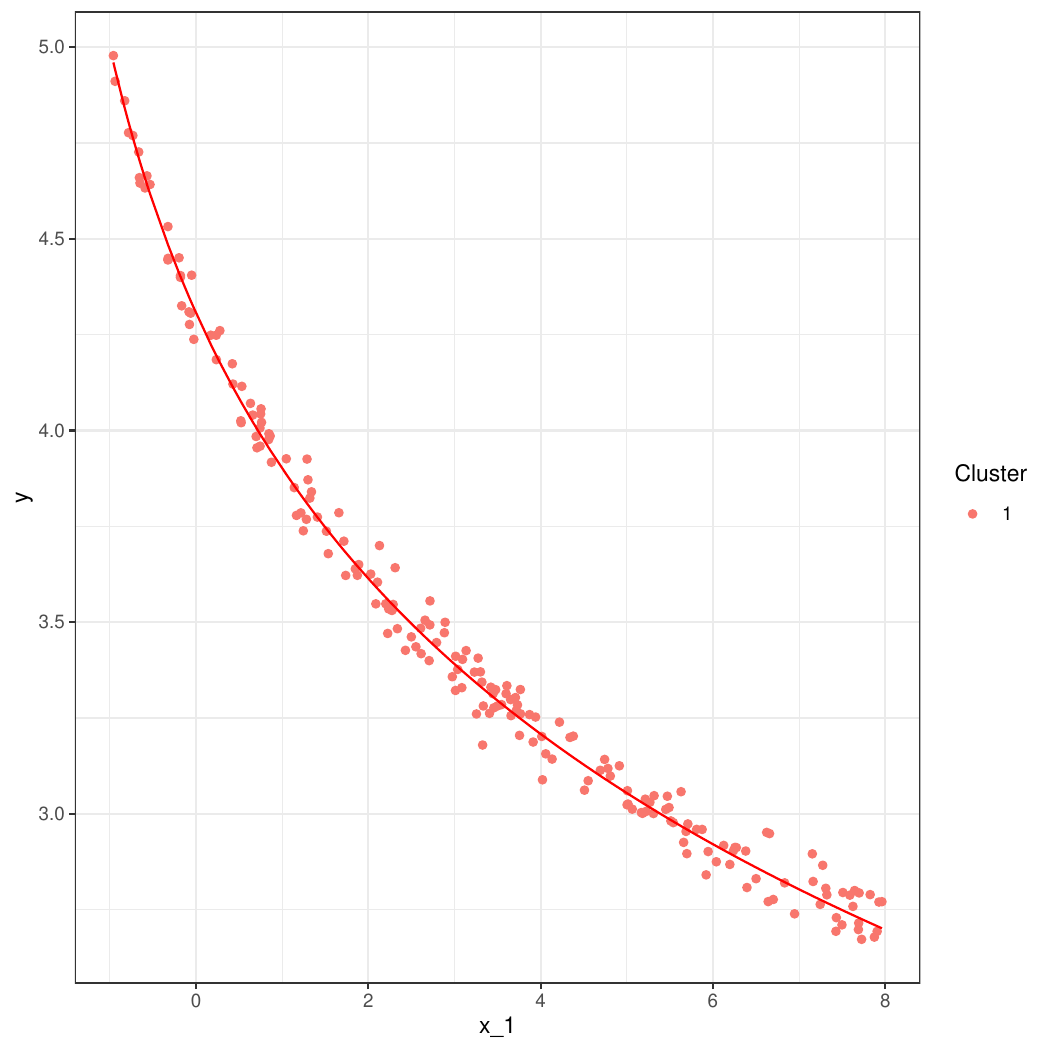}\label{fig:lddp_clusters}}
    \caption{Example 1. Estimated clustering.}
    \label{fig:ex1_clusters}
\end{figure*}

\begin{figure*}[!h]
    \centering
    \subfloat[Joint DP]{\includegraphics[width=0.25\textwidth]{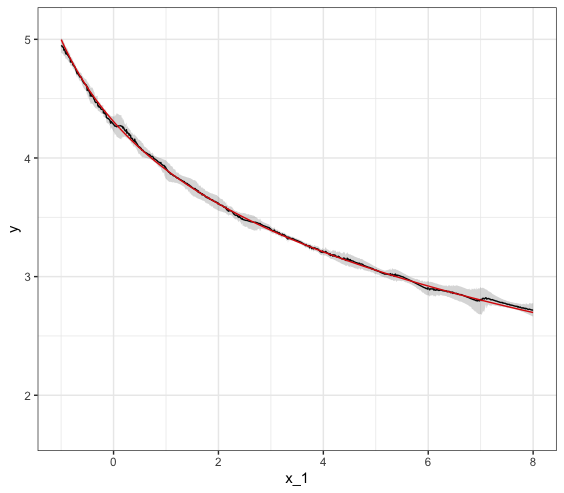}\label{fig:ex1_jointdp_predci}}
    \subfloat[Joint EDP]{\includegraphics[width=0.25\textwidth]{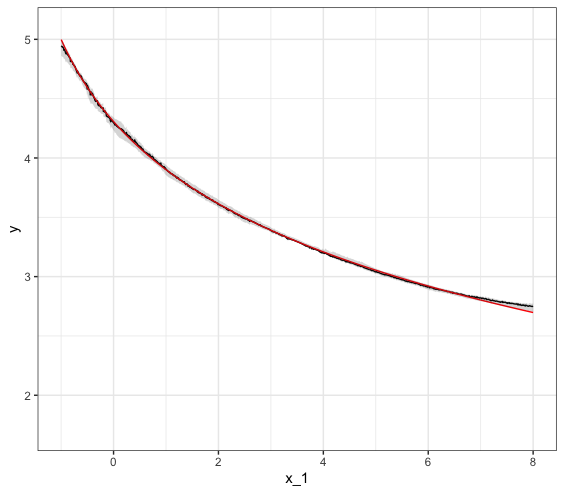}\label{fig:jointedp_predci}}
    \subfloat[NW]{\includegraphics[width=0.25\textwidth]{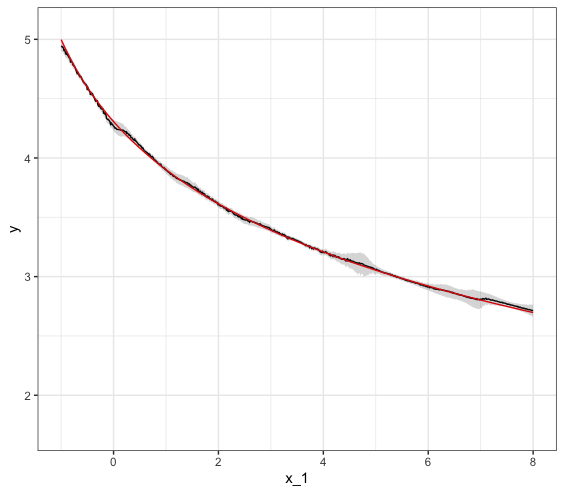}\label{fig:nwreg_predci}}\\
    \subfloat[LSBP]{\includegraphics[width=0.25\textwidth]{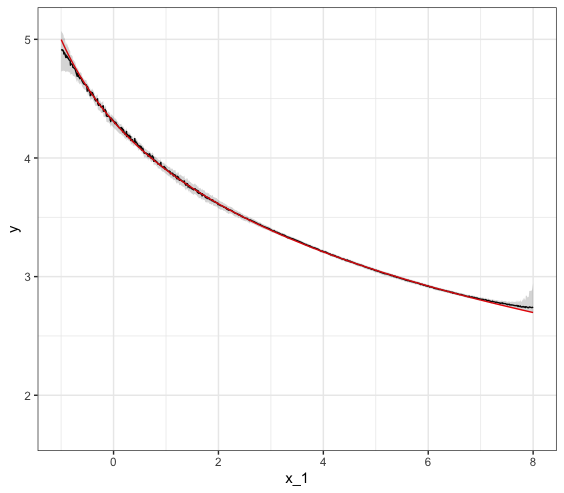}\label{fig:lsbp_predci}}
     \subfloat[LSBP-NS]{\includegraphics[width=0.25\textwidth]{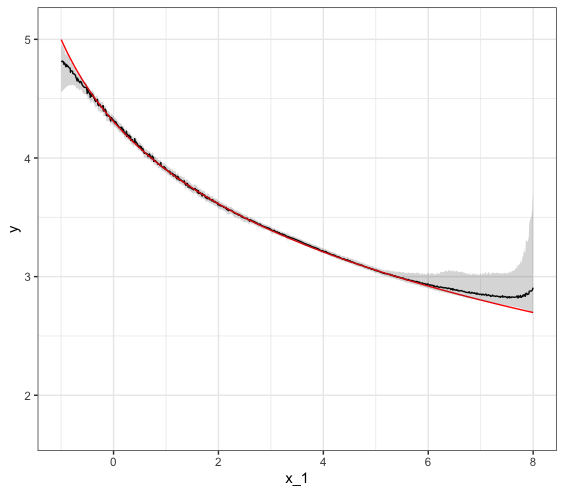}\label{fig:lsbp_predci}}
      \subfloat[LDDP-BS]{\includegraphics[width=0.22\textwidth]{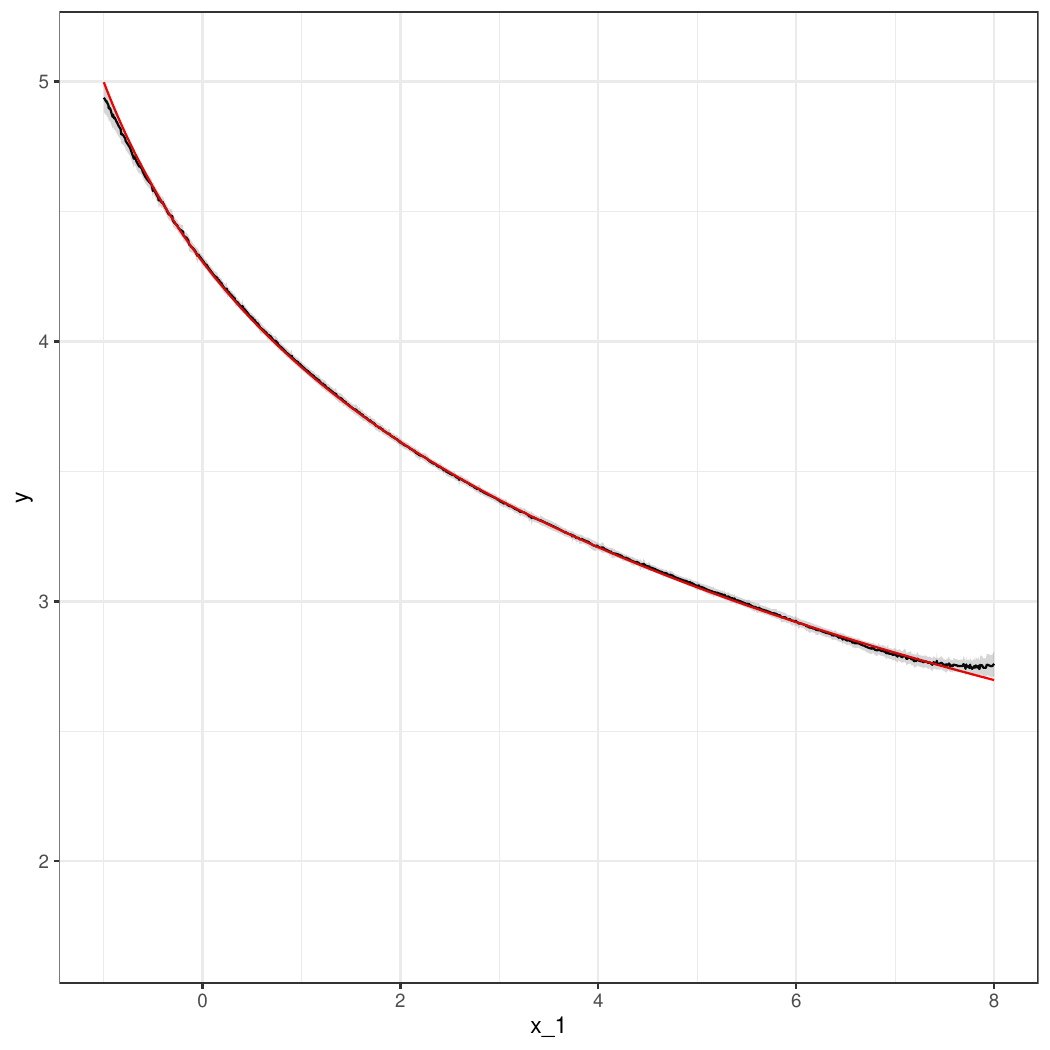}\label{fig:lddpbs_predci}}
   \subfloat[LDDP]{\includegraphics[width=0.22\textwidth]{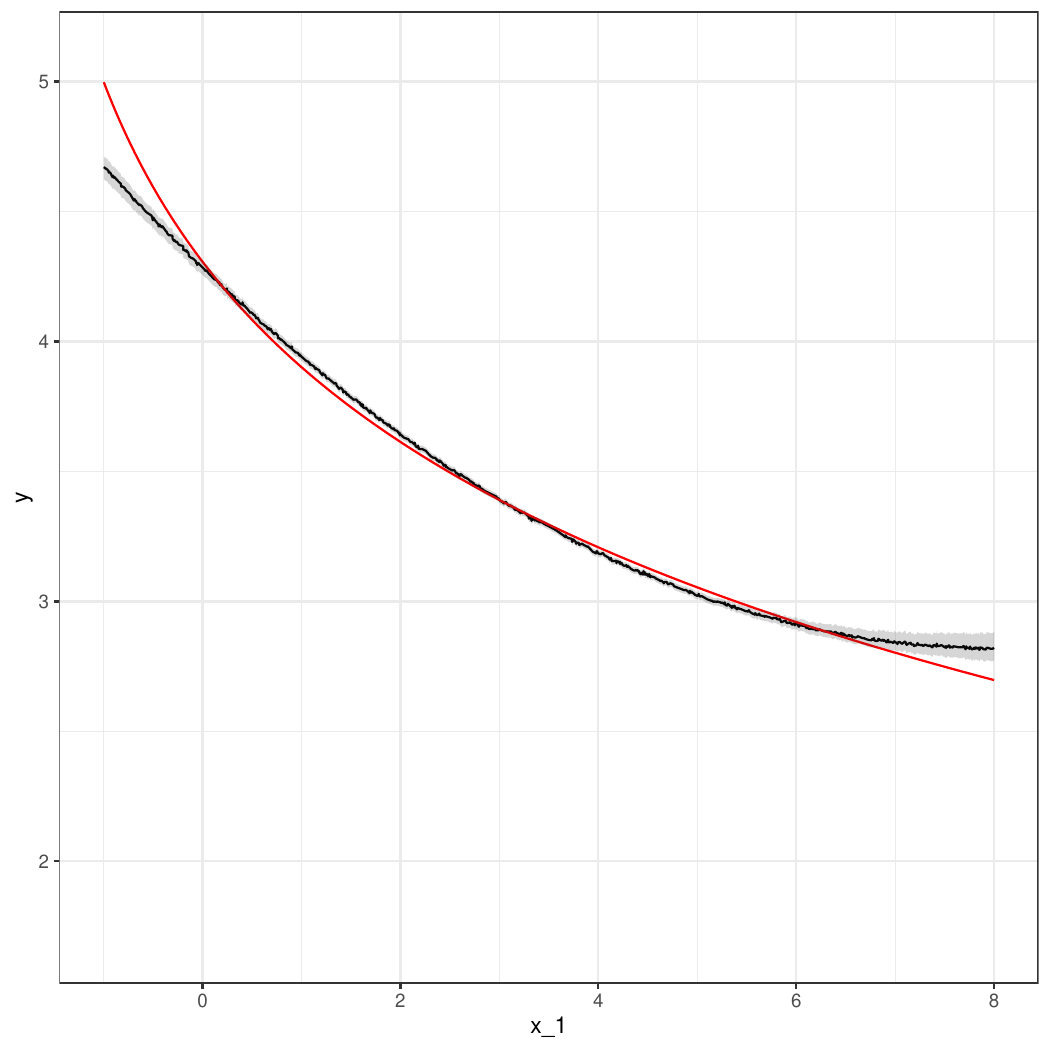}\label{fig:lddp_predci}}
      \caption{Example 1. Predictive regression function with pointwise 95\% CIs.}
    \label{fig:ex1_pred}
\end{figure*}

\begin{figure*}[!h]
    \centering
    \subfloat[Joint DP]{\includegraphics[width=0.25\textwidth]{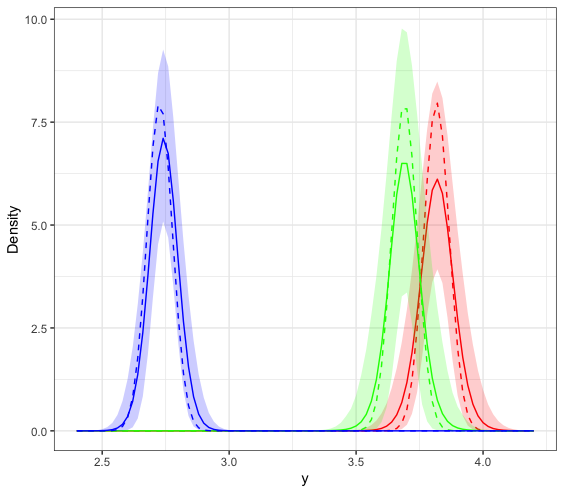}\label{fig:ex1_jointdp_densities}}
    \subfloat[Joint EDP]{\includegraphics[width=0.25\textwidth]{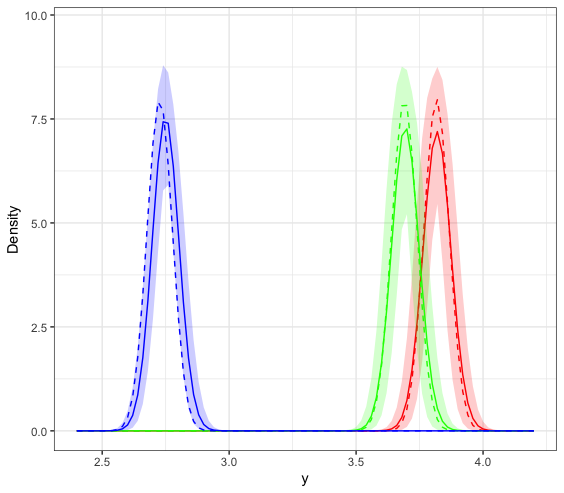}\label{fig:jointedp_densities}}
    \subfloat[NW]{\includegraphics[width=0.25\textwidth]{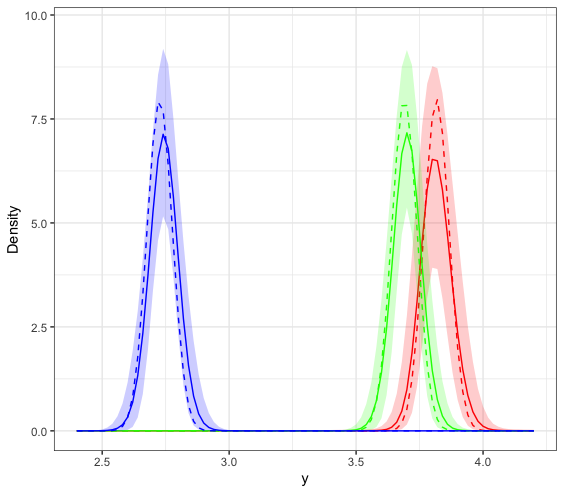}\label{fig:nwreg_densities}}\\
    \subfloat[LSBP]{\includegraphics[width=0.25\textwidth]{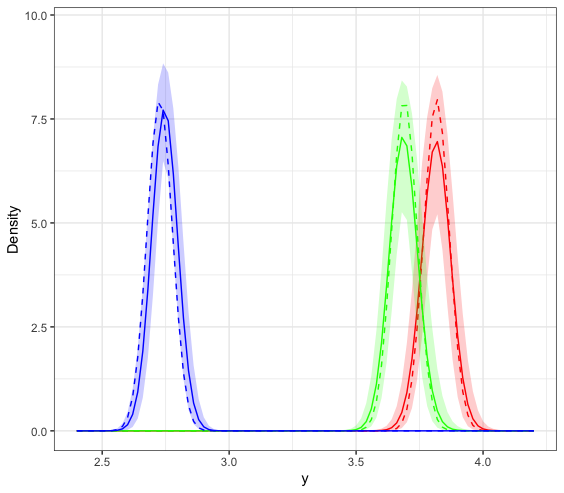}\label{fig:lsbp_densities}}
        \subfloat[LSBP-NS]{\includegraphics[width=0.25\textwidth]{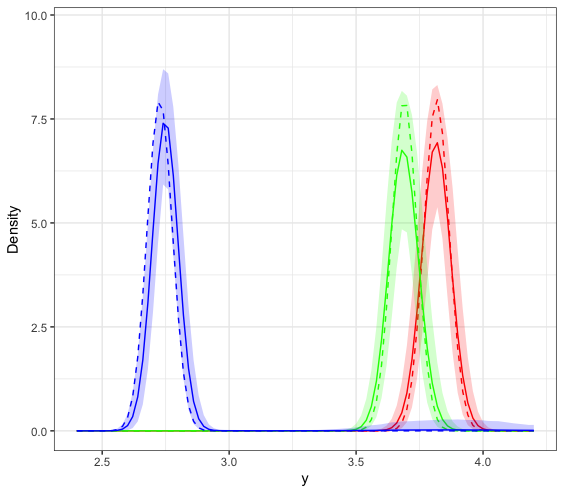}\label{fig:lsbp_densities}}
     \subfloat[LDDP-BS]{\includegraphics[width=0.22\textwidth]{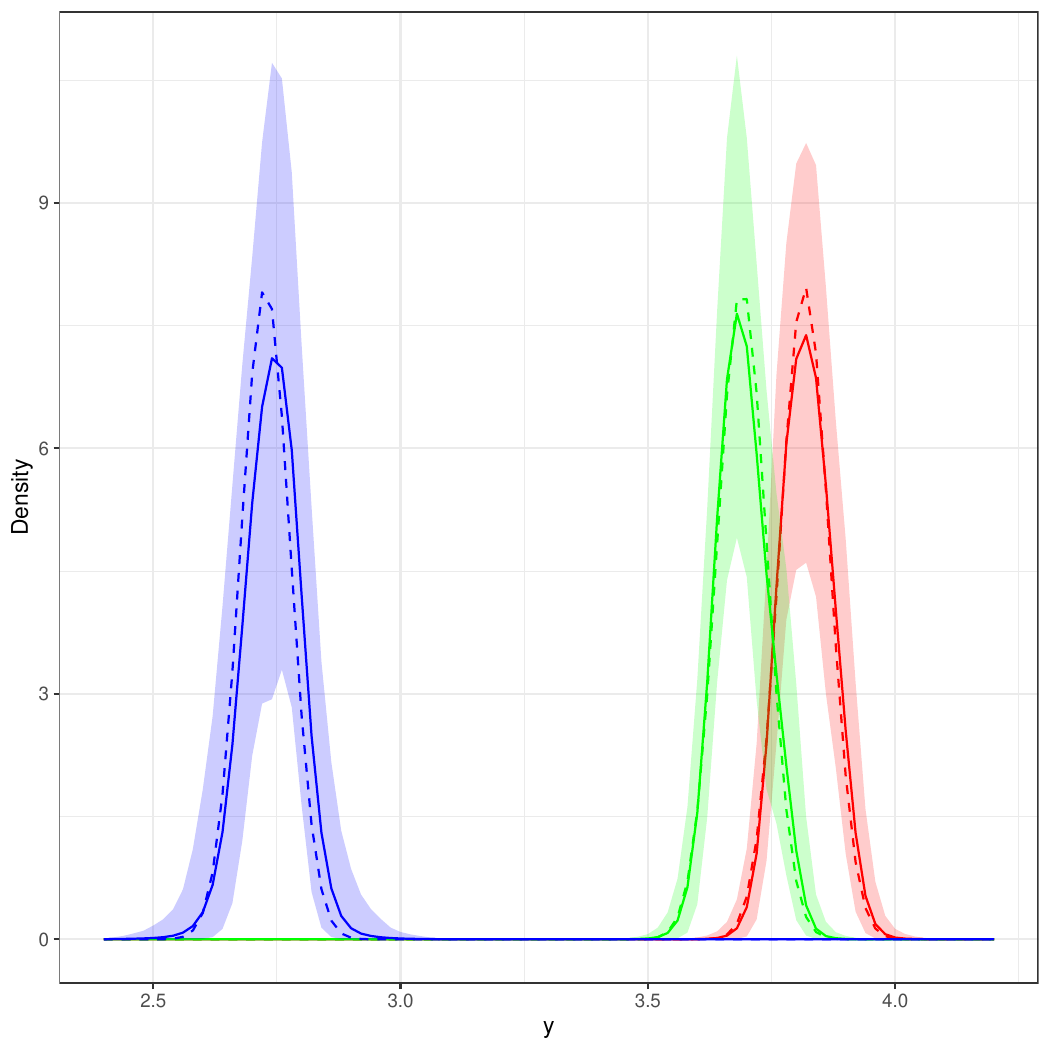}\label{fig:lddpbs_densities}}
 \subfloat[LDDP]{\includegraphics[width=0.22\textwidth]{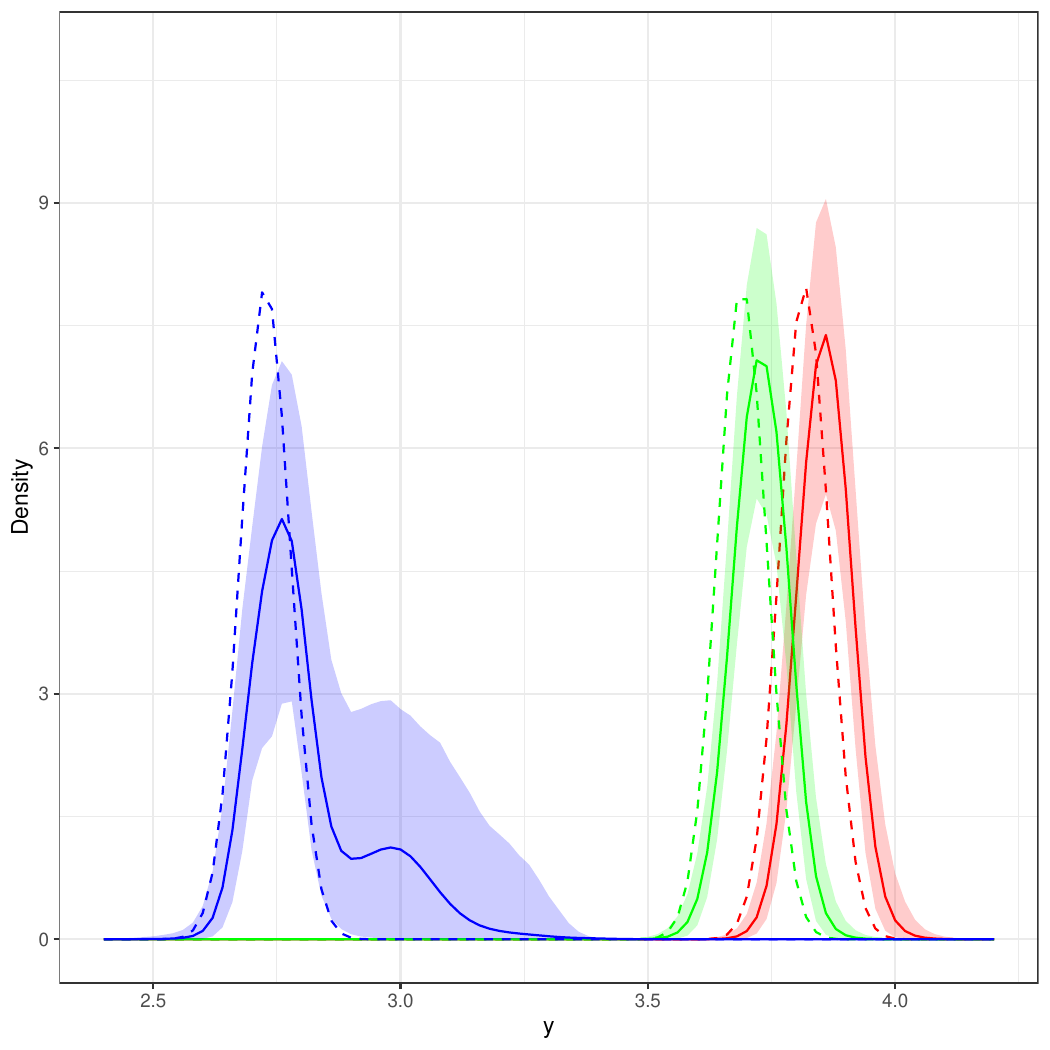}\label{fig:lddp_densities}}
    \caption{Example 1. Predictive density regression with 95\% pointwise CIs for three new covariate values.}
    \label{fig:ex1_dens}
\end{figure*}

\begin{table}[!h]
    \centering
    \begin{tabular}{c|cccc}
          Model & Regression Err & Density Err & Coverage & CI length \\ \hline
          Joint DP & 0.0149 & 0.2600 & 1 & 0.0943 \\
          Joint EDP & 0.0149 & 0.1669 & 0.9438  & 0.0594\\
          NW & 0.0140  & 0.2313 & 0.9750  & 0.0716 \\
          LSBP & 0.0137  & 0.1456 & 0.9888   & 0.0570\\
          LSBP-NS & 0.0433  & 0.2005 & 0.8575   & 0.1216\\
          LDDP-BS & 0.0117 & 0.1522 & 0.9563 & 0.0369\\
          LDDP & 0.0655 & 0.4972 & 0.2838 & 0.0459
    \end{tabular}
    \caption{Summary of results for Example 1. The \textit{regression error} computes the root mean square error between the predictive regression function and the truth; the \textit{density error} reports the $\ell_1$ distance between the predictive and true conditional density  averaged across all test points; the \textit{empirical coverage} of the predictive regression function is the fraction of times the 95\% credible interval  (CI) contains the truth; and the \textit{average CI length} of the predictive regression function reflects the average length of the 95\% CI.  }
    \label{tab:ex1}
\end{table}

When the aim is density regression, a potential downside of the joint approach is that inference is based on the joint likelihood. 
Specifically, if the distribution of the covariates is complex, the marginal distribution of $x$ may drive inference and a large number of distinct mixture components (clusters) will typically be needed to approximate the complex marginal density of $x$, even if the conditional distribution may be more well behaved and described by fewer components. 

As a consequence, inference on conditional parameters is less efficient, which in turn may produce degraded predictive performance and unnecessarily wide CIs due to the smaller sample sizes within each cluster. 

To illustrate this, we simulate $n=200$ data points with $p=2$ covariates; $y$ only depends on $x_1$ and the relationship is relatively well-behaved:
\begin{align*}
    y_i = 5-\log(x_{i,1}+2) +\epsilon_i, \quad \epsilon_i \iidsim \Norm(0, 0.05^2).
\end{align*}
However, the distribution of the covariates is complex:
\begin{align*}
    x_{i,1} &\iidsim \Unif(-1,8),\\
    x_{i,2} \mid x_{i,1} & \overset{ind}\sim \Norm\left( (x_{i,1}-3.5)^2/3-1,0.05^2 \right).
\end{align*}
For all models, the evaluation metrics are reported in Table \ref{tab:ex1}, a point estimate of the clustering \cite{wade2018bayesian} is provided in Figure \ref{fig:ex1_clusters}, and the predictive regression function and conditional densities along with 95\% pointwise CIs are shown in Figures \ref{fig:ex1_pred} and \ref{fig:ex1_dens}, respectively. 

In this setting, the single-weights LDDP-BS performs the best. Indeed, only a single cluster is required for this model with flexible atoms, leading to estimates based on large sample/cluster sizes,  efficient predictions (smallest errors) and improved uncertainty quantification (tightest intervals at the desired coverage). However, if the atoms are not flexible enough, as for the case of linear atoms in the LDDP, predictive inference is poor.

For the joint model, the complex covariate distribution leads to over-partitioning and small clusters (Figures \ref{fig:ex1_jointdp_clusters1} and \ref{fig:ex1_jointdp_clusters2}), with all MCMC samples having between 11 and 17 clusters. This causes a (slight) drop in the predictive performance and larger uncertainty, which is visibly evident in Figures \ref{fig:ex1_jointdp_predci} and \ref{fig:ex1_jointdp_densities}. The two-level clustering of the joint EDP, which allows for a  smaller number of clusters for the conditional density with further nested clusters for the covariates, helps to rectify this behavior. At the outer level of partitions, 90\% of the MCMC samples from the joint EDP have two to three clusters to approximate the nonlinear regression function. Further improvements are possible by employing nonlinear regression kernels, such as the GP regression kernel in \cite{pmlr-v108-gadd20a}. 

The models with dependent weights probabilistically partition the covariate space for local linear approximation of the nonlinear regression function. The LSBP performs the best among these models, and the results are slightly worse than LDDP-BS, as more clusters are required but better than the joint DP in all metrics. In this example, the stick-breaking formulation with linearity (LSBP) outperforms the stick-breaking formulation with natural cubic splines (LSBP-NS), as the partitioning of the covariate space required is relatively simple for this data; the extra flexibility of LSBP-NS is not necessary and increases the parameter space leading to worse predictive inference. We note that the estimated clustering for the LSBP and LSBP-NS is not shown because the \texttt{R} package accompanying the article does not store and return the latent allocation variables in the MCMC output. 

Empirical priors are employed, which is an important, often overlooked aspect, and we highlight in the Appendix \ref{sec:lddp_bs_prior} how vague priors degrade predictions in this example (for conciseness, focusing on a single model, the LDDP-BS).    

\subsection{Example 2: Drawbacks of the Conditional Approach with Dependent Atoms}

\begin{figure*}[!h]
    \centering
    \subfloat[Joint DP]{\includegraphics[width=0.3\textwidth]{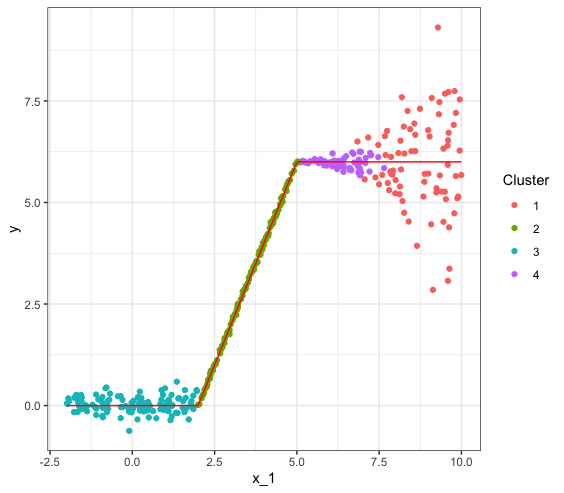}\label{fig:ex2_jointdp_clusters}}
    \subfloat[Joint EDP]{\includegraphics[width=0.3\textwidth]{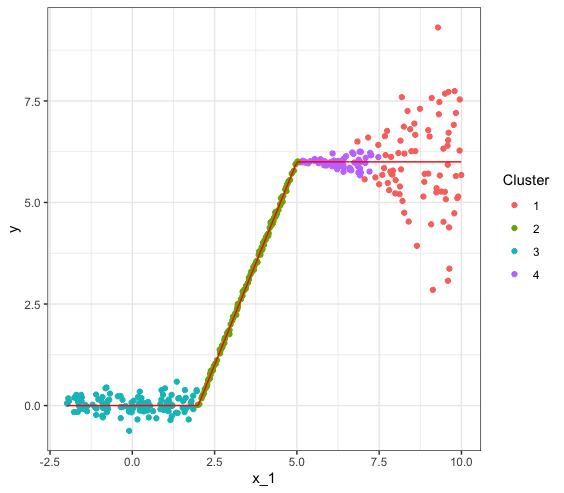}\label{fig:ex2_jointedp_clusters}}
    \subfloat[NW]{\includegraphics[width=0.3\textwidth]{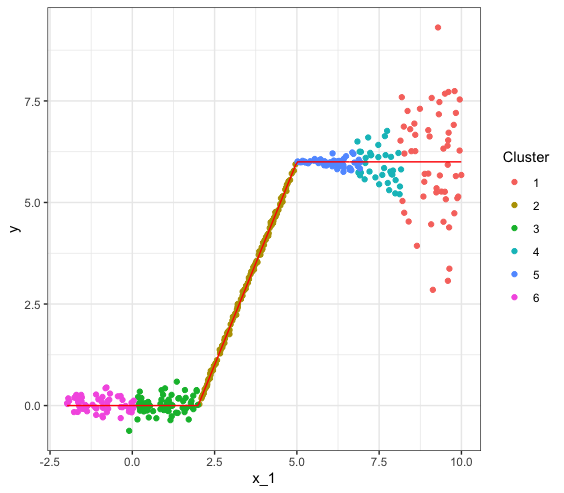}\label{fig:nwreg_clusters}}\\  
      \subfloat[LDDP-BS]{\includegraphics[width=0.26\textwidth]{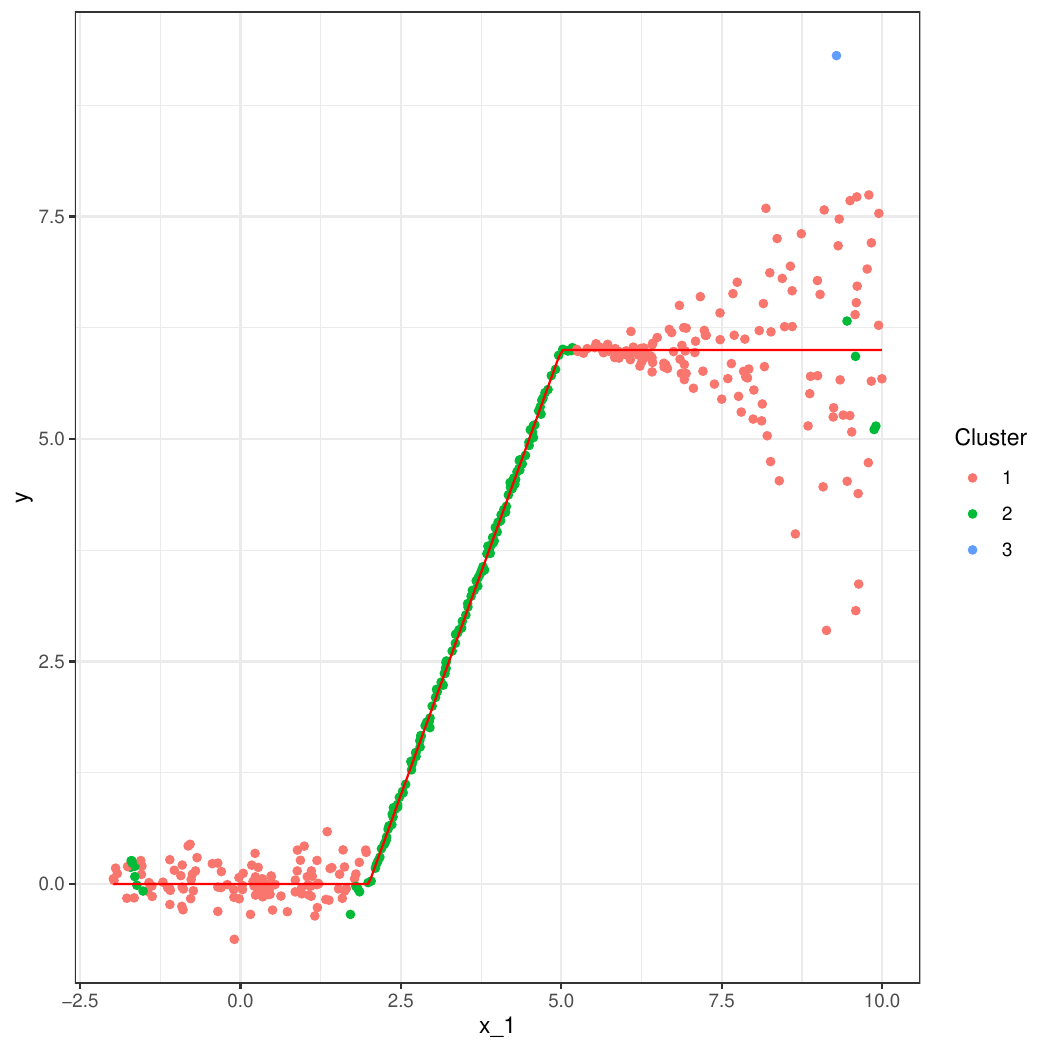}\label{fig:ex2_lddpbs_clusters}}
    \subfloat[LDDP]{\includegraphics[width=0.26\textwidth]{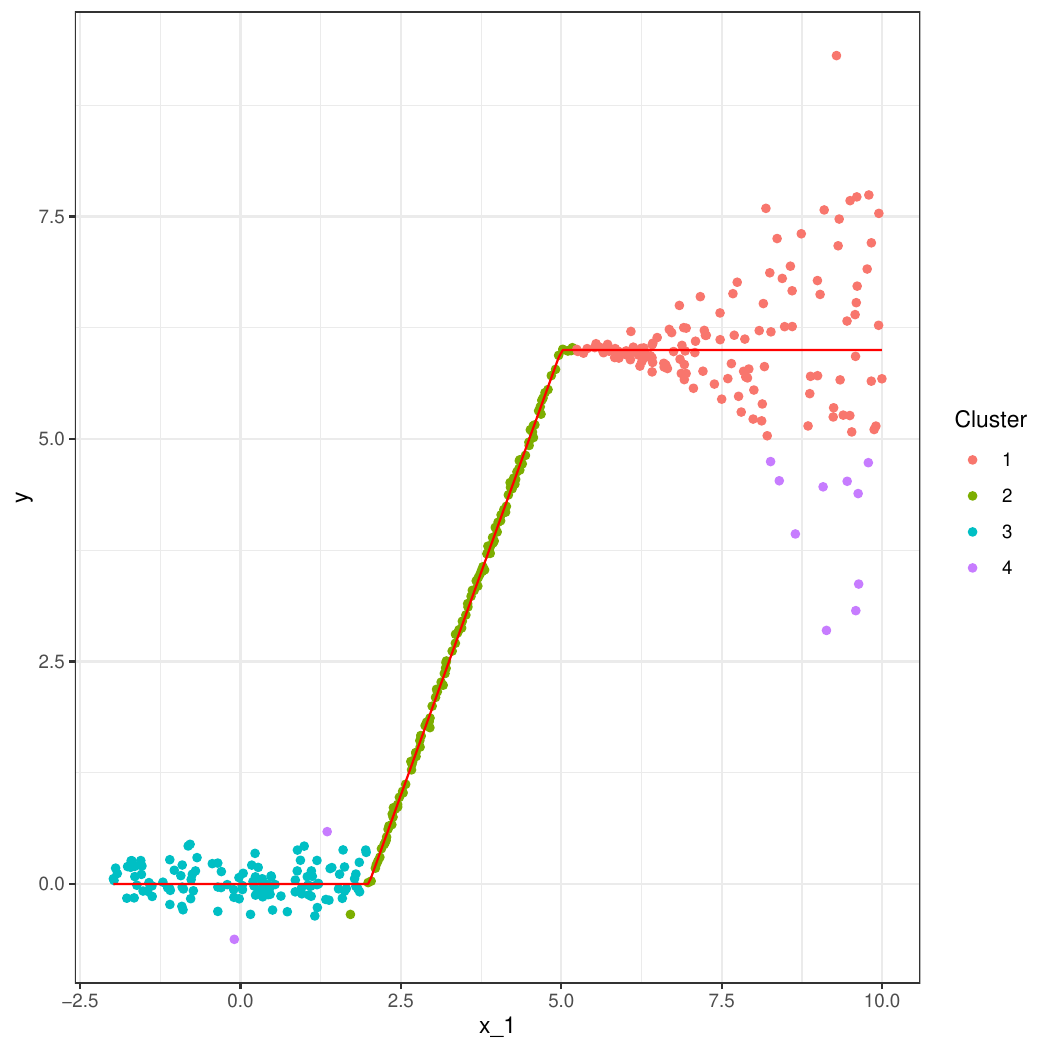}\label{fig:ex2_lddp_clusters}}   
    \caption{Example 2. Estimated clustering}
    \label{fig:ex2_clusters}
\end{figure*}

\begin{figure*}[!h]
    \centering
    \subfloat[Joint DP]{\includegraphics[width=0.25\textwidth]{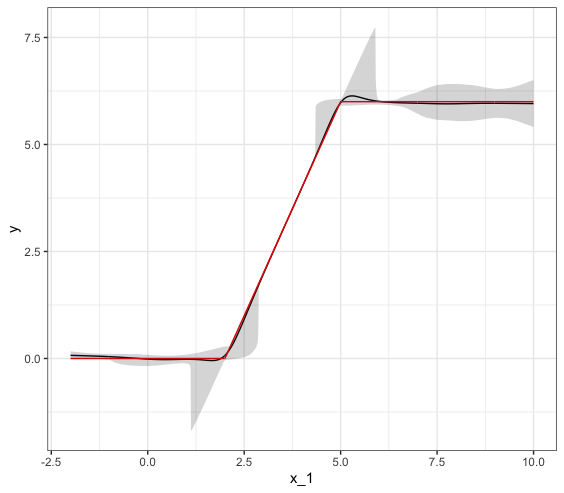}\label{fig:jointdp_predci}}
    \subfloat[Joint EDP]{\includegraphics[width=0.25\textwidth]{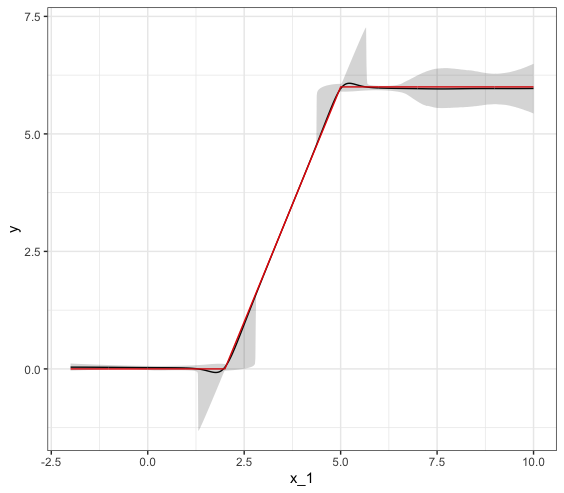}\label{fig:jointedp_predci}}
    \subfloat[NW]{\includegraphics[width=0.25\textwidth]{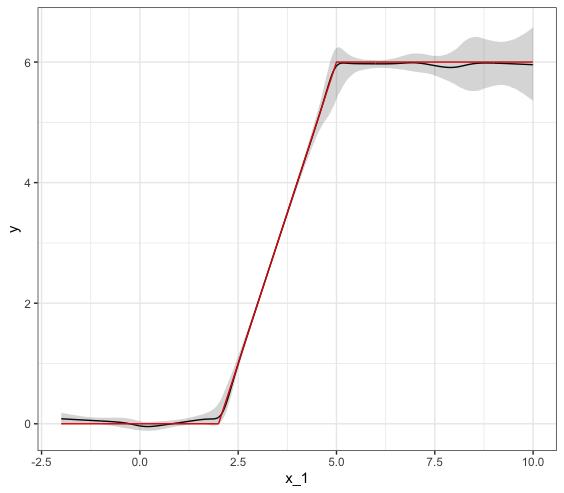}\label{fig:nwreg_predci}}\\
    \subfloat[LSBP]{\includegraphics[width=0.25\textwidth]{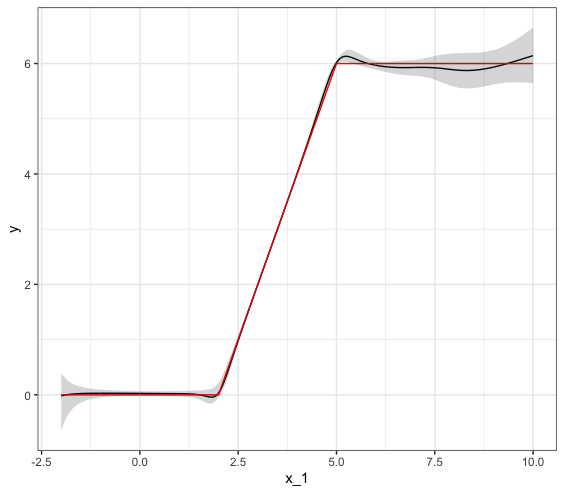}\label{fig:lsbp_predci}}
    \subfloat[LSBP-NS]{\includegraphics[width=0.25\textwidth]{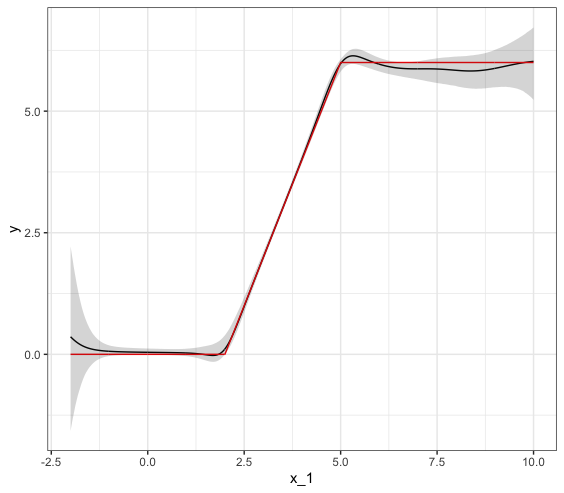}\label{fig:lsbp_predci}}
    \subfloat[LDDP-BS]{\includegraphics[width=0.22\textwidth]{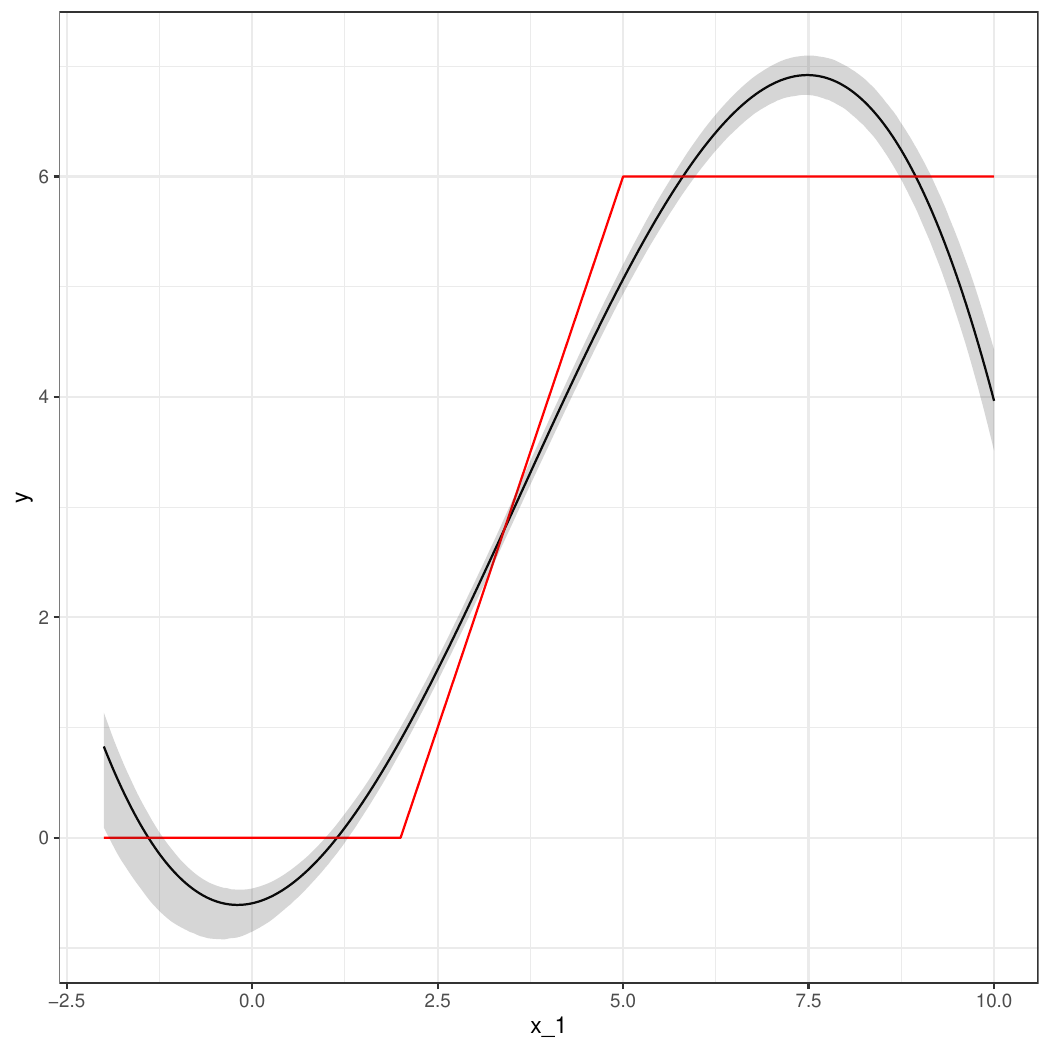}\label{fig:ex2_lddpbs_predci}}
    \subfloat[LDDP]{\includegraphics[width=0.22\textwidth]{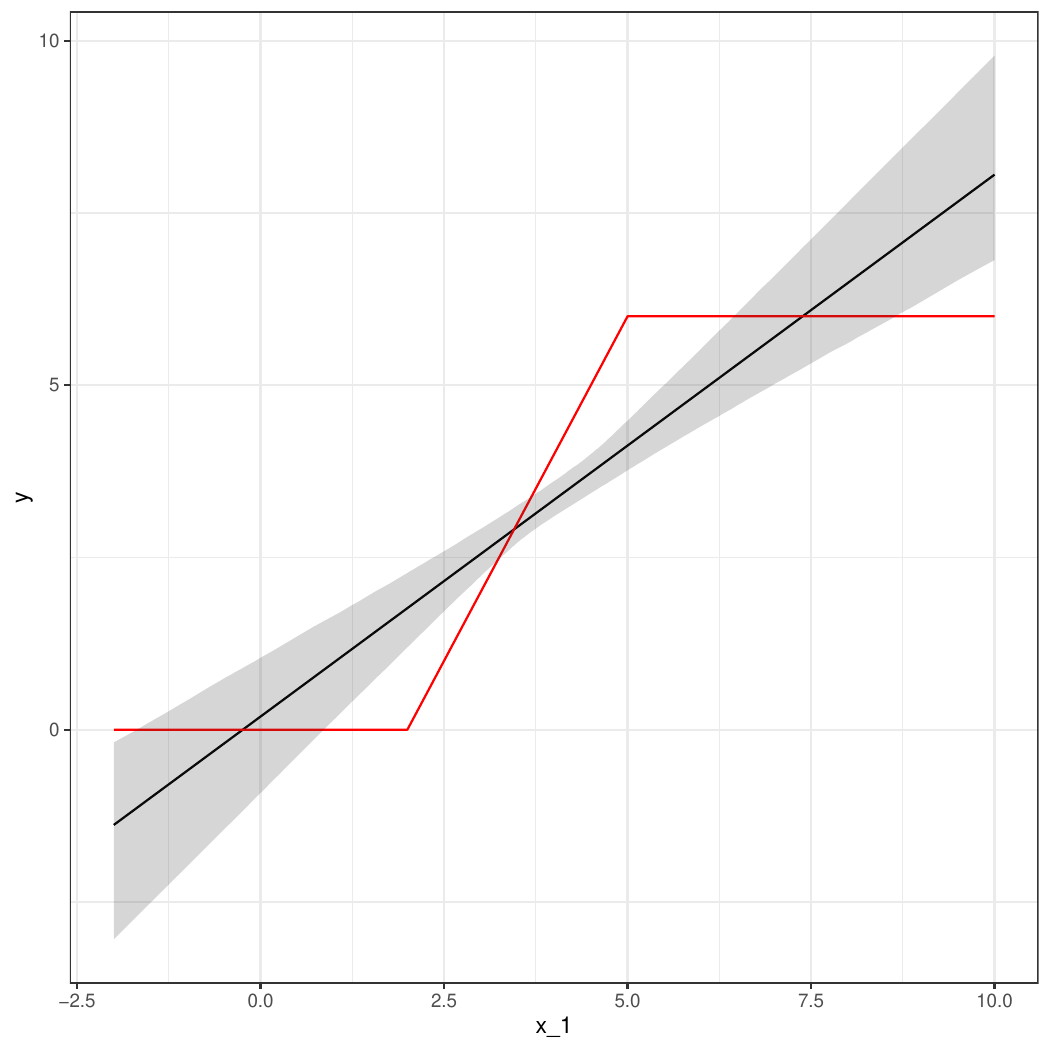}\label{fig:ex2_lddp_predci}}
    \caption{Example 2. Predictive regression function with pointwise 95\% CIs.}
    \label{fig:ex2_pred}
\end{figure*}

\begin{figure*}[!h]
    \centering
    \subfloat[Truth]{\includegraphics[width=0.25\textwidth]{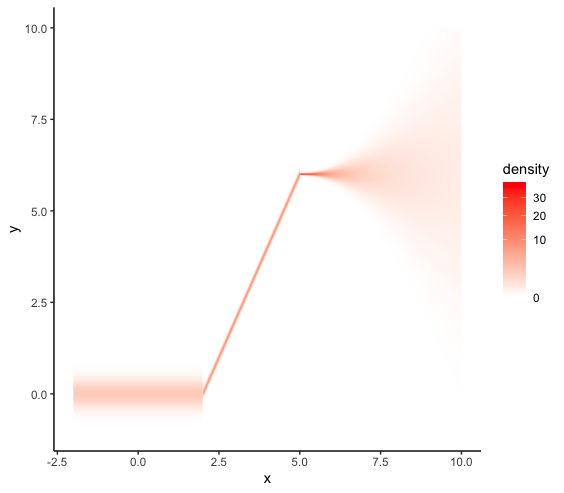}\label{fig:pred}}
    \subfloat[Joint DP]{\includegraphics[width=0.25\textwidth]{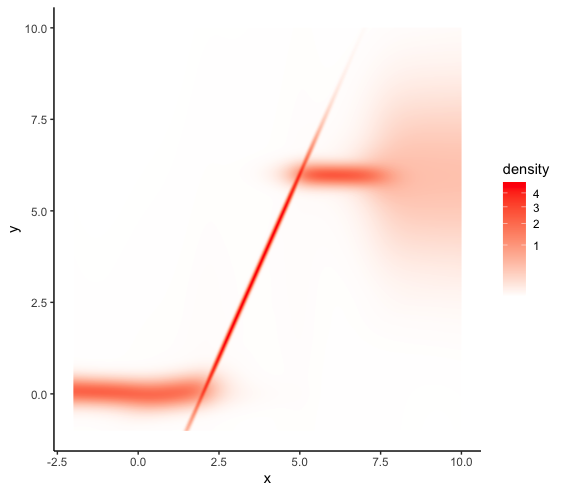}\label{fig:jointdp_densities}}
    \subfloat[Joint EDP]{\includegraphics[width=0.25\textwidth]{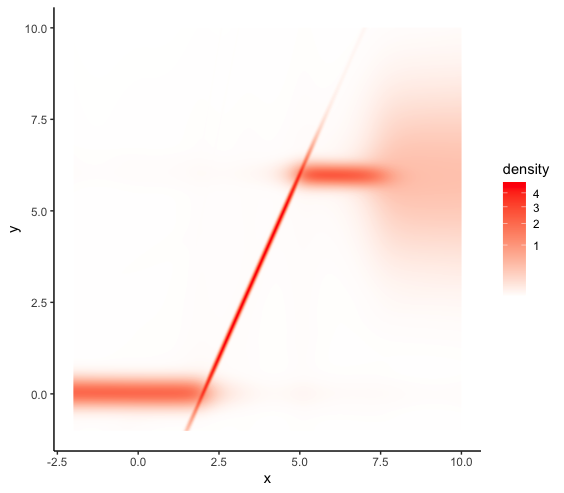}\label{fig:jointedp_densities}}
    \subfloat[NW]{\includegraphics[width=0.25\textwidth]{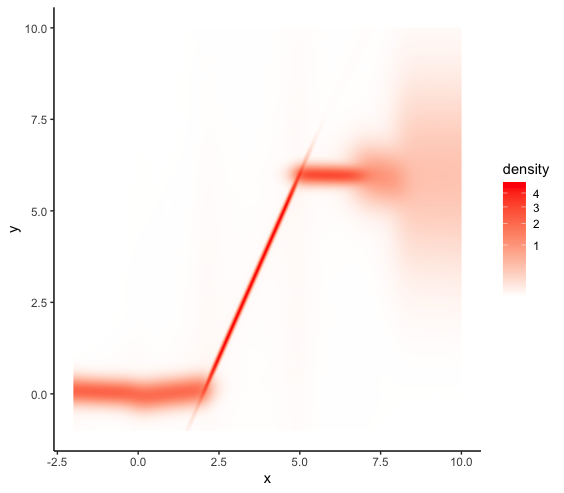}\label{fig:nwreg_densities}}\\
    \subfloat[LSBP]{\includegraphics[width=0.25\textwidth]{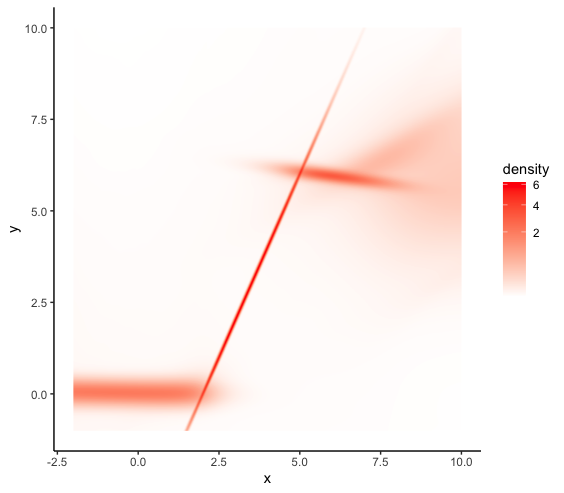}\label{fig:lsbp_densities}}
    \subfloat[LSBP-NS]{\includegraphics[width=0.25\textwidth]{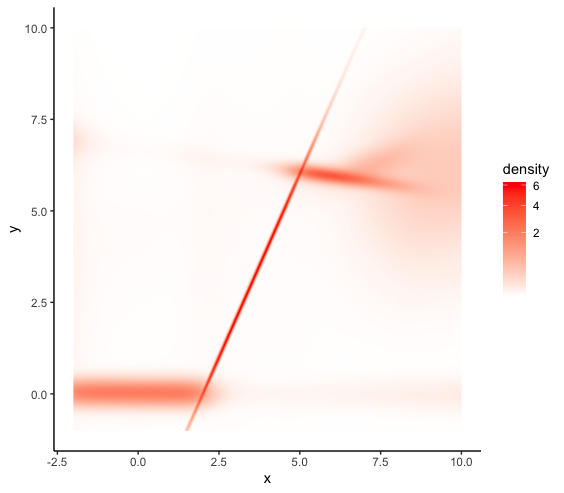}\label{fig:lsbp_densities}}
     \subfloat[LDDP-BS]{\includegraphics[width=0.22\textwidth]{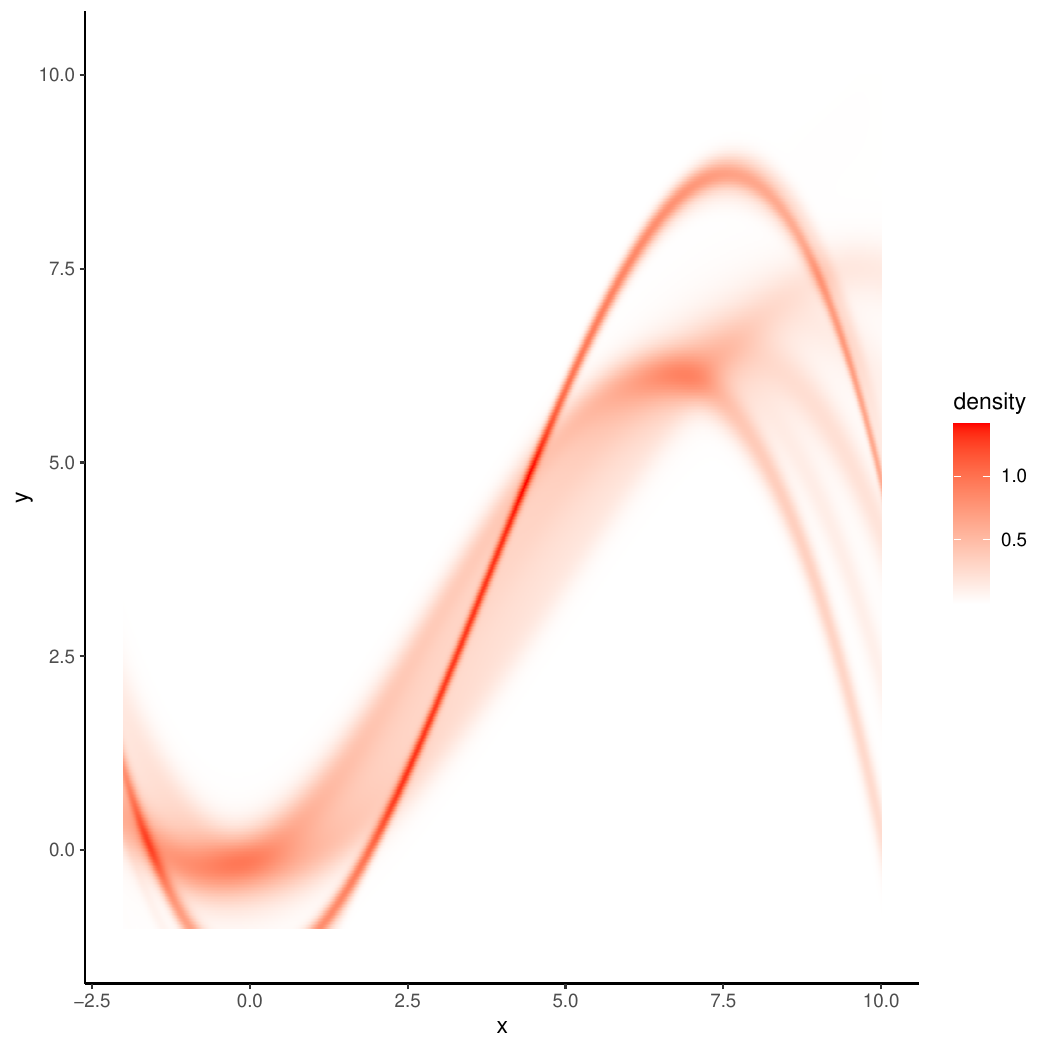}\label{fig:ex2_lddpbs_densities}}
      \subfloat[LDDP]{\includegraphics[width=0.22\textwidth]{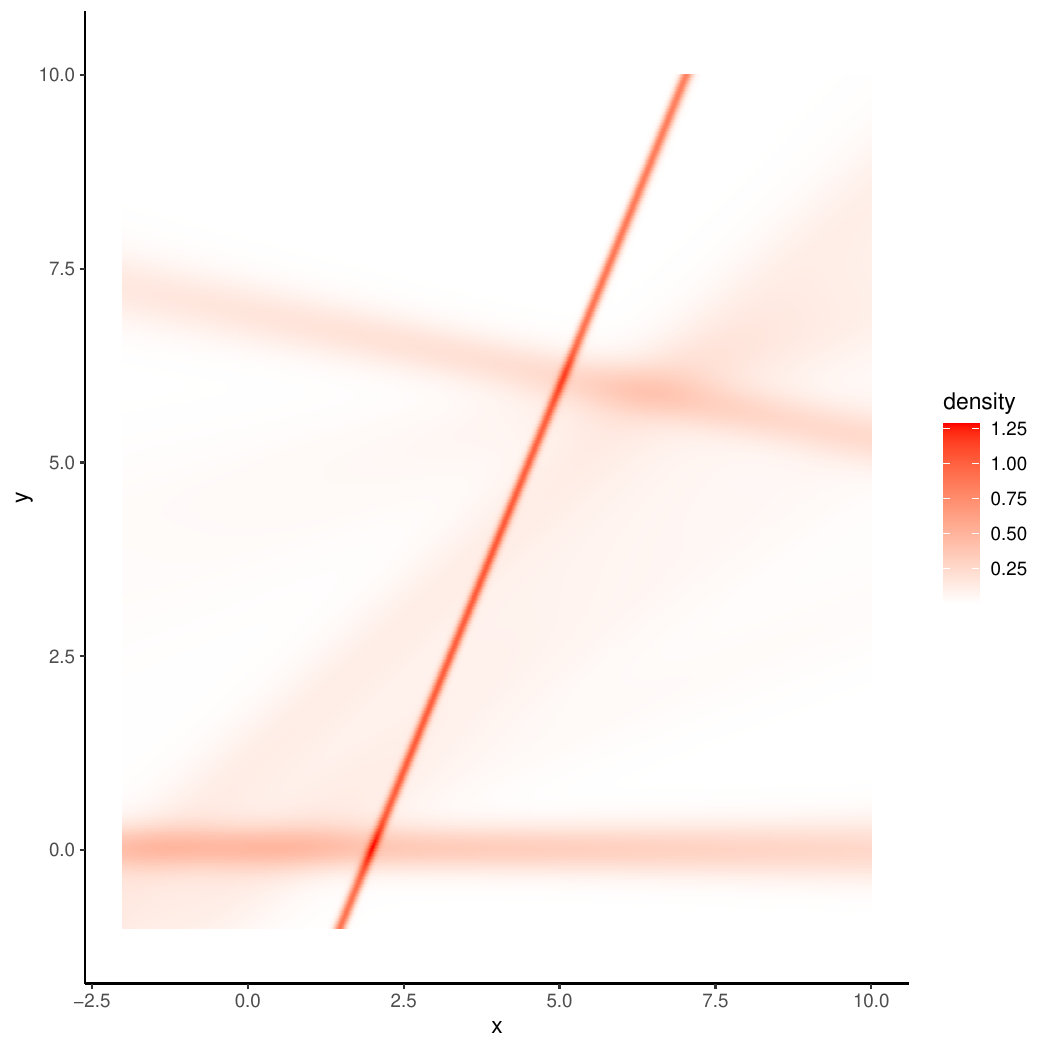}\label{fig:ex2_lddp_densities}}
    \caption{Example 2. Heatmap of the true and estimated density regression.}
    \label{fig:ex2_densities}
\end{figure*}

\begin{table}[!h]
    \centering
    \begin{tabular}{c|cccc}
          Model & Regression Err & Density Err & Coverage & CI length \\ \hline
          Joint DP & 0.0457   & 0.4085    & 1 & 0.5303 \\
          Joint EDP & 0.0344  & 0.3988 & 1 & 0.4263 \\
          NW & 0.0403 & 0.3714 & 0.9534  & 0.3288 \\
          LSBP & 0.0594 & 0.3081 & 0.9351 & 0.2830 \\
          LSBP-NS & 0.0946 & 0.3691 & 1 & 0.4425 \\
          LDDP-BS & 0.6104 & 1.1064 & 0.1799 & 0.3954\\
          LDDP & 1.0517 & 1.2338 & 0.4355 & 1.5792
    \end{tabular}
    \caption{Summary of results for Example 2. 
    }
    \label{tab:ex2}
\end{table}

Practitioners should be aware of a crucial limitation of the conditional approach with dependent atoms: if the specified regression kernel and dependent atoms are not sufficiently flexible to recover the true dependence, poor predictions may result. A simple example of this was provided in \cite{WWP12}, where linear atoms are considered yet the true regression function is quadratic, resulting in extremely poor predictions that lie outside of  the range of the data. While this can be resolved by using more flexible atoms, in this example, we highlight that even with widely-used flexible choices (e.g. splines or Gaussian processes), such issues may still arise. In particular, we assume $n=400$ data points are simulated as:
\begin{align*}
    y_i &= m(x_i) + \epsilon_i,\\
    m(x) &= \left\lbrace \begin{array}{ll}
       0  &  \text{if } x \leq 2,\\
        2x-4 & \text{if } 2<x\leq 5,\\
        6 & \text{if } x> 5,
    \end{array}\right. \\
    \epsilon_i &\overset{ind}\sim \left\lbrace \begin{array}{ll}
       \Norm(0,0.2^2)  &  \text{ if } x \leq 2,\\
        \Norm(0,0.05^2)& \text{ if } 2<x\leq 5,\\
        \Norm(0,(x-5)^2/15+0.01)& \text{ if } x> 5,
    \end{array}\right.\\
    x_i&\overset{iid}\sim \Unif(-2,10).
\end{align*}
Notice that the error distribution changes with $x$; in addition, the true regression function is nonstationary.  
Again, evaluation metrics are reported in Table \ref{tab:ex2}, the estimated clustering is depicted in Figure \ref{fig:ex2_clusters}, and the predictive regression function and heatmap of the conditional densities are shown in Figures \ref{fig:ex2_pred} and \ref{fig:ex2_densities}, respectively.

For the single-weights models with dependent atoms,  flexibility in the mean function is not sufficient to recover the covariate-dependent variance in this example. To fit the data, the estimated partition structure in Figures \ref{fig:ex2_lddpbs_clusters} and \ref{fig:ex2_lddp_clusters} depends on $x$; however, predictions across each cluster are averaged regardless of the covariate value, resulting in quite poor predictive inference in the regression function, density estimates, and uncertainty, which is visibly evident in Figures \ref{fig:ex2_lddpbs_predci}, \ref{fig:ex2_lddp_predci}, \ref{fig:ex2_lddpbs_densities}, and \ref{fig:ex2_lddp_densities}. This highlights that when employing single-weight models, it is important to examine the partition structure a posteriori to ensure it does not depend on $x$, which however may be challenging when $x$ is multivariate and of mixed nature. 

Instead, both the joint models and the models with dependent weights are able to adapt to recover the challenges of this dataset. Again, the LSBP provides a small improvement   compared with LSBP-NS, and the additional flexibility in the cubic spline formulation of the LSBP-NS is unnecessary.

\subsection{Example 3: Drawbacks of the Conditional Approach with Dependent Weights}

\begin{figure*}[!h]
    \centering
    \subfloat[Joint DP]{\includegraphics[width=0.3\textwidth]{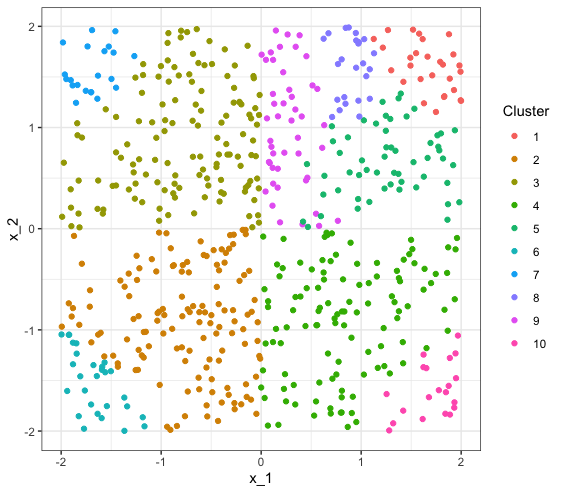}\label{fig:ex3_jointdp_clusters}}
    \subfloat[Joint EDP]{\includegraphics[width=0.3\textwidth]{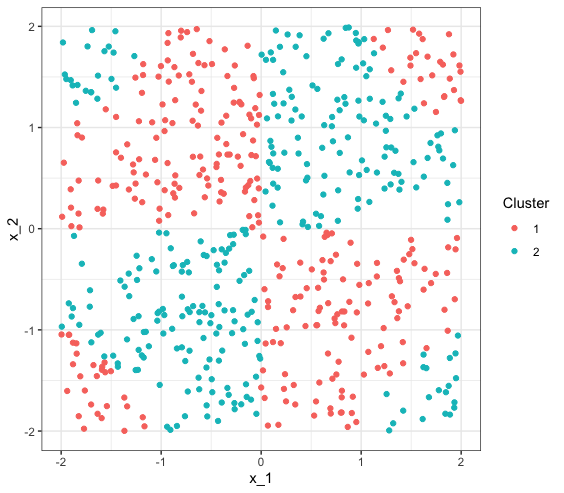}\label{fig:ex3_jointedp_clusters}}
    \subfloat[NW]{\includegraphics[width=0.3\textwidth]{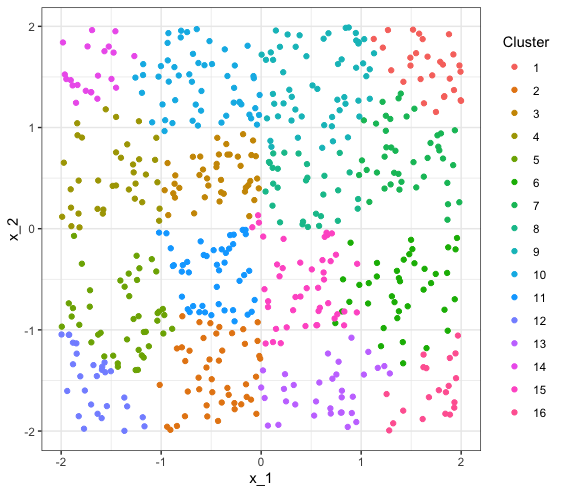}\label{fig:nwreg_clusters}}\\
     \subfloat[LDDP-BS]{\includegraphics[width=0.26\textwidth]{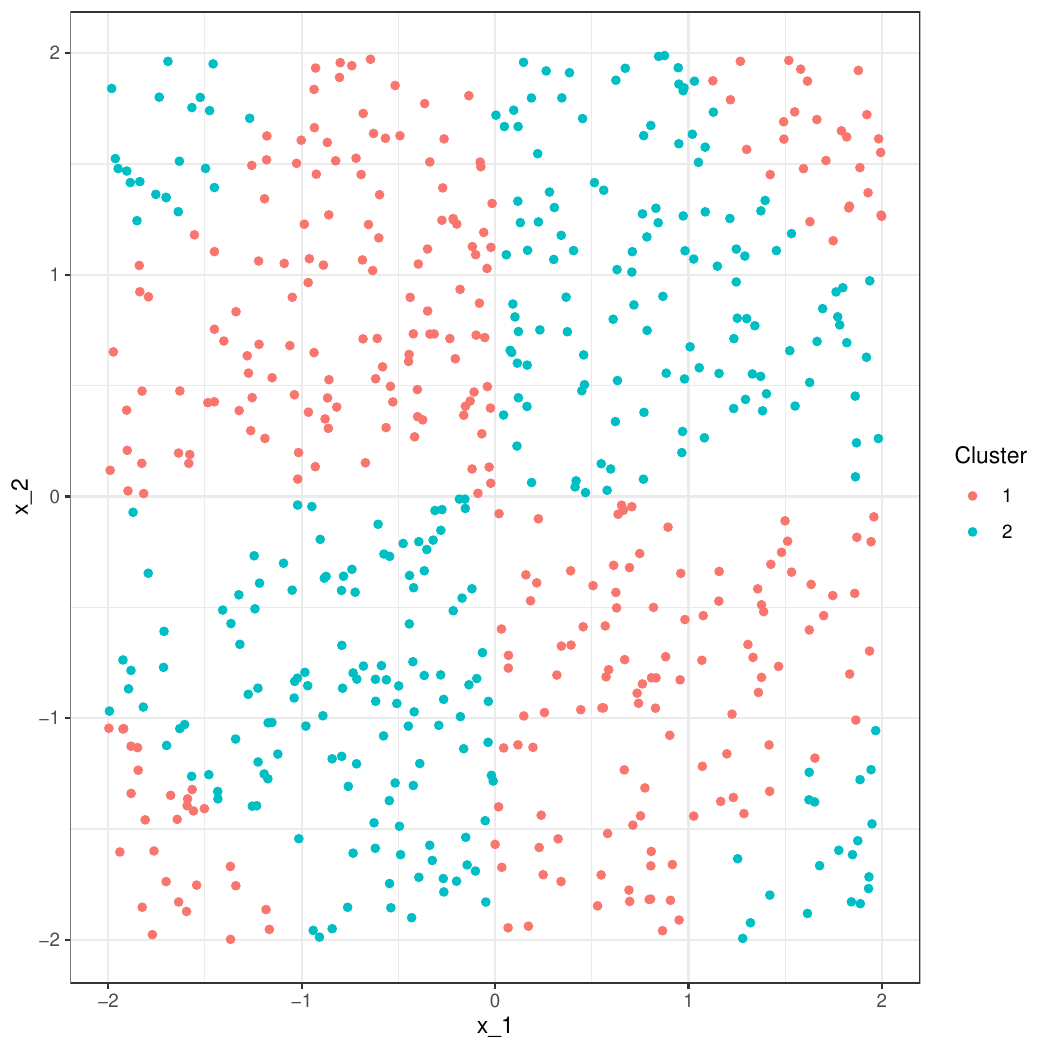}\label{fig:ex3_lddpbs_clusters}}
    \subfloat[LDDP]{\includegraphics[width=0.26\textwidth]{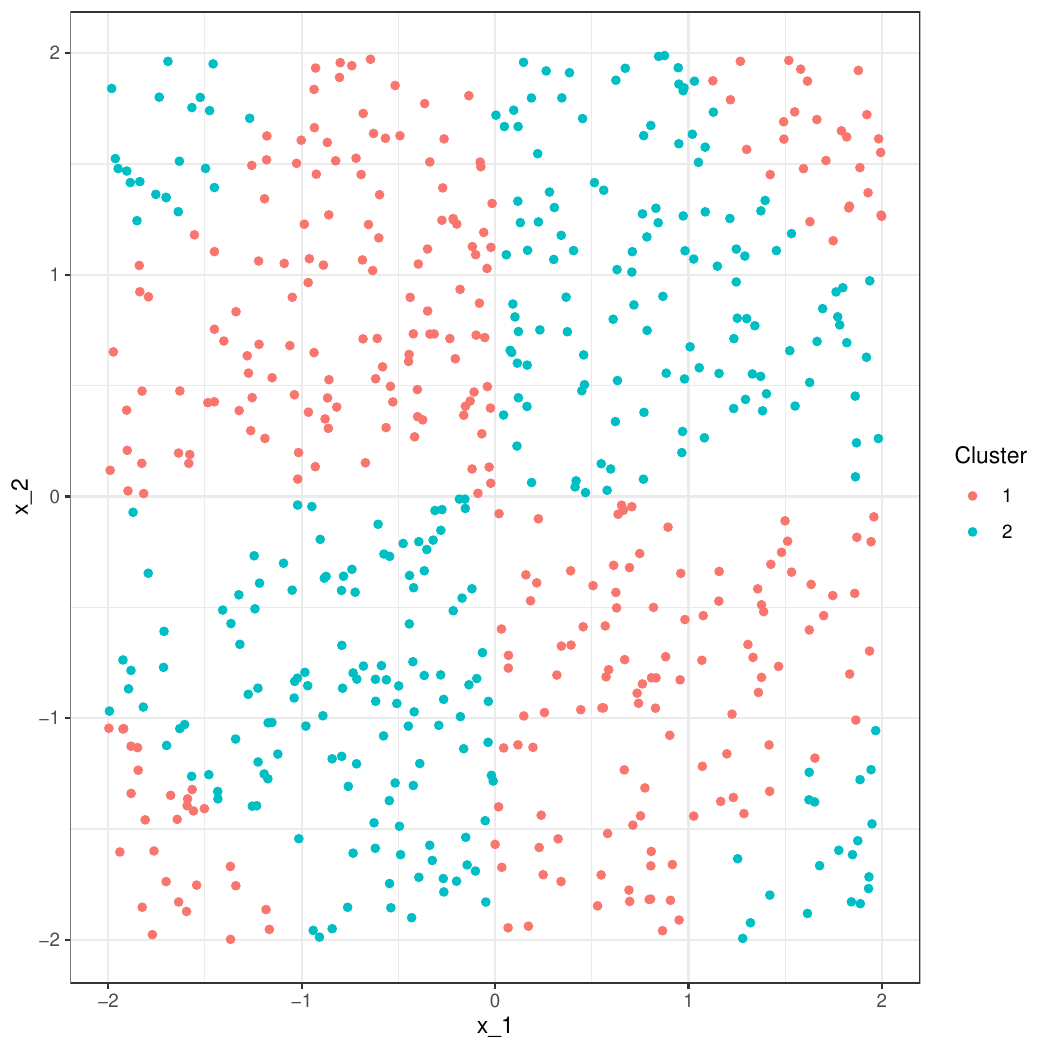}\label{fig:ex3_lddp_clusters}}
    \caption{Example 3. Estimated clustering}
    \label{fig:ex3_clusters}
\end{figure*}

\begin{figure*}[!h]
    \centering
    \subfloat[Truth]{\includegraphics[width=0.25\textwidth]{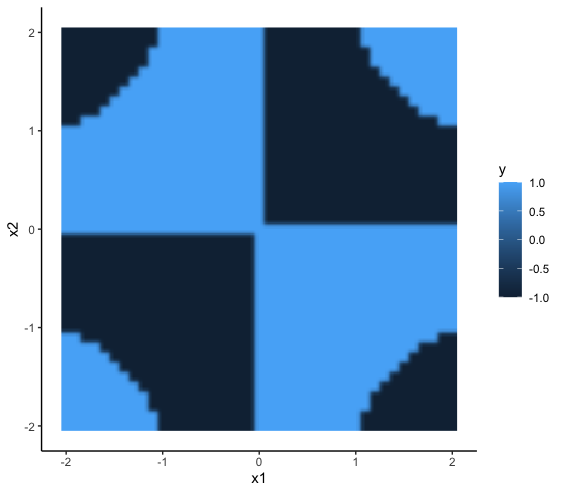}\label{fig:jtrue_pred}}
    \subfloat[Joint DP]{\includegraphics[width=0.25\textwidth]{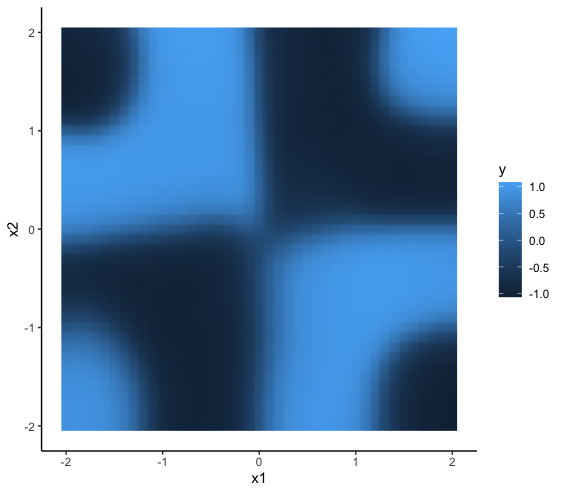}\label{fig:jointdp_pred}}
    \subfloat[Joint EDP]{\includegraphics[width=0.25\textwidth]{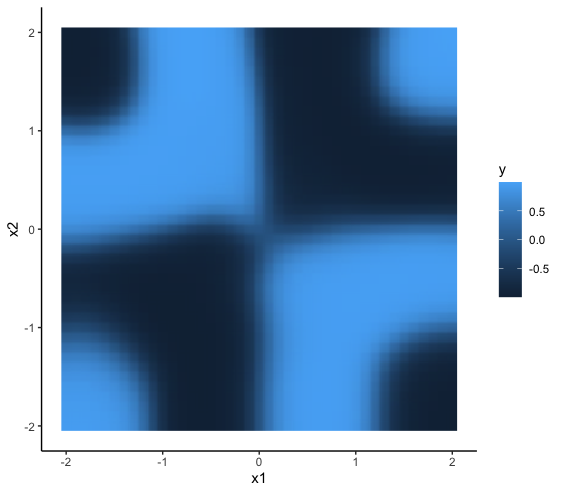}\label{fig:jointedp_predci}}
    \subfloat[NW]{\includegraphics[width=0.25\textwidth]{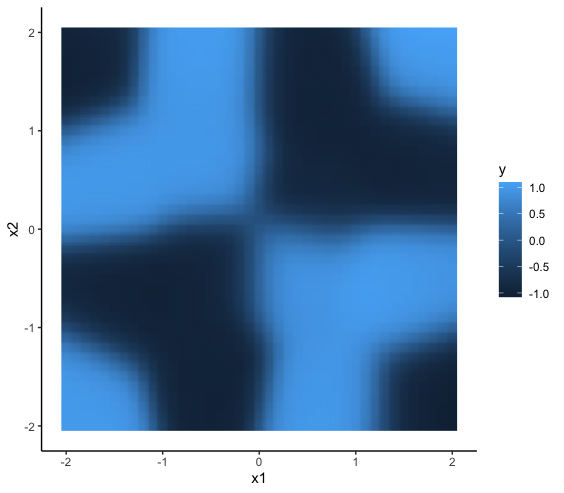}\label{fig:ex3_nwreg_pred}}\\
    \subfloat[LSBP]{\includegraphics[width=0.25\textwidth]{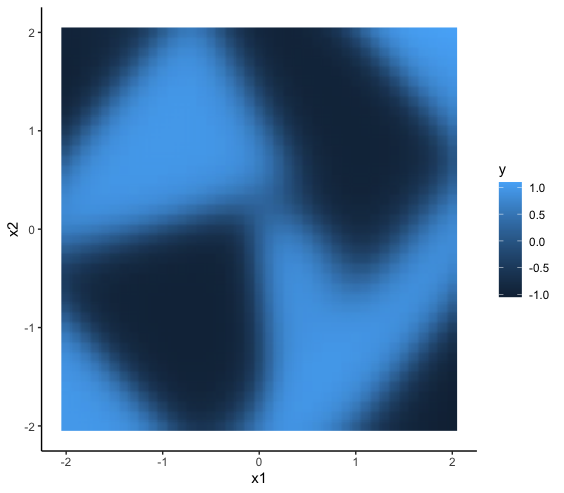}\label{fig:lsbp_predci}}
    \subfloat[LSBP-NS]{\includegraphics[width=0.25\textwidth]{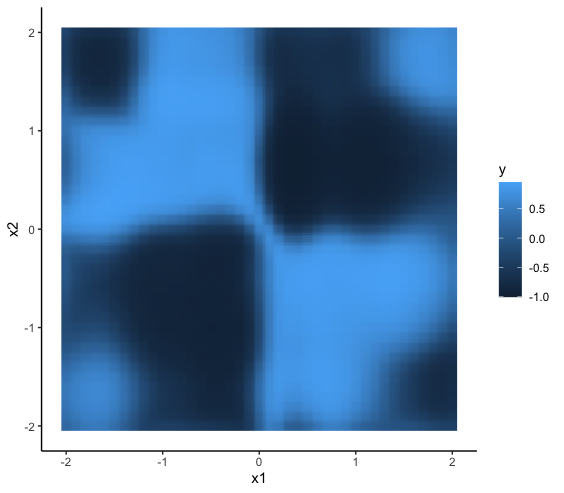}\label{fig:lsbp_predci}}
    \subfloat[LDDP-BS]{\includegraphics[width=0.22\textwidth]{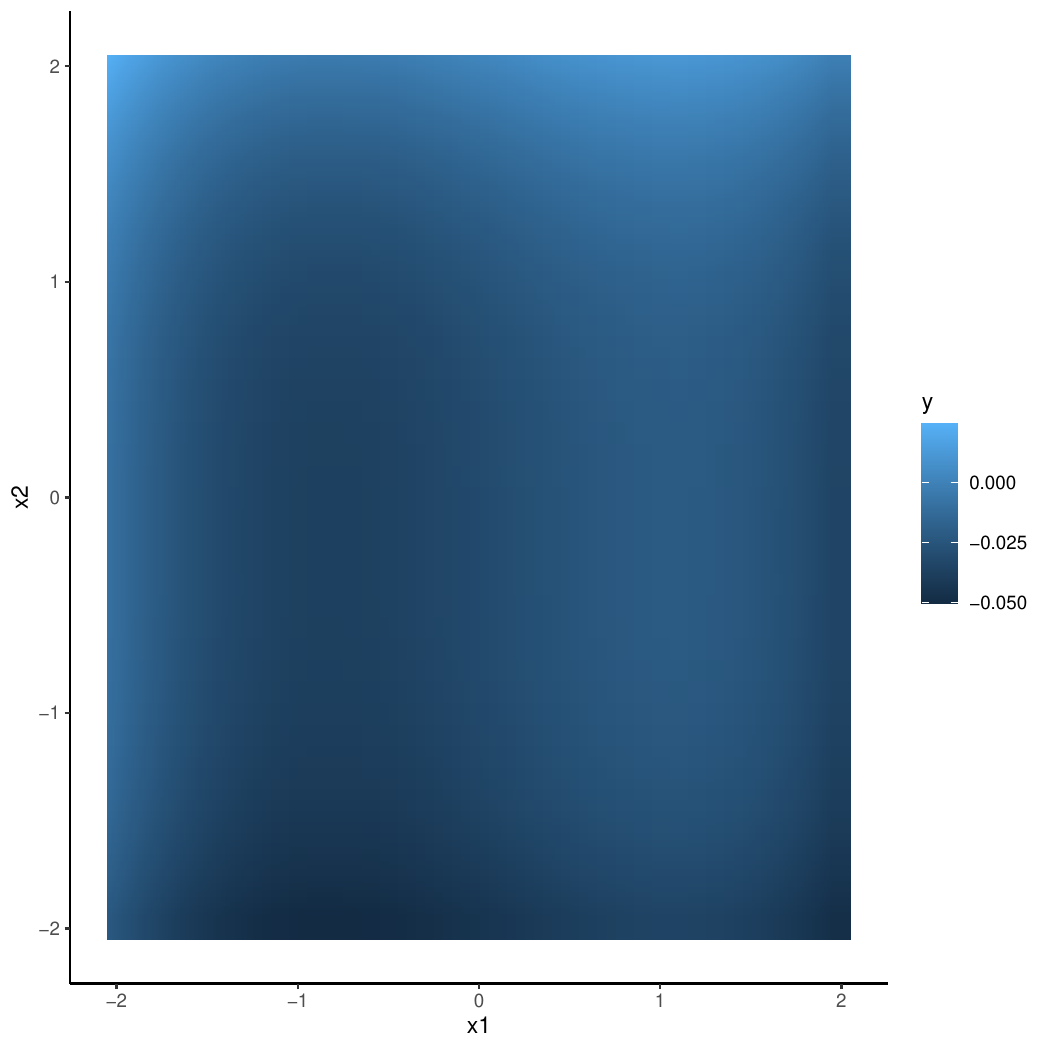}}\label{fig:lddp_predci}
    \subfloat[LDDP]{\includegraphics[width=0.22\textwidth]{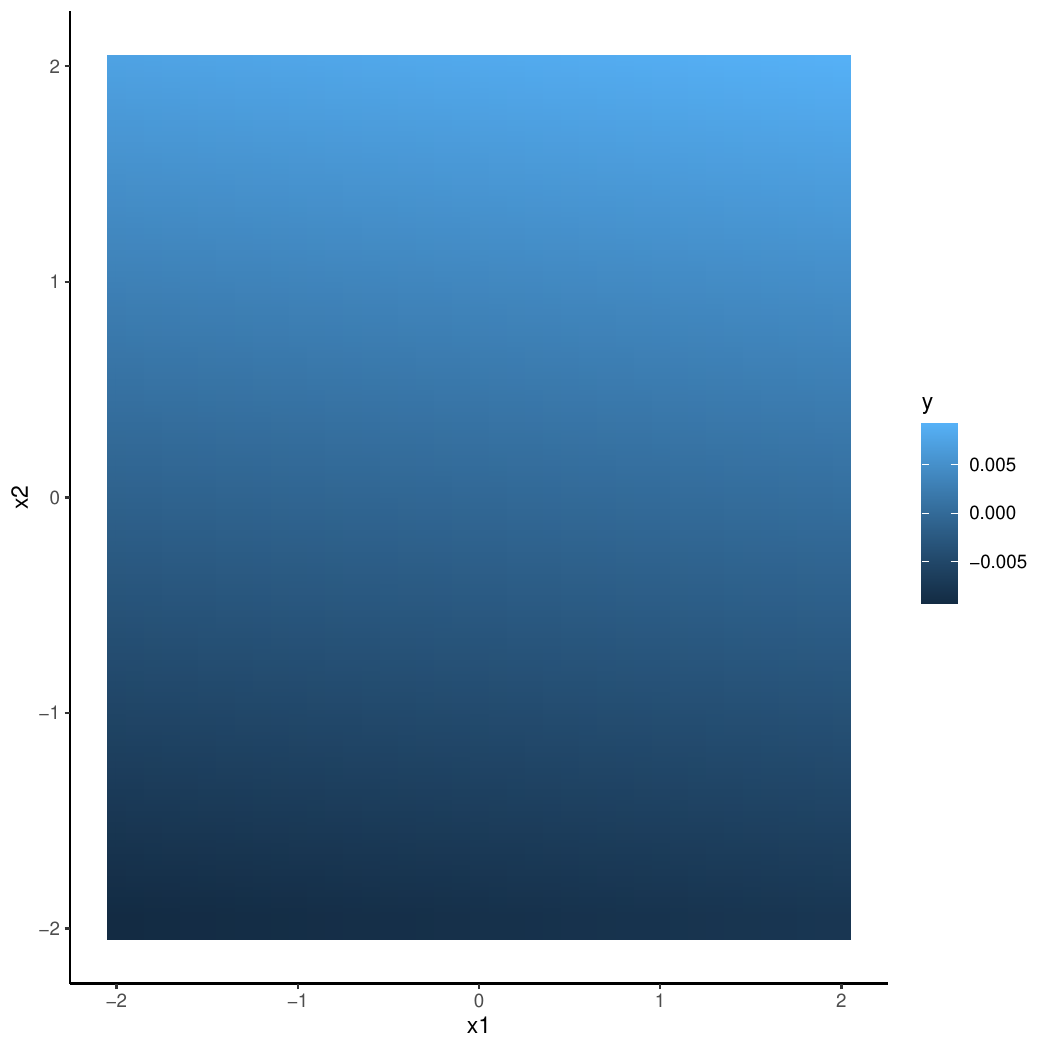}\label{fig:lddp_predci}}
    \caption{Example 3. Heatmap of the true and estimated predictive regression function.}
    \label{fig:ex3_pred}
\end{figure*}

\begin{figure*}[!h]
    \centering
    \subfloat[Joint DP]{\includegraphics[width=0.25\textwidth]{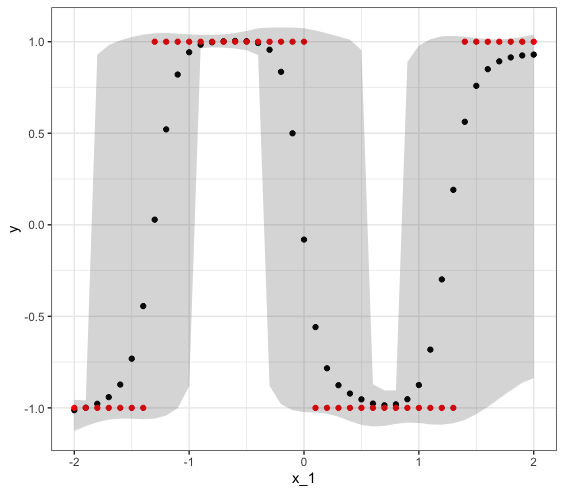}\label{fig:jointdp_predci}}
    \subfloat[Joint EDP]{\includegraphics[width=0.25\textwidth]{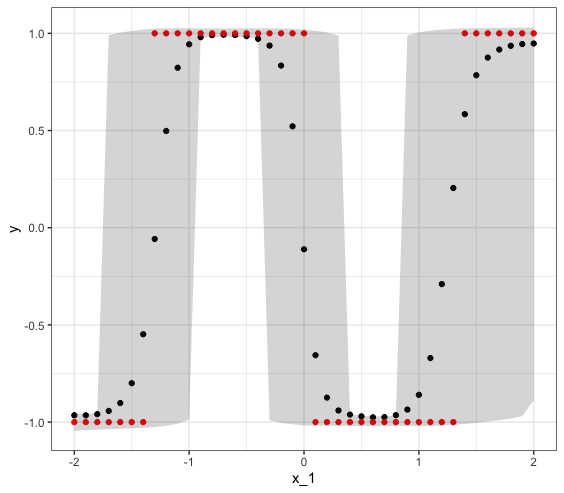}\label{fig:jointedp_predci}}
    \subfloat[NW]{\includegraphics[width=0.25\textwidth]{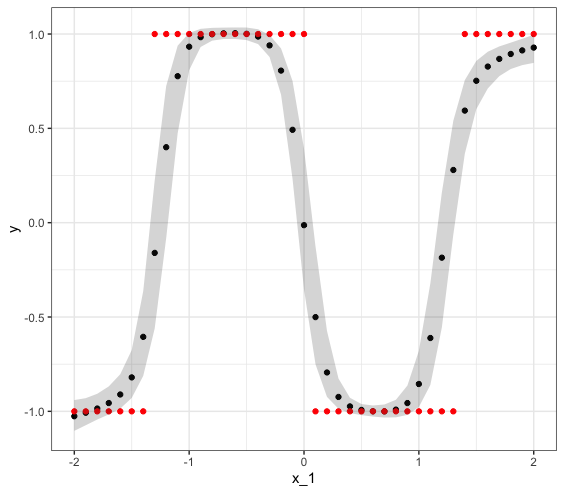}\label{fig:nwreg_predci}}\\
    \subfloat[LSBP]{\includegraphics[width=0.25\textwidth]{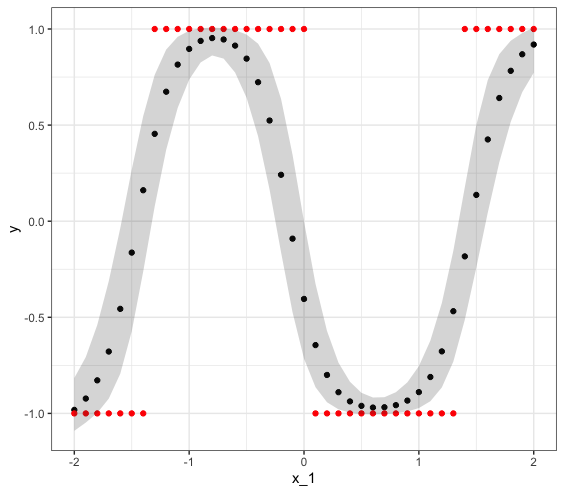}\label{fig:lsbp_predci}}
    \subfloat[LSBP-NS]{\includegraphics[width=0.25\textwidth]{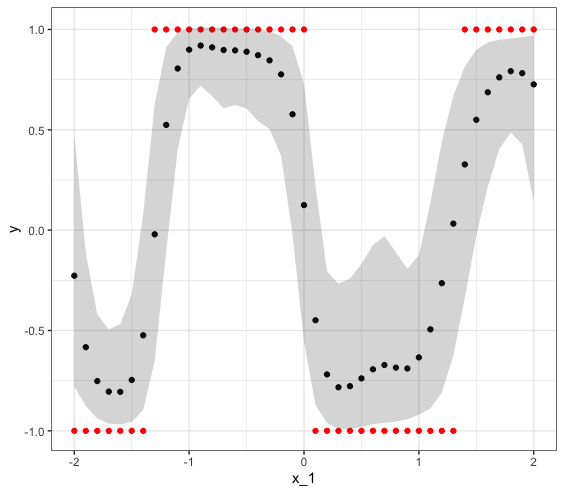}\label{fig:ex3_lsbpns_predci}}
    \subfloat[LDDP-BS]{\includegraphics[width=0.22\textwidth]{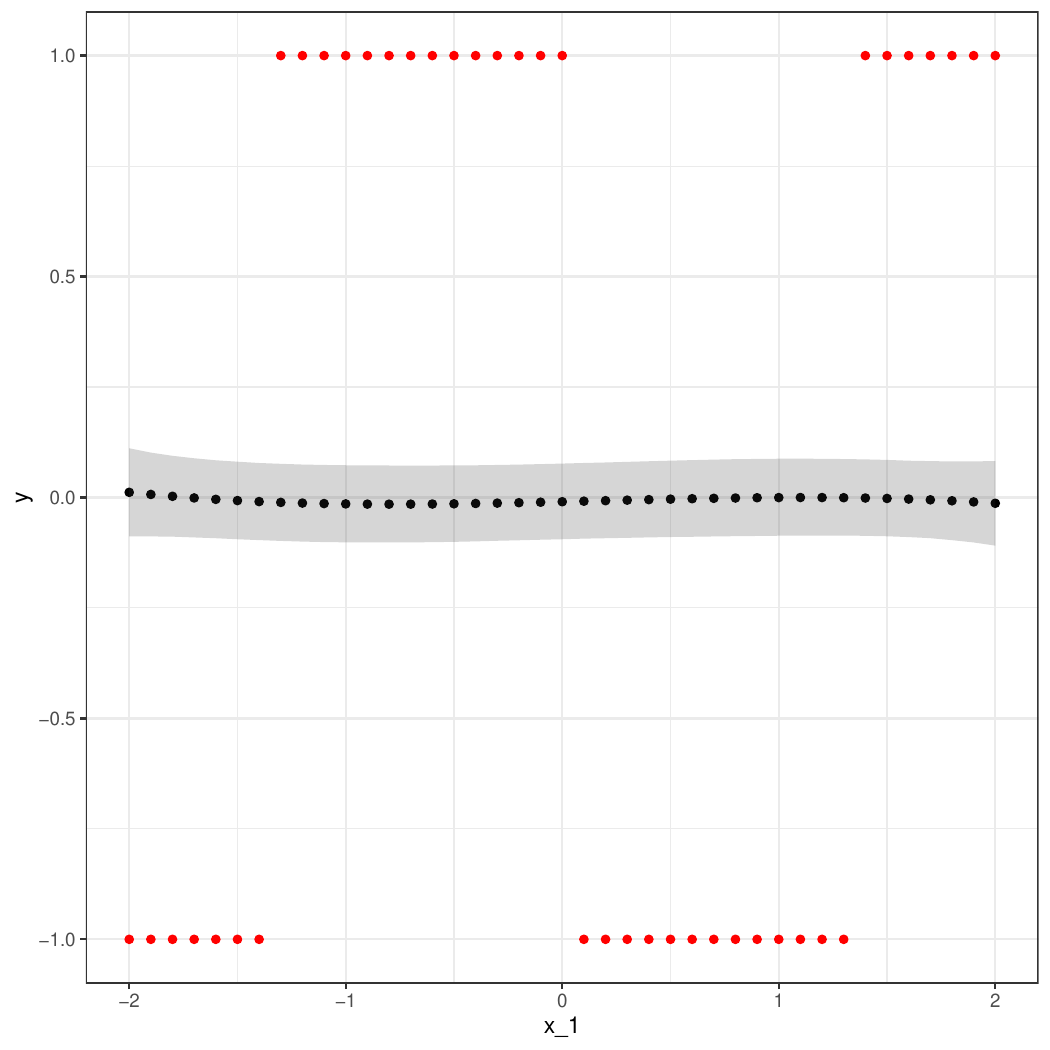}\label{fig:lddpbs_predci}}
    \subfloat[LDDP]{\includegraphics[width=0.22\textwidth]{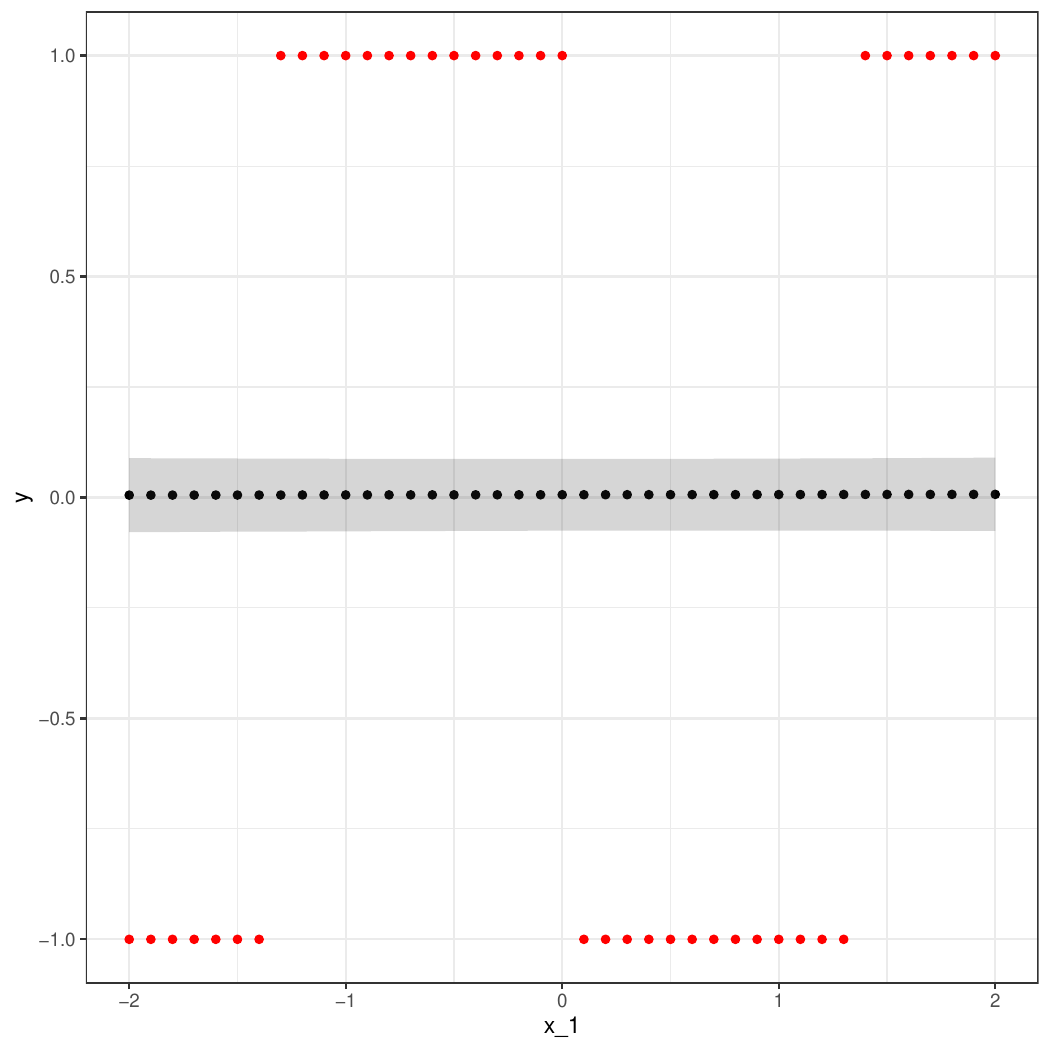}\label{fig:lddp_predci}}
    \caption{Example 3. Slice of the predictive regression function for $x_2 = 1.5$ with 95\% pointwise CIs. Red (black) dots denote the true (estimated) predictive regression function for $x_2 = 1.5$.}
    \label{fig:ex3_predci}
\end{figure*}

\begin{table}[!h]
    \centering
    \begin{tabular}{c|cccc}
          Model & Regression Err & Density Err & Coverage & CI length \\ \hline
          Joint DP & 0.4503  &  0.3865   & 0.9857 & 1.5216\\
          Joint EDP & 0.4336  & 0.2953 & 1 &  1.3619 \\
         NW & 0.4216 & 0.3834 & 0.4372   & 0.3317 \\
         LSBP & 0.5396 & 0.4345 & 0.2427 & 0.5274 \\
        LSBP-NS & 0.4903 & 0.4045 & 0.1999 & 0.7513 \\
        LDDP-BS & 1.001 & 1.0312 & 0 & 0.1760 \\
        LDDP & 1.0000 &1.0210 & 0 & 0.1634   
    \end{tabular}
    \caption{Summary of results for Example 3. 
    }
    \label{tab:ex3}
\end{table}

Dependent weights probabilistically partition the covariate space into regions where the local regression kernels provide a good fit, and thus are a natural choice. However, unlike the joint model and the single-weight approaches, powerful inference tools constructed for Bayesian mixture models can not be straightforwardly  used and bespoke, often expensive algorithms are required. Moreover, in popular stick-breaking constructions, dependence is defined at the level of nonlinear transformation of the weights. 
This makes it difficult to understand the implied dependence structure  between the weights and covariates, and in turn, choosing the hyperparameters and functional shapes required can be challenging.

To investigate this, we generate $n=600$ data points as follows:
\begin{align*}
 y_i &= m(x_i) + \epsilon_i, \quad \epsilon_i \overset{iid}\sim \Norm(0, 0.1^2), \\
 m(x_i) &= \left\lbrace \begin{array}{cc}
     1 & \sin(x_{i,1} x_{i,2} \pi/2) \leq 0 \\
     -1 & \text{ otherwise}
 \end{array} \right. ,
\end{align*}
with
$$x_{i,j} \overset{iid}\sim \Unif(-2,2), \text{  for } j=1,2.$$ Again, evaluation metrics are reported in Table \ref{tab:ex3}, the estimated clustering is depicted in Figure \ref{fig:ex3_clusters}, and a heatmap of the predictive regression function and a slice of the predictive regression function at $x_2= 1.5$ with 95\% pointwise CIs are shown in Figures \ref{fig:ex3_pred} and \ref{fig:ex3_predci}, respectively.

First, focusing on the models with dependent weights, we observe that the LSBP with linear dependence in the stick-breaking proportions does not perform well, and the LSBP-NS with more flexible dependence defined through the spline basis expansion provides significant improvements in predictions. In this case, we selected four knots at suitably chosen quantiles for both covariates, and explored increasing to seven knots (not shown for conciseness), which slightly decreased predictive performance.  
Moreover, as further described in the Appendix,  the prior on the parameters involved in the stick-breaking proportions is especially relevant. For all three experiments, we have employed the same multivariate normal prior on the logit stick-breaking regression coefficients with zero mean and diagonal covariance matrix with diagonal $(100, 10,\ldots, 10)$. With this prior choice, predictions appear to be overly smoothed across components and uncertainty quantification, with empirical coverage at 0.1999, is poor (Figure \ref{fig:ex3_lsbpns_predci}). Increasing the prior variance, drastically improves prediction and uncertainty quantification (Appendix \ref{sec:lsbp_prior}) in this example, but we note that it worsens the results in the previous two examples. For the model with normalized weights, we utilize Gaussian kernels, with the prior on the location and scale set empirically based on the mean and variance of the covariates. The  regression function is estimated well (Figure \ref{fig:ex3_nwreg_pred}), but the empirical coverage of 0.4372 is again too low, and, similar to the LSBP-NS, alternate prior choices, e.g. that encourage larger scale parameters, may help to improve predictive inference. 
Thus, the results highlight how selecting both the form of the nonlinear dependence  and the prior on the parameters  in the stick-breaking proportions can greatly affect predictions and uncertainty, yet due to lack of interpretability, determining these in practice can be challenging. 

The joint DP model requires many clusters (Figure \ref{fig:ex3_jointdp_clusters}) due to the complex joint relationship, while the estimated clustering of the joint EDP contains only two clusters (Figure \ref{fig:ex3_jointedp_clusters}) that reflect the data-generating mechanism. This leads to improved predictive performance for the joint EDP. However, although the empirical coverage, of 0.9857 and 1 for the joint DP and EDP models, respectively, achieves the desired level, the intervals, with an average length of 1.5216 and 1.3619, respectively, are too wide. Again, this may improve with alternate prior choices that encourage smaller within cluster variance.  

Lastly, predictive performance is extremely poor for the single-weight models. Although the estimated clustering in Figures \ref{fig:ex3_lddpbs_clusters} and \ref{fig:ex3_lddp_clusters} reflects the data-generating mechanism, it is covariate-dependent.  As explained in Example 2, the cluster-specific predictions are then averaged regardless of the covariate value, resulting in the observed poor prediction. We note that the LDDP-BS performance did not improve with the use of more interior knots for both $x_1$ and $x_2$ and, indeed, two model selection criteria (the widely applicable information criterion and the log pseudo marginal likelihood) favour the specification with no interior knots over specifications that considered an increasing number of interior knots, placed at the quantiles of the covariates.

\section{Concluding Remarks}

\begin{figure*}[!h]
    \centering
\includegraphics[scale=0.5]{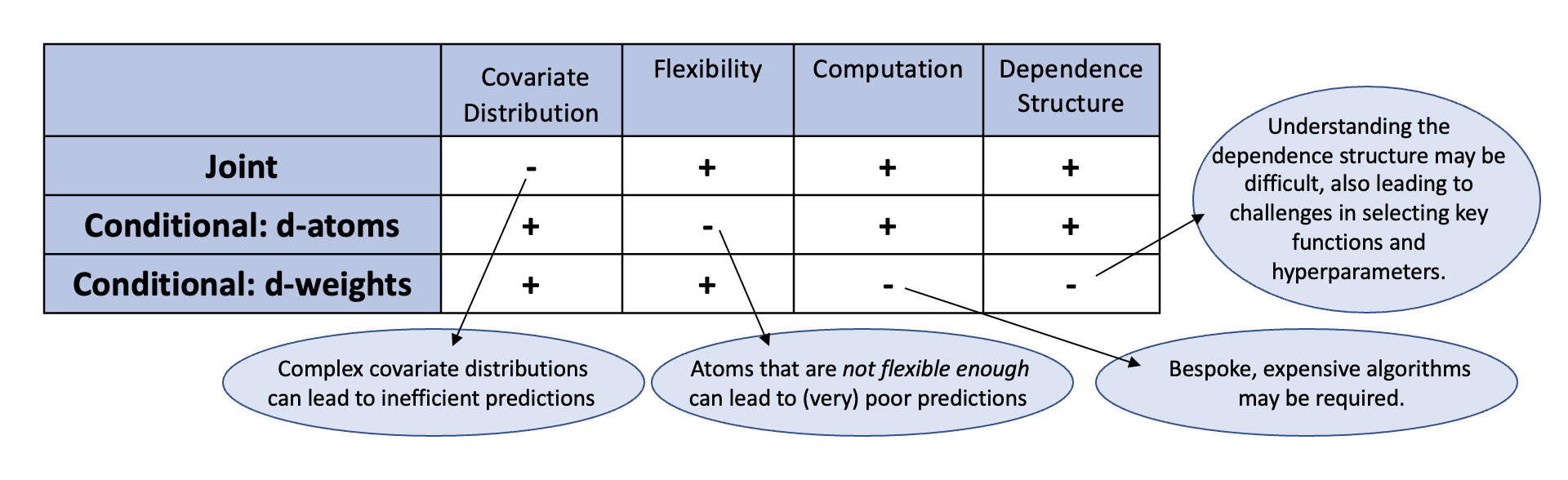}
    \caption{A comparison of the three general approaches for Bayesian density regression.}
    \label{fig:comp}
\end{figure*}

Bayesian dependent mixture models provide flexible density regression to capture many challenges and complexities of modern data. 
Such models are numerous and we have broadly categorized them into three main types: 1) joint models, 2) conditional models with single weights and dependent atoms, and 3) conditional models with flexible weights. Another important class is comprised of models based on covariate-dependent random partition models or urn schemes. In a specific case, such models correspond to the joint model, but in general, they are in the flavor of models with dependent weights. In addition, within each model type, the number of model and prior choices is large, and deciding among them can be challenging. By careful examination of the effects of such choices on prediction and through pragmatic comparisons, we have shed light on the advantages and disadvantages of the different models in order to guide practitioners in their choice.

First, the joint modeling approach 
has the advantage of computational simplicity and performs well in practice from a predictive perspective. 
The drawbacks are shown in our experiments; specifically, when the joint density is complex, this can lead to over-partitioning and small clusters, producing (slightly) less efficient estimates, degraded predictive performance, and unnecessarily wide credible intervals. Using a more flexible prior choice, such as the enriched Dirichlet process, which allows for two-level clustering, can help to rectify this behavior.

The conditional approach, on the other hand, has the advantage of modeling the conditional density directly, which can lead to improved prediction. 
When single weights are assumed, computations are straightforward, making use of standard tools for mixture models. However, as shown in our experiments, flexibility in the atoms is crucial for flexible density regression.  
Yet, increasing flexibility in the atoms, increases the computational burden and interpretations may become increasingly difficult. 
Moreover, a critical limitation highlighted in our experiments is that (extremely) poor prediction may result, when the atoms are insufficiently flexible and, in order to fit the data, the estimated partition structure depends on the covariates. In this case, the predictions across each cluster are averaged regardless of the covariate value. We emphasize that when using such models in practice, it is important to examine the partition aposteriori to ensure it does not depend on the covariates. However, this may be challenging in modern datasets, with e.g. multivariate and mixed covariates. Developing tools to quantify and test for dependence between the random partition and covariates is  an open direction of future research. 

The conditional approach with covariate-dependent weights is flexible and performs well across our experiments. As for the joint model, these models imply a covariate-dependent partitioning of the data, which can greatly improve prediction. However, unlike the joint model, posterior inference is based directly on the conditional likelihood of interest. Drawbacks include burdensome computations, and 
a lack of interpretation of the dependence structure in the weights, especially for the widely-used stick-breaking constructions, which amplifies the difficulty in selecting the functional shapes and hyperparameters required.  
In fact, the logit stick-breaking process is among the top performing models in our experiments, if the order of the splines expansion and prior on the stick-breaking coefficients are selected \textit{appropriately}. Indeed, performance can change drastically based on these choices, and in general, we find that higher orders are beneficial when the relationship between the partition structure and the covariates is anticipated to be complex.  In addition, large prior variances encourage larger  stick-breaking coefficients, and thus, sharper boundaries between clusters and less smoothing across components. Given the importance on prediction and due to the lack of interpretation and empirical specification, selection of both the order and prior variances, either through information criteria or hierarchically, would greatly aid practitioners. Of course, such choices must be made for each covariate dimension and therefore, may be more challenging in multivariate settings. On the other hand, conditional models with normalized weights define the dependence structure directly on the weights, leading to greater interpretation and the possibility of subjective or empirical specification of certain parameters. However, the intractable normalizing constant in the definition of the weights makes these models the most computationally expensive. Finally, to conclude, in Figure \ref{fig:comp} we provide a schematic summary comparison of the three approaches.

Returning to the data presented in Section \ref{sec:intro} and analyzed in \cite{wade2022colombian}, we can apply the guidance and lessons learned from the analysis and experiments. The exploratory analysis suggests a mildly nonlinear mean regression between the age-at-event responses and the continuous covariate (age at interview), non-Gaussianity as well as variance and tail behavior that changes with covariates. Such challenges may be problematic for the single-weights models, and at the very least, the LDDP should not be used. The joint models are better suited to this setting, and allow imputation of missing covariate values. The conditional models with dependent weights are also appropriate, considering the number of responses and covariates, and indeed this is the strategy followed in \cite{wade2022colombian}, who use normalized weights. Moreover, both the joint and dependent weight models can provide flexible density regression by building on linear regression models, making it easier to adapt the models to account for the censoring and constraints in the responses.

Moving towards the next steps, high-dimensional datasets are becoming increasingly abundant and pose computational \citep[see Chapter 5 of][]{Fruhwirth2019} and theoretical \cite{Kiran2023} challenges to mixture models. For instance, in the unconditional case, \cite{Kiran2023} noted that care is needed in specifying both the kernel and the base measure for the atoms in high-dimensions, otherwise the posterior can degenerate on extreme clustering structures. This led the authors to propose a class of latent factor mixture models that is amenable to scalable inference and can avoid the pitfalls of high-dimensionality under mild assumptions. Within the models we have reviewed, the enriched DP model is a simple adaptation of the joint model to deal with its shortcomings in high-dimensions and computations remain relatively simple. However, since the number of $x$-kernels is likely to be large in high-dimensions, computations may become burdensome for increasing $p$. This effect clearly depends on the dataset and further work is needed to explore it. A possible extension for future research is to combine the enriched DP mixture model with dimension reduction techniques. On the other hand, the model based on normalized weights is methodologically attractive, but may not be well suited to large high-dimensional problems for computational reasons. In particular, although exact posterior sampling is available via the introduction of latent variables, the number of latent variables required increases with $p$. Further work is needed to explore the behavior of the model and algorithm in high-dimensions and, if needed, to develop possible extensions in this setting. Lastly, for the LDDP models based on a B-splines expansion of the continuous covariates, in large $p$ settings, the number of regression coefficients per component will be very large and so some form of dimension reduction or variable selection is mandatory.


\begin{appendix}
\section*{Appendix}

\subsection{Empirical prior specification}\label{sec:lddp_bs_prior}
\begin{figure*}[!h]
\centering
 \subfloat[Predictive regression function]{\includegraphics[width=0.25\textwidth]{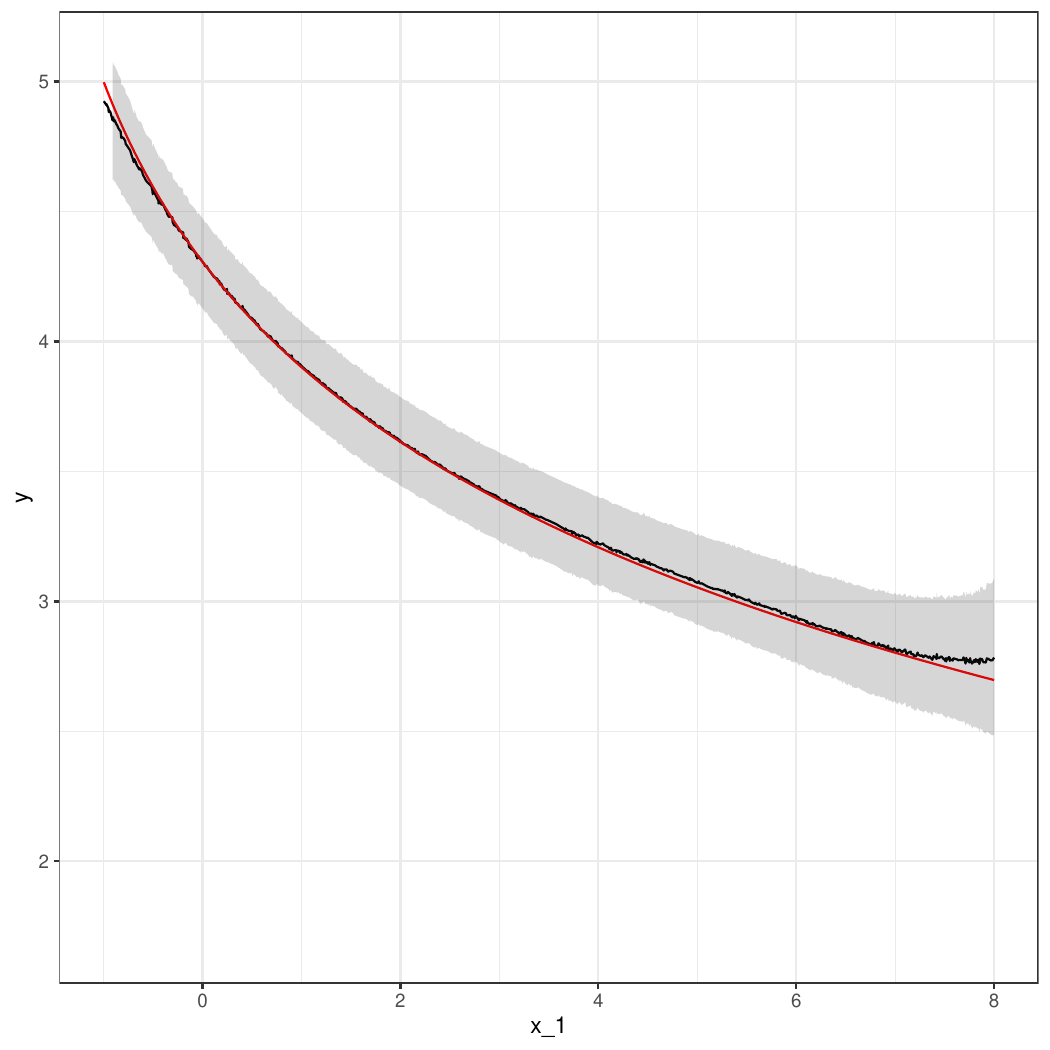}}\label{fig:lddpbs_densities}
 \subfloat[Predictive density regression]{\includegraphics[width=0.25\textwidth]{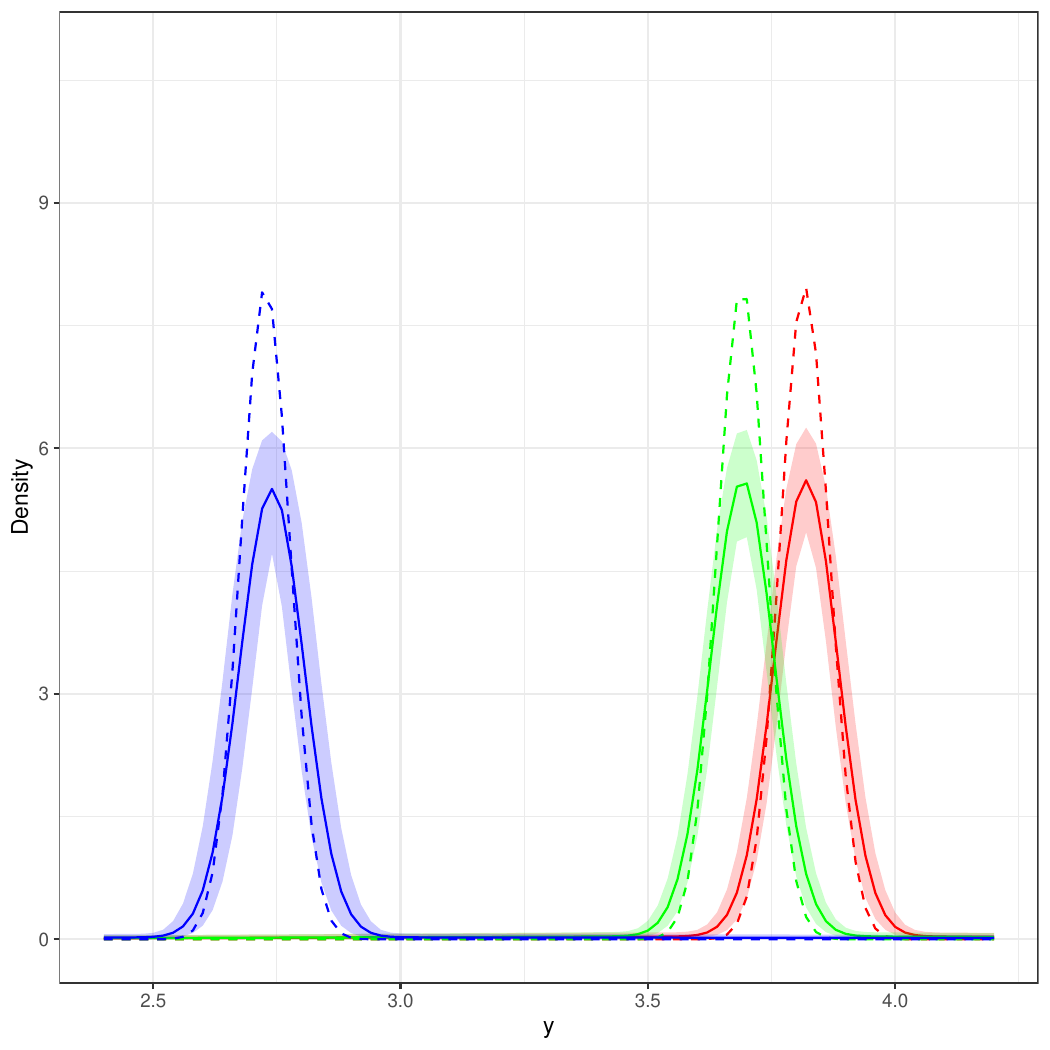}}\label{fig:lddpbs_densities}
  \caption{Example 1. Predictive regression function and density regression for the LDDP-BS with a `non-informative' prior specification for the parameters of the baseline measure.}
   \label{fig:ex1_vague_prior}
\end{figure*}

We illustrate the impact that the choice of the hyperparameter values can have on the results. We concentrate on the LDDP-BS formulation in Example 1, which performed the best in this scenario.  In all three examples, we have used
\begin{equation*}
\tilde{\mu}_j(x) = \tilde{\beta}_j^{\prime}\lambda(x),
\end{equation*}
where $\lambda(x)$ is the cubic B-splines basis formulation, with no interior knots. A conjugate baseline measure was considered
\begin{equation*}
(\tilde{\beta}_j,\tilde{\sigma}^{-2}_j) \overset{\text{iid}}\sim\text{N}(m,S)\Gamma(a,b),
\end{equation*}
with conjugate hyperpriors
\begin{equation*}
m\sim\text{N}(m_0,S_0),\quad S^{-1}\sim\text{Wishart}(\nu,(\nu\Psi)^{-1}).
\end{equation*}
Here $a$ and $b$ denote, respectively, the shape and rate parameters of the gamma distribution. Hyperparameters $m_0$ and $\Psi$ must be chosen to represent the prior belief about the regression coefficients associated to each mixture component and about their covariance matrix, respectively, whereas $S_0$ and $\nu$ are chosen to represent the confidence in the prior belief of $m_0$ and $\Psi$, respectively. In all results presented in Section \ref{sec:comparison}, the data (response and covariates) were analyzed in the original scale and the following data-driven hyperparameter values were specified
\begin{align*}
 m_0 & = \widehat{\beta},\quad S_0=\widehat{\Sigma},\quad \nu = Q + 2, \quad  \Psi = 30\widehat{\Sigma},\\
 a & = 2,\quad b=\widehat{\sigma}/2,
\end{align*}
where $\widehat{\beta}$ and $\widehat{\sigma}$ are the least squares estimates from fitting the linear model $y_i = \tilde{\beta}^{\prime}\lambda(x_i) +\sigma\varepsilon_i$, where $\mathbb{E}(\varepsilon_i)= 0$ and $\text(\varepsilon_i) = 1$, and $\widehat{\Sigma}$ is the estimated covariance matrix of $\widehat{\beta}$. For the case of two continuous covariates, both modeled via a cubic B-splines basis expansion with no interior knots, $Q$ is equal to $7$. In addition, we have also standardized both the response and the covariates (by subtracting their mean and dividing by their standard deviation) and considered the following hyperparameter values (on the standardized scale)
\begin{align*}
 m_0 & = 0_Q,\quad S_0=10 I_Q,\quad \nu = Q + 2, \quad  \Psi = I_Q,\\
 a & = 2,\quad b=0.5.
\end{align*}
We further set, in both cases, $\alpha = 1$ and used the blocked Gibbs sampler of \cite{IJ} capping the number of mixture components to $20$. The graphical results for this second configuration of hyperparameter values are presented in Figure \ref{fig:ex1_vague_prior}. As can be observed, the pointwise $95\%$ credible band for the predictive regression function is much more wider and the estimated conditional density functions do not recover the corresponding true ones so well as when considering the data-driven prior. This is obviously also reflected in the computed regression and density errors. Under this `non-informative' prior configuration, the root mean squared error between the predictive regression function and the truth is $0.0206$,  the empirical coverage of the predictive regression function is 1, the average CI length of the predictive regression function is $0.3710$, and the $\ell_1$ distance between the predictive and true conditional density averaged across all test points is $0.3663$. By comparison, the corresponding values when considering the data-driven hyperparameter values are, respectively, $0.0117$, $0.9563$, $0.0369$, and $0.1522$. 

\subsection{Prior specification in stick-breaking dependent weights}\label{sec:lsbp_prior}

\begin{figure*}[!h]
    \centering
    \subfloat[Truth]{\includegraphics[width=0.25\textwidth]{plots/ex3_true_pred_v2.png}\label{fig:jtrue_pred}}
    \subfloat[LSBP]{\includegraphics[width=0.25\textwidth]{plots/ex3_lsbp_pred_v2.png}\label{fig:lsbp_predci}}
    \subfloat[LSBP - Prior 2]{\includegraphics[width=0.25\textwidth]{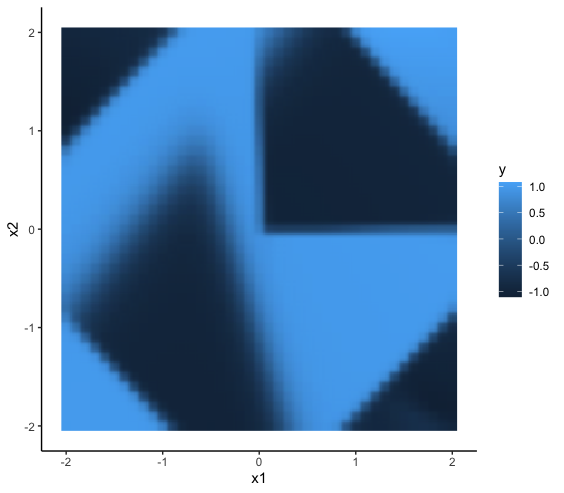}\label{fig:lsbp_predci}}
    \subfloat[LSBP - Prior 3]{\includegraphics[width=0.25\textwidth]{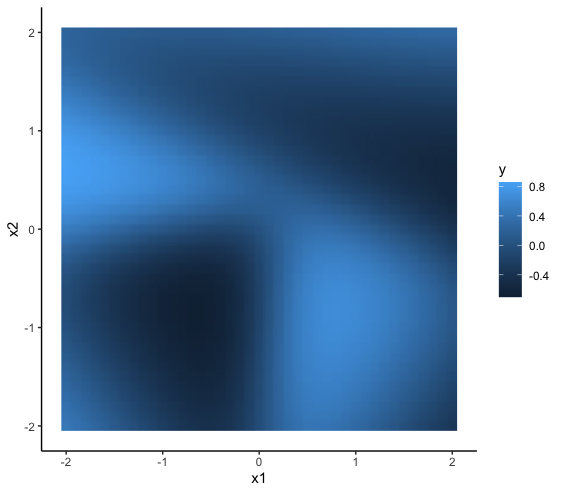}\label{fig:lsbp_predci}}\\
     \subfloat[LSBP]{\includegraphics[width=0.25\textwidth]{plots/ex3_lsbp_pred_wci_v2.png}\label{fig:lsbp_predci}}
    \subfloat[LSBP - Prior 2]{\includegraphics[width=0.25\textwidth]{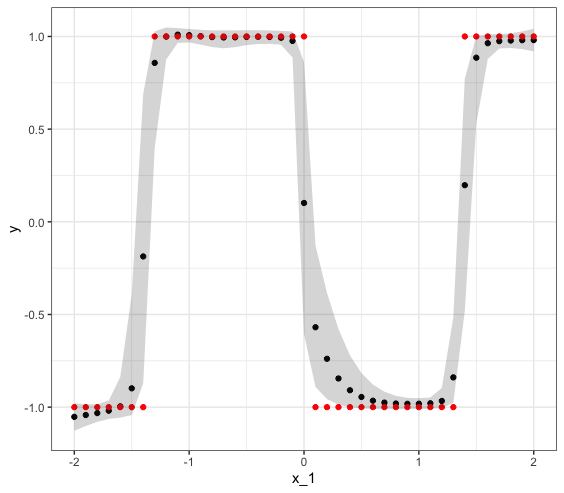}\label{fig:lsbp_predci}}
    \subfloat[LSBP - Prior 3]{\includegraphics[width=0.25\textwidth]{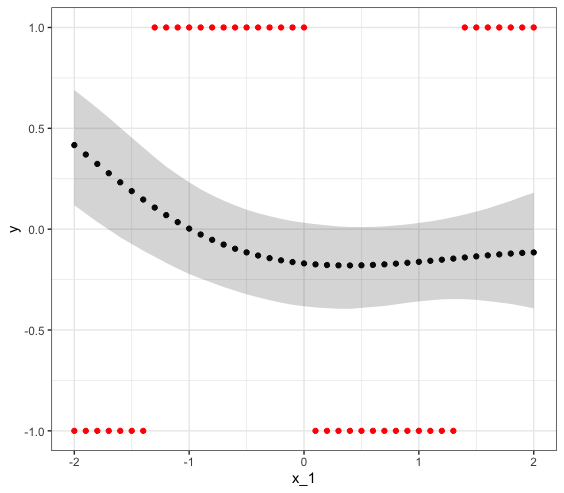}\label{fig:lsbp_predci}}
    \caption{Example 3. Heatmap of the true and estimated predictive regression function (top row) and slice at $x_2=1.5$ with CI (bottom row) for the LSBP with three different prior choices for the stick-breaking parameters. Red (black) dots denote the true (estimated) predictive regression function for $x_2 = 1.5$.}
    \label{fig:ex3_lsbp}
\end{figure*}

\begin{figure*}[!h]
    \centering
    \subfloat[Truth]{\includegraphics[width=0.25\textwidth]{plots/ex3_true_pred_v2.png}\label{fig:jtrue_pred}}
    \subfloat[LSBP-NS]{\includegraphics[width=0.25\textwidth]{plots/ex3_lsbpns5_pred_v2.png}\label{fig:lsbp_predci}}
    \subfloat[LSBP-NS - Prior 2]{\includegraphics[width=0.25\textwidth]{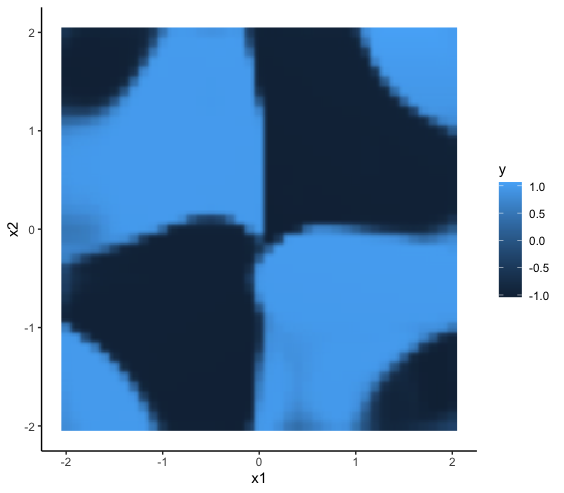}\label{fig:lsbp_predci}}
    \subfloat[LSBP-NS - Prior 3]{\includegraphics[width=0.25\textwidth]{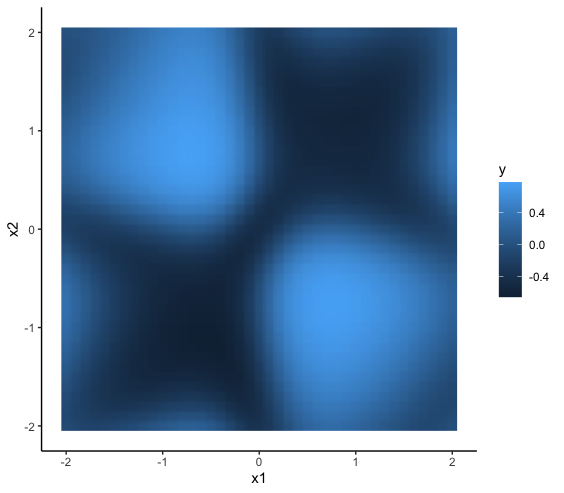}\label{fig:lsbp_predci}}\\
     \subfloat[LSBP-NS]{\includegraphics[width=0.25\textwidth]{plots/ex3_lsbpns5_pred_wci_v2.png}\label{fig:lsbp_predci}}
    \subfloat[LSBP-NS - Prior 2]{\includegraphics[width=0.25\textwidth]{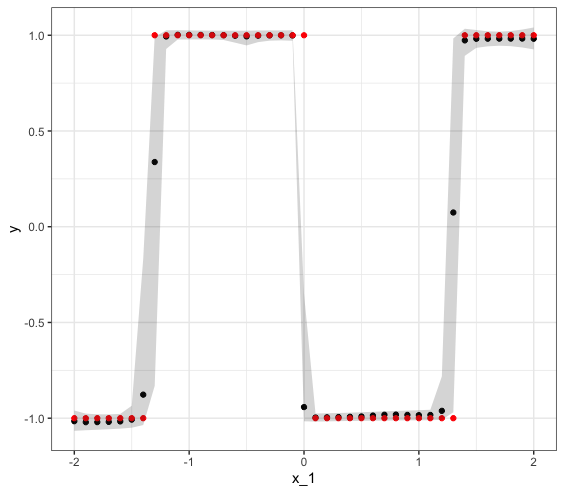}\label{fig:lsbp_predci}}
    \subfloat[LSBP-NS - Prior 3]{\includegraphics[width=0.25\textwidth]{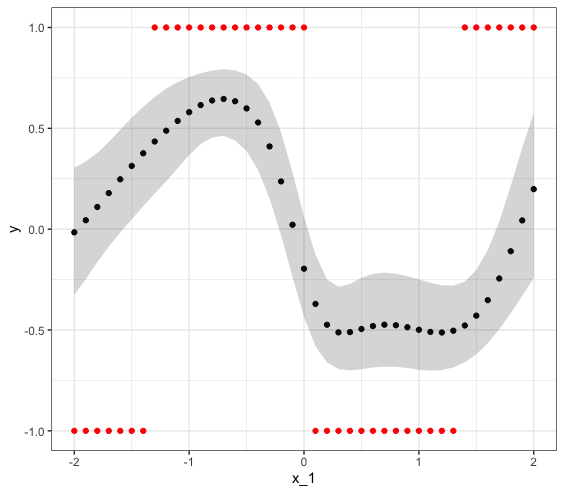}\label{fig:lsbp_predci}}
    \caption{Example 3.  Heatmap of the true and estimated predictive regression function (top row) and slice at $x_2=1.5$ with CI (bottom row) for the LSBP-NS with three different prior choices for the stick-breaking parameters. Red (black) dots denote the true (estimated) predictive regression function for $x_2 = 1.5$.}
    \label{fig:ex3_lsbpns}
\end{figure*}

Prior specification for the parameters involved in the stick-breaking construction of the dependent weights can be challenging due to difficulties in interpreting the effects of these parameters. To highlight this, we explore three different choices of priors for the coefficients in the logit stick-breaking process. Recall that in the logit stick-breaking, the stick-breaking proportions are defined as:
\begin{align*}
    v_j(x) = l(\tilde{b}_j^\prime \lambda(x)),
\end{align*}
where $\lambda(x)$ is simply $(1,x^\prime)^\prime$ in the linear construction (LSBP) or is the natural cubic spline basis with $4$ knots at suitably chosen quantiles in LSBP-NS. The prior for $\tilde{b}_j$ is a multivariate normal, with zero mean in all prior settings and a diagonal covariance matrix with diagonal $(100,10,\ldots, 10)$ in the first prior (P1), $(10^4,\ldots, 10^4)$ in the second case (P2), and  $(1,\ldots, 1)$ in the third case (P3). The results, visualized in Figure \ref{fig:ex3_lsbp} for the LSBP and Figure \ref{fig:ex3_lsbpns} for the LSBP-NS and summarized in Table \ref{tab:ex3_lsbp}, clearly highlight how the performance of the model changes drastically across the different prior choices.   

\begin{table}[!h]
    \centering
    \begin{tabular}{c|ccc}
          Model & Regression Err  & Coverage & CI length \\ \hline
          LSBP & 0.5396  & 0.2427 & 0.5274 \\
          LSBP (P2) & 0.5007   & 0.6597 & 0.3567  \\
          LSBP (P3) & 0.8079 & 0 &  0.4817 \\ \hline          
        LSBP-NS & 0.4903  & 0.1999 & 0.7513 \\LSBP-NS (P2) & 0.3653  & 0.9447  &0.3689 
        \\LSBP-NS (P3) & 0.7866  &  0 & 0.5062
    \end{tabular}
    \caption{Summary of results for Example 3 for the LSBP and LSBP-NS with three different prior choices for the stick-breaking parameters. 
    }
    \label{tab:ex3_lsbp}
\end{table}

\end{appendix}

\bibliographystyle{imsart-number} 
\bibliography{main}

\end{document}